\documentclass[11pt]{amsart}
\pdfoutput=1

\usepackage{amssymb}
\usepackage{amsthm}
\usepackage{amsfonts}
\usepackage{amsmath}
\usepackage{mathrsfs}
\usepackage{pmboxdraw}
\usepackage{verbatim} 
\usepackage{graphicx}
\usepackage{color}
\usepackage[colorlinks=true, citecolor=blue, filecolor=black, linkcolor=black, urlcolor=black]{hyperref}
\usepackage{cite}
\usepackage[normalem]{ulem}
\usepackage{subcaption}
\usepackage{bbm}
\usepackage{bm}
\usepackage{mathtools}
\usepackage{todonotes}
\usepackage{multirow,makecell}  
\usepackage{tablefootnote}

\usepackage{xparse}

\ExplSyntaxOn
\NewDocumentCommand{\MeijerG}{smmmm}
 {
  \IfBooleanTF{#1}
   {
    \vic_meijerg:nnnnnn { #2 } { #3 } { #4 } { #5 } { small } { }
   }
   {
    \vic_meijerg:nnnnnn { #2 } { #3 } { #4 } { #5 } { } { \; }
   }
 }

\seq_new:N \l__vic_meijerg_args_in_seq
\seq_new:N \l__vic_meijerg_args_out_seq

\cs_new_protected:Nn \vic_meijerg:nnnnnn
 {
  \seq_set_split:Nnn \l__vic_meijerg_args_in_seq { | } { #3 }
  \seq_clear:N \l__vic_meijerg_args_out_seq  
  \seq_map_inline:Nn \l__vic_meijerg_args_in_seq
   {
    \seq_put_right:Nn \l__vic_meijerg_args_out_seq
     {
      \begin{#5matrix} ##1 \end{#5matrix}
     }
   }
  G\sp{#1}\sb{#2}
  \left(
  \seq_use:Nn \l__vic_meijerg_args_out_seq { #6\middle|#6 }
  #6\middle|#6
  #4
  \right)
 }
\ExplSyntaxOff

\newcommand{\RN}[1]{%
	\textup{\uppercase\expandafter{\romannumeral#1}}%
}

\def\Re{ \mathrm{Re}}
\def\Im{ \mathrm{Im}}

\def\OO{{\rm O}}

\newcommand{\sgn}{\operatorname{sgn}}

\newcommand{\bx}{{\bf x}}

\newcommand{\dd}{\mathrm d}

\newtheorem{proposition}{Proposition}[section]
\newtheorem{lemma}[proposition]{Lemma}
\usepackage{enumerate}
\newcommand{\Tr}{\operatorname{Tr}}

\newcommand{\cC}{\mathcal C}
\newcommand{\cK}{\mathcal K}
\newcommand{\cL}{\mathscr L}
\newcommand{\Def}{\mathfrak D}
\newcommand{\supp}{\operatorname{supp}}

\theoremstyle{plain}
\newtheorem*{thm*}{Theorem}
\newtheorem{thm}{Theorem}[section]

\newtheorem*{prop*}{Proposition}
\newtheorem*{lem*}{Lemma}
\newtheorem{rem}[thm]{Remark}

\theoremstyle{definition}
\newtheorem*{eg*}{Example}
\newtheorem*{egs*}{Examples}
\newtheorem*{Q*}{Question}

\theoremstyle{remark}
\newtheorem*{rmk*}{Remark}
\newtheorem*{rmks*}{Remarks}

\numberwithin{equation}{section}

\usepackage{color}
\definecolor{Blue}{rgb}{0.00, 0.00, 1.00}
\definecolor{Red}{rgb}{1.00, 0.00, 0.00}
\newcommand{\red}{\color{Red}}

\hypersetup{
    colorlinks=true,       
    linkcolor=red,          
    citecolor=blue,        
    filecolor=magenta,      
    urlcolor=cyan           
}

\newcommand{\nn}{\nonumber}
\newcommand{\be}{\begin{equation}}
\newcommand{\ee}{\end{equation}}
\newcommand{\bea}{\begin{eqnarray}}
\newcommand{\eea}{\end{eqnarray}}

\newcommand{\q}{{\bf q}}

\newcommand{\ii}{\mathrm{i}} 

\begin{document}
\title[Linear statistics in the Coulomb gas: cumulants beyond rotational invariance]{
Linear statistics in the Coulomb gas: cumulants beyond rotational invariance
}






\author{Paul Bourgade}
\address{Paul Bourgade: Courant Institute, New York University}
\email{bourgade@cims.nyu.edu}

\author{Pierre Le Doussal}
\address{Pierre Le Doussal: Laboratoire de Physique de l'\'Ecole Normale
Sup\'erieure, CNRS, ENS \& PSL University, Sorbonne Universit\'e,
Universit\'e de Paris, 75005 Paris, France}
\email{pierre.ledoussal@phys.ens.fr}

\author{Gr\'egory Schehr}
\address{Gr\'egory Schehr: Sorbonne Universit\'e, Laboratoire de Physique
Th\'eorique et Hautes Energies, CNRS UMR 7589, 4 Place Jussieu,
75252 Paris Cedex 05, France}
\email{schehr@lpthe.jussieu.fr}



\date{\today}

\begin{abstract}
We consider a classical Coulomb gas of \(N\) particles in dimension \(d\geq2\),
confined by a general external potential \(U\) in the large $N$ limit. In that limit the density of particles has a compact support called the droplet. Here we study smooth linear
statistics \({\mathcal L}_N=\sum_{i=1}^N f({\bf x}_i)\) at large \(N\). While the leading variance of \({\mathcal L}_N\) is by now well understood in many settings, much less is known about higher cumulants. When the test function 
$f({\bf x})=f(|{\bf x}|)$ and the potential are rotationally invariant, these cumulants were computed recently at leading order. In this work we
derive  explicit formulas for the leading third and fourth cumulants for
general smooth droplets, without assuming that either the potential or the test
function is radial.  
The answers are expressed as intrinsic boundary integrals
over the equilibrium droplet. The third cumulant depends on the cube of the
jump of the normal derivative of the harmonic extension of \(f\), while the
fourth cumulant involves in addition the exterior Dirichlet-to-Neumann 
operator, the mean curvature of the boundary, and normal derivatives of the
background density. Our derivation uses the tilted potential
\(U_s=U+s f/\beta\), the associated deformation of the droplet, and shape
derivatives of the Dirichlet energy. This deformation has a natural interpretation
as the infinitesimal motion of the free boundary in the so-called {\it obstacle-problem}
formulation of the equilibrium measure. Our formulas extend previous studies by  Wiegmann and Zabrodin, in two dimensions for $\beta=2$. 
In their setting, we provide a rigorous proof of the limiting third and fourth cumulants,  supporting 
the proposed formulas for general Coulomb gases. We specialize the general formulas to the disk, the
half-plane and the sphere, and we check them against exactly solvable
non-radial deformations, including translated elliptic droplets in \(d=2\) and
spheroidal deformations in \(d=3\). 
\end{abstract}

\maketitle
\par\medskip
\tableofcontents

\section{Introduction} \label{sec:introduction}

\subsection{Overview and related works}
\label{sec:overview-related-works}

One-component Coulomb gases occupy a central place at the interface of statistical mechanics,
probability, analysis, and random matrix theory. They describe particles with
logarithmic ($d=2$) or Newtonian repulsion ($d=3$) confined by an external potential, and appear
naturally in invariant random matrix ensembles, non-Hermitian random matrices,
the complex Ginibre ensemble, random normal matrices, one-component plasmas,
and related models of quantum Hall droplets and trapped fermions
\cite{Mehta2004,Forrester2010,Ginibre1965,
RiderVirag,AmeurHedenmalmMakarov,LebleSerfaty,BBNY19,SaffTotik1997,Serfaty_book,Dauxois_review,Lewin_review,Fermions_RMT,rotating_fermions,Byun_Review}. From the analytic point of
view, the equilibrium measure of the gas is described by logarithmic or
Coulomb potential theory with external field.  In the large $N$ limit its support is compact and called the droplet \cite{SaffTotik1997,SerfatySerra,
Friedman1982,CaffarelliSalsa2005}, called $D_0$ in the following.

A basic way to probe the fluctuations of a Coulomb gas is through linear
statistics. For a smooth test function \(f({\bf x})\), one studies the random variable \({\mathcal L}_N=\sum_{i=1}^N f({\bf x}_i)\)
obtained by summing \(f\) over the particle positions. Smooth linear statistics are governed to leading order at large $N$ by a central limit theorem and Gaussian
fluctuations. The variance is of order $O(1)$, and takes the form of a Dirichlet energy, for instance in dimension $d=2$ it is given by 
\be {\red\label{varianceoverview}} 
{\rm Var} \, {\mathcal L}_N =  \frac{1}{2 \pi \beta} \int_{\mathbb{R}^2} d^2 {\bf x} \, |\vec \nabla f^{D_0}({\bf x})|^2 + o(1) \;,
\ee 
where the function $f^{D_0}$ is the harmonic extension of $f$, extending $f$ outside $D_0$. In $d=2$ it is connected to the Gaussian free field \cite{RiderVirag,AmeurHedenmalmMakarov,LebleSerfaty,
BBNY19,SerfatyAnyd}. 

When the test function $f$ is the indicator of a domain, one
obtains the full counting statistics of the number of particles in that
domain. In that case the test function is not smooth and the properties of ${\mathcal L}_N$ are different. 
Counting statistics and mesoscopic or intermediate
statistics have been studied extensively in determinantal processes and random
normal matrices, including disk counting statistics, annular statistics,
critical regimes, spectral gaps, and recent universality results
\cite{Charlier2022,CharlierLenells2023,AmeurCharlierCronvallLenells2023,
AmeurCharlierCronvallLenells2024,AmeurCharlierCronvall2025,
FenzlLambert2022,MarzoMolagOrtegaCerda2026}. Related questions for real and
symplectic Ginibre ensembles and for two-dimensional Coulomb gases at special
values of \(\beta\) have also been investigated in
\cite{AkemannByunEbkeSchehr2023,AkemannByunEbke2023,ByunLee26}.

Most of this literature concerns the first two cumulants, or regimes in which
the limiting field is Gaussian. Higher cumulants are subleading in the
large-\(N\) expansion, but they contain the first non-Gaussian corrections to
the fluctuations. For Coulomb gases with rotationally invariant confinement,
explicit formulas for higher cumulants of smooth linear statistics were
obtained in \cite{OurCoulomb}. This was achieved by computing the expectation value $\langle e^{-Ns{\mathcal L}_N}\rangle_U$ obtained from the equilibrium energy in a tilted potential \(U_s=U+s f/\beta\). The cumulants are then obtained by taking derivatives with respect to $s$. 
In that setting, rotational invariance implies that the perturbed droplet $D_s$ remains a ball and the calculation is tractable.
The
resulting formulae show that the higher cumulant of order $q >2$ scales as $\sim N^{2-q}$
in any dimension, with an amplitude which depends only on the boundary values of \(f\) and its derivatives at the edge of $D_0$. Closely related crossover questions, from smooth
statistics to microscopic or counting statistics, were studied in
\cite{LeDoussalSchehrMicroscopic2025}. Higher cumulants and large deviations
have also been analyzed in one-dimensional Coulomb and Riesz gases
\cite{LeDoussalSchehrRieszCumulants,ForresterRieszGegenbauer,
FlackMajumdarSchehrTruncatedOCP,FlackMajumdarSchehrFCSOCP,
FlackMajumdarSchehrJelliumVariance,RigasJelliumCovariance,BeenakkerRiesz}.

While the formula for the variance \eqref{any} is valid for a general confining potential $U$, and a large class of smooth test functions $f$,
computing the higher cumulants for general $U,f$ and $D_0$ is a much harder task. Indeed, it requires computing the deformations of the (non-spherical) droplet $D_s$ as $s$ varies around $s=0$. For a general
external potential, the droplet may have an arbitrary smooth shape, and a
perturbation of the potential moves the whole boundary in a nonlocal way. 
In determinantal or free-fermion models, more
general shapes can sometimes be treated using kernel asymptotics,
semiclassical analysis, or Widom-type formulas
\cite{SmithLeDoussalMajumdarSchehr2021,SmithLeDoussalMajumdarSchehr2022,
FenzlLambert2022,DeleporteLambert2024,CharlesEstienne2020,
OblakLapierreMoosaviStephanEstienne2024,AkemanDuits2026}. For Coulomb gases at general
\(\beta\), by contrast, explicit higher-cumulant formulas beyond rotational
invariance were not available.

The aim of the present paper is to derive such formulas. We obtain the leading
third and fourth cumulants of smooth linear statistics for Coulomb gases in
arbitrary dimension \(d\geq2\), for general smooth droplets $D_0$ and {\it without}
assuming rotational invariance. The answer is expressed intrinsically on the
boundary of the droplet. The third cumulant involves the cube of the jump of
the normal derivative of the harmonic extension of \(f\), while the fourth
cumulant also involves the exterior Dirichlet-to-Neumann operator, the mean
curvature, and normal derivatives of the background density. 
Our approach is to use a formula, obtained from the so-called 
obstacle-problem \cite{Friedman1982,CaffarelliSalsa2005,LebleSerfaty,SerfatySerra}, for the boundary velocity of the droplet \(D_s\)
as $s$ varies. In $d=2$ we derive the same results using conformal maps,
Julia's variational formula, and the Hilbert transform
\cite{Julia1922,Schiffer1946,BergmanSchiffer1953,Pommerenke1992,Nehari1952}.

{For $d=2$ and $\beta=2$, the corresponding three- and four-point connected
correlations were derived by Wiegmann and Zabrodin
\cite[equations~(42) and~(44)]{WiegmannZabrodin2003}, using functional
variations of the large-$N$ two-point function and Hadamard's formula for the
exterior Green function. Here we differentiate the Dirichlet energy using
the obstacle-problem boundary velocity, which also applies in higher
dimensions. Section~\ref{sec:rigorous} gives a rigorous proof of the two
cumulant asymptotics under the analytic assumptions stated there.
For the large-$N$ expansion of the two-dimensional Dyson-gas free energy,
see also \cite{ZabrodinWiegmann2006Dyson}.}

The paper is organized as follows. In Section
\ref{sec:cg-linear-statistics} we recall the Coulomb-gas setting, the
definition of linear statistics, and the harmonic-extension formula for the
variance. In Section \ref{sec:mainresults} we state the main results: the
general third and fourth cumulant formulas, followed by their specializations
to the disk, the half-plane, and the sphere. Section
\ref{sec:method-overview} explains the main steps of the method. The detailed
shape-derivative derivation is given in Section
\ref{sec:derivation-general}, with the third and fourth derivatives treated in
Sections \ref{sec:third-derivative} and \ref{sec:fourth-derivative},
respectively, and the reduction to simple geometries in Section
\ref{sec:special-geometries}. Section \ref{sec:conformal-tools} gives an
independent two-dimensional derivation using conformal tools, first for the
disk and then for a general simply connected droplet.
Section~\ref{sec:rigorous} proves the third and fourth cumulant
asymptotics for $d=2$, $\beta=2$, and real-analytic data near the droplet. Remark~\ref{anapp:general-beta} gives a sufficient condition on the partition
function for general $d$ and $\beta$.
The appendices collect
background and checks: exact checks for translated elliptic and spheroidal
droplets in Appendices \ref{app:elliptictranslationcheck} and
\ref{app:spheroidcheck}, the one-dimensional Coulomb case in Appendix
\ref{app:1d}, the obstacle-problem formulation in Appendix
\ref{sec:obstacle}, Julia's formula for the inverse map in Appendix
\ref{app:inverseJulia}, the Green-function version of Hadamard's formula in
Appendix \ref{app:greenhadamard}, the variation of the surface element in
Appendix \ref{app:jacobian}, a Fourier check in the half-plane in Appendix
\ref{app:halfspacefouriercheck}, a perturbative construction of harmonic
extensions near the disk in Appendix \ref{app:perturbativeharmonicdisk}, and a
Hilbert-transform identity in Appendix \ref{app:Hilbertidentity}.

\subsection{Coulomb gas and linear statistics} \label{sec:cg-linear-statistics} 
We consider a $d$-dimensional Coulomb gas of $N$ particles at positions $\bx_1,\dots,\bx_N$ in
the presence of an external potential. At equilibrium in the canonical ensemble 
at inverse temperature $\beta$ the joint probability distribution function (PDF) of the positions 
${\mathcal P}(\bx_1,\ldots,\bx_N)$
is given by
\begin{align}
  {\mathcal P}(\bx_1,\ldots,\bx_N) = \frac{1}{Z_{N,U}} 
  e^{- \beta {\mathcal E}(\bx_1,\ldots,\bx_N)}
  \,,\label{eq:pdf} 
\end{align}
where 
the total energy reads
\begin{align} \label{energy}
  {\mathcal E}(\bx_1,\ldots,\bx_N) = \sum_{1\leq i<j\leq N} V_d(|\bx_i-\bx_j|) + N \sum_{k=1}^N U(\bx_k)\, , 
\end{align}
and $Z_{N,U}$ is the partition function (with implicit dependence on $\beta$).
Here $V_d(x)$ is the Coulomb interaction potential in $d$ dimensions
\begin{align}\label{Coulomb_int}
  V_d(x) = \left\{\begin{array}{ll}
    \frac{x^{2-d}}{d-2} & d\neq 2\,,\\
    -\log(x) &d=2\,,
  \end{array}\right.
\end{align}
which satisfies the Poisson equation, i.e.,
\bea \label{Poisson}
\Delta V_d(|\bx|) = -\Omega_d \, \delta(\bx) \quad, \quad \Omega_d = \frac{2\pi^{d/2}}{\Gamma(d/2)} \; ,
\eea
where $\Omega_d$ is the area of the unit sphere embedded in dimension $d$ and $\Delta$ the Laplacian.
In Eq. (\ref{energy}) $U(\bx)$ is an external potential, which is assumed to be smooth and confining, so that the particles remain in the vicinity of the origin.
Note that in Eq. (\ref{energy}) the effective potential is actually $N\,U(\bx)$, such that the equilibrium average density of particles has a support of size of order $O(1)$ in the limit $N \to \infty$. Finally, $Z_{N,U}$ in Eq. (\ref{eq:pdf}) denotes the partition function of the system. An important quantity is the mean empirical density defined as
\be 
\bar \rho_N(\bx) = \left\langle \frac{1}{N} \sum_{i=1}^N \delta(\bx-\bx_i) \right\rangle_{U}
\ee 
where $\langle \cdots \rangle_U$ denotes the average over the joint PDF in (\ref{eq:pdf}).
For later purposes it is convenient to indicate the external potential as a subscript. 

In the limit $N \to \infty$, the mean empirical measure converges weakly to the equilibrium measure, $\bar\rho_N(\bx)\,d^d\bx\rightharpoonup\rho_{\rm eq}(\bx)\,d^d\bx$. The support of the equilibrium
density is called the droplet, denoted by $D_0$, and one has
\be \label{rho_eq}
\rho_{\rm eq}(\bx) = 
\begin{cases} 
&\frac{1}{\Omega_d} \Delta U(\bx) \quad, \quad {\rm for} \; \bx \in D_0 \\
&0 \quad  \quad, \quad \hspace*{1.1cm} {\rm for} \;  \bx \notin D_0 \;.
\end{cases}
\ee 
Determining the shape of the droplet $D_0$ for a general $U(\bx)$ is a non-trivial problem,
whose solution is known only in a few cases. Note that the support is such that $\Delta U(\bx) \geq 0$ 
for all $\bx \in D_0$.

%

In this paper we are interested in the linear statistics, i.e., the fluctuations of ${\mathcal L}_N$ defined as
\begin{align}
  {\mathcal L}_N = \sum_{i=1}^N f(\bx_i)\,,\label{eq:lin}
\end{align}
where $f(\bx)$ is a given arbitrary but sufficiently smooth function (see below for more precise conditions). In particular, the class of functions considered here does not include indicator functions, which appear in counting statistics problems. 
While the first moment is easily obtained to leading order at large $N$ as 
\be \label{firstmoment}
\langle {\mathcal L}_N \rangle \simeq N \int_{\mathbb{R}^d} d^d {\bx} \, \rho_{\rm eq}(\bx) f(\bx)
\ee
computing the higher-order cumulants is much more difficult. 
There is however a simple formula for the variance in the large $N$ limit.
Before displaying it, let us define, 
for any function $f$ defined on $\overline D$ 
(or alternatively on $\mathbb R^d$), the function $f^{D}$, called harmonic extension, as
\be
f^D(\bx)=f(\bx) \quad \text{for } \bx\in D,\qquad
\Delta f^D(\bx)=0 \quad \text{for } \bx\in D^c,\qquad
f^D|_{\partial D}=f|_{\partial D}.
\ee
where $D^c$ denotes the complement of the domain $D$ in $\mathbb{R}^d$.
Thus $f^D$ is the function which coincides with $f$ in $D$ and whose restriction to $D^c$ is the harmonic extension of the boundary data $f|_{\partial D}$, with boundedness and zero total flux at infinity, i.e., 
$
\lim_{R\to\infty}\int_{|\bx|=R}\partial_r f^D\,dS=0.
$

Then the formula for the variance in $d=2$ is known from 
the context of random normal matrix ensembles \cite{RiderVirag,AmeurHedenmalmMakarov}, and 
for general two-dimensional Coulomb gases 
\cite[Thm.~1]{LebleSerfaty}, \cite{BBNY19}. It reads 
\be \label{any} 
{\rm Var} \, {\mathcal L}_N =  \frac{1}{2 \pi \beta} \int_{\mathbb{R}^2} d^2 {\bf x} \, |\vec \nabla f^{D_0}({\bf x})|^2 + o(1) \;,
\ee 
where $\vec \nabla f^{D_0}({\bf x})$ denotes the two-dimensional gradient 
of $f^{D_0}({\bf x})$. {The formula is valid} for general, i.e. possibly non-rotationally invariant, $U({\bf x})$ and $f({\bf x})$,
under some assumptions of regularity. This formula  extends to dimension $d \geq 2$
with $\frac{1}{2 \pi}$ replaced by $\frac{1}{\Omega_d}$. 
Rigorous higher-dimensional CLT results are more delicate; see \cite{SerfatyAnyd} for related results and extensions
under an assumption of no phase-transition for the partition function.



For general $\beta$ and $d$, formulas for
the higher cumulants were obtained in \cite{OurCoulomb} when both the
potential and the test function are rotationally invariant,
i.e. $U(\bx)=U(|\bx|)$ and $f(\bx)=f(|\bx|)$.
In that case the equilibrium density is given by
\be 
\rho_{\rm eq}(\bx) = \rho_{\rm eq}(|\bx|) \quad , \quad  \rho_{\rm eq}(r) = \frac{1}{\Omega_d} \frac{1}{r^{d-1}}(r^{d-1} U'(r))' \;,
\ee 
where primes denote derivatives. In the simplest case the droplet is a ball, of radius $R_0$ fixed by the normalisation condition
\be 
\int_{\mathbb{R}^d} d^d {\bx} \, \rho_{\rm eq}(\bx) = 1 \quad , \quad U'(R_0) R_0^{d-1} = 1 \;.  \label{norma}
\ee
Let us recall the expression of the variance from \eqref{any}
\be
{\rm Var}\,{\mathcal L}_N
=
\frac{1}{\beta}\int_0^{R_0} dr\, r^{d-1} \, [f'(r)]^2 + o(1) \, .
\ee

Remarkably, it was found that the higher cumulants only depend on the function $f'(r)$
and on its higher derivatives, evaluated exactly at $r=R_0$, i.e. at the {\it boundary} of the droplet. For instance, the third cumulant was found to be given, at leading order for large $N$, by the formula
\be \label{cum3rot}
\langle {\mathcal L}_N^3 \rangle_c \simeq \frac{1}{\beta^2 N} 
\frac{R_0^{d-1} f'\left(R_0\right)^3}{
   U''\left(R_0\right)+\frac{d-1}{R_0^d}} \;,  
\ee 
where the normalization condition \eqref{norma} has been used. 
It behaves as $O(1/N)$ at large $N$, in contrast to the second cumulant which is $O(1)$.
More generally it was found that the $q$-th order cumulant behaves at large $N$ as
\be 
\langle {\mathcal L}_N^q \rangle_c \simeq \kappa_q(R_0) \, N^{- (q-2) } 
\ee 
with explicit expressions for the coefficients $\kappa_q(R_0)$.  
\\

\subsection{Main results} \label{sec:mainresults} 

The aim of the present paper is to obtain formulas for some of the higher cumulants in 
the case where the linear statistics function $f( \bx )$ is {\it not spherically symmetric}.
We have obtained very general formulas for the third and fourth cumulants in space dimension $d \geq 2$. They become quite explicit when specialized to the case where the droplet $D_0$ has a simple geometry, for instance a disk in $d=2$, a ball in $d=3$, or a half-space for $d=2,3$.

\subsubsection{Third cumulant: general formula}

To present the general result we first introduce some notations. 
Let $\partial D_0$ be the boundary of the droplet. 
For any function $g(\mathbf{x})$ whose gradient $\vec \nabla g(\mathbf{x})$
has a normal jump discontinuity across $\partial D_0$, one defines the normal jump
\bea 
 [\partial_n g](\mathbf{x}) &=&
 [\vec \nabla g](\mathbf{x})\cdot\mathbf n(\mathbf{x})
=
\partial_n^+ g(\mathbf{x})
-
\partial_n^- g(\mathbf{x}) 
\\
& =& 
\lim_{\epsilon\downarrow 0}
\vec \nabla g(\mathbf{x}+\epsilon\mathbf n(\mathbf{x}))\cdot\mathbf n(\mathbf{x})
-
\lim_{\epsilon\downarrow 0}
\vec \nabla g(\mathbf{x}-\epsilon\mathbf n(\mathbf{x}))\cdot\mathbf n(\mathbf{x}) 
\quad , \quad \mathbf{x}\in\partial D_0  \;. \label{def_dn+}
\eea 
where here and below $\mathbf n(\mathbf{x})$ is the unit normal on $\partial D_0$, 
pointing outward from $D_0$.

Our result for the third cumulant is then the following integral over the boundary of the droplet
\be
\langle {\mathcal L}_N^3\rangle_c
\simeq
-\frac{1}{N}
\frac{1}{\beta^2\Omega_d}
\int_{\partial D_0} dS\,
\frac{\left([\vec \nabla f^{D_0}](\mathbf{x})\cdot\mathbf n(\mathbf{x})\right)^3}
{\Delta U(\mathbf{x})} \, .
\ee
This formula is formally valid for any sufficiently regular droplet $D_0$ in any dimension $d\geq 2$.
For $d=2$ and $\beta=2$, it is the
equal-test-function specialization of
\cite[equation~(42)]{WiegmannZabrodin2003}.
We establish it rigorously for $f$ real-analytic in a neighborhood of the droplet,  in  Section~\ref{sec:rigorous}, see Proposition~\ref{anapp:prop}.

\subsubsection{Third cumulant: disk }

This formula can be evaluated more explicitly in the case of a rotationally invariant potential in $d=2$, $U(\bx) = U(|\bx|)$
such that the droplet $D_0$ is a disk of radius $R_0$, determined by $U'(R_0) R_0 = 1$. 
In polar coordinates the linear statistics function takes the form 
$f(\bx)= f(r,\phi)$. For any function $g(\phi)$ on the circle of radius $R_0$
we define its Fourier decomposition
\be 
g(\phi) = \sum_{n \in \mathbb{Z}} e^{i\,n\phi} \hat{g}_n \quad , \quad 
\hat g_n = \int_0^{2 \pi} \frac{d\phi}{2 \pi} g(\phi) e^{- i n \phi} 
\ee
together with its Hilbert transform $h(\phi)$
\be 
 h = H g  \quad \Longleftrightarrow \quad \hat h_n = \widehat{(H g)}_n  = - i \,{\rm sign}(n)\, \hat g_n 
\ee 
Note that if $g(\phi)$ is real then $h(\phi)$ is real too.
Our result for the third cumulant is the following formula
\be \label{cum3circleintro}
\langle {\mathcal L}_N^3 \rangle_c \simeq
\frac{1}{N \beta^2 R_0^2 \Delta U} \int_0^{2 \pi}  \frac{d\phi}{2 \pi} 
\big(R_0 \partial_{r} f(r,\phi)|_{r=R_0}   + ( H \partial_\phi {\sf f})(\phi)\big)^3     ,
\ee 
where ${\sf f}(\phi) := f(R_0,\phi)$ is the linear statistics function on the circle, and $\Delta U= U''(R_0) + \frac{U'(R_0)}{R_0}$. It is simple to check that this
formula reproduces Eq. \eqref{cum3rot} in the case where $f(r,\phi)=f(r)$. 
This formula passes a few more tests based on the very few cases
where exact solutions of the droplet problem exist. For a simple example, Ginibre $U(r)=r^2/2$ and
$f(r,{\phi})=f_0(r)+2 f_n(r)\cos(n \phi)$ with $n\ge1$, it gives 
\be
\langle{\mathcal L}_N^3\rangle_c
\simeq
\frac{1}{2N\beta^2}
\left[
f_0'(1)^3
+
6f_0'(1)\left(f_n'(1)+n f_n(1)\right)^2
\right].
\ee

\subsubsection{Third cumulant: half-space }
Next we present a result where the droplet $D_0$ is the half-plane $x<0$,
in Cartesian coordinates $(x,y)$ in $d=2$. Its boundary $\partial D_0$ is the vertical line $x=0$. 
We interpret this as a formal flat-boundary limit, with local particle density
$N\tilde U''(x)/(2\pi)$ on $x<0$. A potential $U(x,y)=\tilde U(x)$ independent
of $y$ does not itself define the confined finite-$N$ ensemble
\eqref{eq:pdf}. We take $f(x,y)$ compactly supported.
Our result for the third cumulant is 
\be \label{cum3lineintro}
\langle {\mathcal L}_N^3 \rangle_c \simeq \frac{1}{2 \pi N \beta^2 \tilde U''(0)} \int_{\mathbb{R}} 
dy \, \left( \partial_x f(x,y)|_{x=0} + H \partial_y {\sf f} \right)^3    
\ee 
where ${\sf f}(y)=f(0,y)$. 
Here the Fourier decomposition is defined as ${\sf f}(y)= \int_{\mathbb{R}} \frac{dq}{2 \pi} 
\hat{{\sf f}}_q e^{i q y}$
and $\widehat{H{\sf f}}_q=-i\,{\rm sgn}(q)\,\hat{\sf f}_q$, so that 
$(H \partial_y {\sf f})(y)=\int_{\mathbb{R}} \frac{dq}{2 \pi} |q| \hat{\sf f}_q e^{i q y}$. 
Although \eqref{cum3lineintro} is obtained by an independent derivation,
one can see that it is consistent with \eqref{cum3circleintro} 
in the limit of a very large radius of curvature $R_0 \to +\infty$
(with $\phi \sim y/R_0$ and $x=R_0 (\cos \phi - 1) \simeq - y^2/(2 R_0)$). 
A nice feature of the calculation leading to \eqref{cum3lineintro} is that 
it can be easily extended to any space dimension $d$, for 
a droplet with a boundary being the hyperplane $x=0$. The same formula holds with $2\pi$ replaced by $\Omega_d$ and $dy$ by $d^{d-1}{\bf y}$, for $y \to {\bf y}$ a $d-1$ vector, and $(H \partial_y {\sf f})(y)$
replaced by $\int_{\mathbb{R}^{d-1}} \frac{d^{d-1} {\bf q}}{(2 \pi)^{d-1}} |{\bf q}| \hat{{\sf f}}_{\q} e^{i {\bf q} \cdot {\bf y}}$
with $f(0,{\bf y})={\sf f}({\bf y})= \int_{\mathbb{R}^{d-1}} \frac{d^{d-1} {\bf q}}{(2 \pi)^{d-1}}
\hat{{\sf f}}_{\q} e^{i {\bf q} \cdot {\bf y}}$.

\subsubsection{Third cumulant: 3D sphere }

In $d=3$, for a rotationally invariant potential $U(r)$ and a spherical droplet $D_0$ of radius determined by $R_0^2U'(R_0)=1$, the explicit formula for the third cumulant reads in spherical coordinates $(r, \theta, \phi)$ 
\be \label{cum3sphere}
\langle {\mathcal L}_N^3\rangle_c
\simeq
\frac{1}{N}
\frac{R_0^2}{4\pi\beta^2\Delta U(R_0)}
\int_{S^2} d\Omega\,
\left[
\partial_r f(r,\theta,\phi)|_{r=R_0}
+
\frac{1}{R_0}
\sum_{\ell\geq 1}\sum_{m=-\ell}^{\ell}
(\ell+1)f_{\ell m}Y_{\ell m}(\theta,\phi)
\right]^3
\ee
where $\int_{S^2}d\Omega
=
\int_0^\pi d\theta\,\sin\theta
\int_0^{2\pi}d\phi$, $\Delta U(R_0)=U''(R_0)+\frac{2}{R_0}U'(R_0)$
and 
\be \label{defflm}
f(R_0,\theta,\phi)
=
\sum_{\ell=0}^{\infty}\sum_{m=-\ell}^{\ell}
f_{\ell m}Y_{\ell m}(\theta,\phi) \quad , \quad 
f_{\ell m}
=
\int_{S^2} d\Omega\,
Y_{\ell m}^*(\theta,\phi)\,
f(R_0,\theta,\phi).
\ee 
Here the $Y_{\ell m}(\theta,\phi)$ are the standard 
spherical harmonics, with $\int_{S^2}d\Omega\,
Y_{\ell m}^*(\theta,\phi)Y_{\ell' m'}(\theta,\phi)
=
\delta_{\ell\ell'}\delta_{mm'}$. 

For radial $f$, the sum over $\ell\geq1$ vanishes, and \eqref{cum3sphere} reduces to
\eqref{cum3rot} for $d=3$. 

\subsubsection{Fourth cumulant: general formula}

We first introduce some additional notations. We define the exterior Dirichlet-to-Neumann operator $\Lambda^+$, as
\be
\Lambda^+ h
=
\partial_n^+ u_h,
\qquad
\Delta u_h=0\quad\text{in }D_0^c,
\qquad
u_h|_{\partial D_0}=h,
\ee
with the same zero-flux condition at infinity as for $f^{D_0}$.  {We recall that $\partial_n^+$ is defined in Eq.~(\ref{def_dn+}).}
More schematically one can write it as
$\Lambda^+\left(g|_{\partial D_0}\right)
= \partial_n^+ g^{D_0}$. On the circle, it is related to the Hilbert transform by
 $\Lambda^+ h
=
-\frac{1}{R_0}H\partial_\phi h$.

Next, one denotes by $M({\bf x})=\vec \nabla_{\partial D_0}\cdot{\bf n}({\bf x})$
the mean curvature of the boundary of the droplet $\partial D_0$, defined as the surface divergence
of the outward unit normal. Equivalently, if
$X$ is a local parametrization of $\partial D_0$, and $g_{ij}=\partial_i X\cdot\partial_j X$, 
$h_{ij}=\partial_i{\bf n}\cdot\partial_j X$ are respectively the first
and second fundamental forms, then $M=g^{ij}h_{ij}$.
$M$ is the sum of the principal
curvatures. In particular $M=(d-1)/R_0$ for a ball of radius $R_0$, and
$M=0$ for a flat boundary.

With these definitions our general result for the fourth cumulant
reads
\be
\begin{aligned}
\langle {\mathcal L}_N^4\rangle_c
&\simeq
 \frac{- 1}{\beta^3\Omega_d N^2}
\int_{\partial D_0}dS\,
\left[
3\,\frac{[\partial_n f^{D_0}]^2}{\Delta U}\,
\Lambda^+\!\left(\frac{[\partial_n f^{D_0}]^2}{\Delta U}\right)
+
4\,\frac{[\partial_n f^{D_0}]^3\Delta f}{(\Delta U)^2}
\right.
\\
&\hspace{3.0cm}\left.
+
2 
M \frac{[\partial_n f^{D_0}]^4}{(\Delta U)^2}
+
\frac{[\partial_n f^{D_0}]^4\partial_n\Delta U}{(\Delta U)^3}
\right] .
\end{aligned}
\ee
Here all quantities are evaluated on $\partial D_0$, ${\bf n}$ is the unit normal pointing
outward from $D_0$. It is formally valid for any sufficiently smooth droplet $D_0$ in any dimension $d\geq 2$.
For $d=2$ and $\beta=2$, taking
$f_1=f_2=f_3=f_4=f$ in \cite[equation~(44)]{WiegmannZabrodin2003}
gives this formula, with the exterior normal convention used here.
Proposition~\ref{anapp:prop} below gives its rigorous asymptotics.

\subsubsection{Fourth cumulant: disk}

For a radial $U(r)$ and the droplet $D_0$ being the disk of radius $R_0$ 
the fourth cumulant reads 
\be \label{4cumdiskintro}
\begin{aligned}
\langle {\mathcal L}_N^4\rangle_c
&\simeq
\frac{1}{2\pi\beta^3N^2(\Delta U)^2}
\int_0^{2\pi}d\phi\,
\bigg[
3J(\phi)^2 H\partial_\phi\!\left(J(\phi)^2\right)
+
4R_0J(\phi)^3\Delta f(R_0,\phi)
\\
&\hspace{3.0cm}
-
\frac{
R_0U'''(R_0)+3U''(R_0)+R_0^{-2}
}{\Delta U}
J(\phi)^4
\bigg],
\end{aligned}
\ee
where
\be
\Delta U=U''(R_0)+\frac{1}{R_0^2},
\qquad
J(\phi)
=
\partial_r f(r,\phi)|_{r=R_0}
+
\frac{1}{R_0}H\partial_\phi{\sf f}(\phi),
\qquad
{\sf f}(\phi)=f(R_0,\phi).
\ee

For the simple example, Ginibre $U(r)=r^2/2$ and
$f(r,\phi)=2 f_n(r)\cos(n \phi)$, it gives 
\be
\langle{\mathcal L}_N^4\rangle_c
\simeq
\frac{3}{\beta^3N^2}
\left(f_n'(1)+n f_n(1)\right)^3
\left[2\left(f_n''(1)+f_n'(1)-n^2 f_n(1)\right)
+(n-1)\left(f_n'(1)+n f_n(1)\right)\right].
\ee

For a rotationally invariant $f(r,\phi)=f(r)$ and $J(\phi)
=f'(R_0)$,
one recovers (upon rearranging some terms) the result of \cite{OurCoulomb}
for $d=2$  
\bea  
 \langle {\mathcal L}_N^4 \rangle_c & \simeq & \frac{1}{\beta^{3} N^{2} (\Delta U)^2} \bigg( 
\frac{ \left(2 -R_0^{3} U'''(R_0)\right) f'(R_0)^4 } 
{R_0^2 \Delta U} + 
   f'(R_0)^3 \left( f'(R_0)+4 R_0 f''(R_0)\right)  \bigg) 
\eea

\subsubsection{Fourth cumulant: half-space}

When the droplet is a half-space, in $d=2$, 
with $U(x,y)=\tilde U(x)$ independent of $y$, the fourth cumulant reads
\be \label{cum4lineintro} 
\langle {\mathcal L}_N^4 \rangle_c \simeq \frac{1}{2 \pi \beta^3 N^2 \tilde U''(0)^2}
\int_{-\infty}^{+\infty} dy \, \bigg[ 3 g(y)^2 H\partial_y (g(y)^2)
+
4 g(y)^3 (\Delta f)(0,y)
- \frac{\tilde U'''(0)}{\tilde U''(0)} g(y)^4 \bigg]
\ee 
where 
\bea 
&& g(y) = \partial_x f(x,y)|_{x=0} + (H \partial_y {\sf f})(y) \\
&&  (\Delta f)(0,y) = \partial^2_x f(x,y)|_{x=0} + \partial^2_y {\sf f}(y) 
\quad , \quad {\sf f}(y)=f(0,y)
\eea 

\subsubsection{Fourth cumulant: sphere}

For a rotationally invariant potential $U(r)$ in $d=3$, such that
the droplet is a {ball} of radius $R_0$, with $R_0^2 U'(R_0)=1$,
the fourth cumulant reads 
\be \label{4cumsphereintro}
\begin{aligned}
\langle{\mathcal L}_N^4\rangle_c
&\simeq
\frac{1}{4\pi\beta^3N^2(\Delta U)^2}
\bigg[
3R_0
\sum_{\ell\geq1}\sum_{m=-\ell}^{\ell}
(\ell+1)|q_{\ell m}|^2
\\
&\hspace{1.0cm}
+
\int_{S^2}d\Omega\,
\left(
4R_0^2J^3 (\Delta f)_{r=R_0}
-
\frac{
R_0^2U'''(R_0)
+
6R_0U''(R_0)
+
6 R_0^{-2}
}{\Delta U}
J^4
\right)
\bigg].
\end{aligned}
\ee
where 

\be \Delta U=U''(R_0)+\frac{2}{R_0^3}
\quad , \quad 
J(\theta,\phi)
=
\partial_r f(r,\theta,\phi)|_{r=R_0}
+
\frac{1}{R_0}
\sum_{\ell\geq1}\sum_{m=-\ell}^{\ell}
(\ell+1)f_{\ell m}Y_{\ell m}(\theta,\phi).
\ee
Here the $f_{\ell m}$ were defined in \eqref{defflm} 
and one has introduced the decomposition
\be
J(\theta,\phi)^2
=
\sum_{\ell=0}^{\infty}\sum_{m=-\ell}^{\ell}
q_{\ell m}Y_{\ell m}(\theta,\phi)
\quad , \quad q_{\ell m} = \int_{S^2} d\Omega\,
Y_{\ell m}^*(\theta,\phi)\, J(\theta,\phi)^2. 
\ee
\\

For radial $f$, one has
\be
J=f'(R_0),
\qquad
\Delta f(R_0)=f''(R_0)+\frac{2}{R_0}f'(R_0),
\qquad
\Delta U=U''(R_0)+\frac{2}{R_0^3}.
\ee
and the formula \eqref{4cumsphereintro} recovers the 
general radial formula of \cite{OurCoulomb} for $d=3$, which reads
\be
\left\langle {\mathcal L}_N^4\right\rangle_c
\simeq
\frac{1}{\beta^3N^2}
\left[
\frac{(6-R_0^4U'''(R_0))f'(R_0)^4}{R_0^2(\Delta U)^3}
+
\frac{R_0f'(R_0)^3\left(2f'(R_0)+4R_0f''(R_0)\right)}{(\Delta U)^2}
\right].
\ee
\\

{\bf Remark}.
There exist special potentials for which the droplet is known explicitly, such
as radial potentials, elliptic deformations, droplets with point charges, and
quadrature-domain or polynomial-curve droplets in the normal matrix model
\cite{ByunPlanar23,WiegmannZabrodin2003,ElbauFelder2005,Elbau2007,
EtingofMa2007,TeodorescuBettelheimAgamZabrodinWiegmann2005,
KuijlaarsLopezGarcia2015,BalogMerzi2015,BalogGravaMerzi2015,ABK2021,BY2026}.
Some of these examples can be used to construct
pairs \((U,f)\) for which the leading large-\(N\) asymptotics of \(\chi(s,N)\),
and hence all leading cumulants, can be obtained exactly, allowing for checks of our formulas. In Appendix \ref{app:elliptictranslationcheck} we provide the leading large-$N$ cumulants for an elliptic deformation of the disk, and check that they match our formulas. In higher dimensions,
anisotropic quadratic confinements lead to ellipsoidal droplets, by the
classical fact that the Newtonian potential of a uniformly 
charged ellipsoid is
quadratic inside the ellipsoid \cite{Chandrasekhar1969,Karp1994}.
In Appendix \ref{app:spheroidcheck} we provide a nontrivial check of the $d=3$ formula using an ellipsoidal deformation of the sphere. 
\\

{\bf Remark}. Upon proper interpretation the general formula obtained here
can also be applied in dimension $d=1$ for the linear Coulomb interaction $V_1(x)=- |x|$. This is shown in Appendix \ref{app:1d} where the full generating function $\chi(s,N)$ is computed at large $N$ for an arbitrary smooth function $f(x)$, and the low-order cumulants are extracted from it. There too we restrict to a support being a single interval. These results are compared to existing literature.

\subsection{Main steps of the method} \label{sec:method-overview} 

To obtain these results we first introduce the cumulant generating function (CGF) $\chi(s,N)$
defined as 
\begin{align}
 \chi(s,N) =  \log \langle e^{- N s \, {\mathcal L}_N } \rangle_U = 
 \log \langle e^{- N s \sum_{i=1}^N f(\bx_i)}\rangle_U \,,\label{eq:chidef}
\end{align}
where we recall that the average $\langle \cdots \rangle_U$ is taken with respect to the joint PDF in (\ref{eq:pdf}). 
The cumulants can then be obtained from $\chi(s,N)$ as
\be 
\langle {\mathcal L}_N^q \rangle_c = N^{-q} (- \partial_s)^q  \chi(s,N)|_{s=0} \, .
\ee 
Next, one notes that one can write
\be 
\chi(s,N) = \log \frac{ Z_{N,U_s}}{Z_{N, U}} \quad , \quad 
U_s(\bx) := U(\bx) + \frac{s}{\beta} f(\bx)  \;,
\ee 
where the numerator is the partition function of the Coulomb gas in a shifted external potential
$U_s(\bx)$. Taking the large $N$ limit and differentiating with respect to $s$, see \cite{OurCoulomb} for details, we obtain
our main starting formula 
\be \label{starting} 
- \frac{1}{N^2} \partial_s  \chi(s,N) = \int_{D_s} d^d {\bf x} \, \rho_s({\bf x}) f({\bf x}) \quad , \quad \rho_s(\bx) = \frac{1}{\Omega_d} ( \Delta U(\bx) + \frac{s}{\beta} \Delta f(\bx) ) \quad \text{for} \quad {\bf x} \in D_s 
\ee 
where this equality is understood at leading order in the large $N$ limit, as will be implicit below.
Here $\rho_s(\bx)$ denotes the equilibrium density in the shifted external potential $U_s$
and $D_s$ denotes its support, i.e. the corresponding droplet (and of course $\rho_s$
vanishes outside $D_s$). Note that we have assumed
the commutation of limits $N \to \infty$ and $s \to 0$.  
For $d=2$ and $\beta=2$, Section~\ref{sec:rigorous} proves the
third and fourth cumulant asymptotics without rotational invariance, under
a real-analytic  assumption.
A full proof for general $d$ and $\beta$ remains out of reach of the present
argument. Remark~\ref{anapp:general-beta} states a sufficient condition on
derivatives of the partition function that justifies the required limit
interchange.

Taking $s=0$ in the above formula recovers the leading-order expression for the first moment in \eqref{firstmoment}.
To obtain the higher cumulants we need to take further derivatives of \eqref{starting} 
w.r.t. the variable~$s$. When both $U$ and $f$ are rotationally invariant, it is easy to determine the shifted droplet $D_s$,
which remains a ball as long as no radial gap opens, and it is then possible 
to compute the r.h.s. of \eqref{starting} as a function of $s$ for general
$U(r)$ and $f(r)$, as in \cite{OurCoulomb}. 
Here however determining $D_s$ becomes a much more difficult problem. Fortunately, to 
obtain the cumulants we only need the small variations of $s$. 
In that case, it turns out that the infinitesimal variations of $D_s$ can be obtained
using the solution to the so-called {\it obstacle problem}.
We now recall what is needed for our purpose here; for a more
general exposition of the problem, see e.g. \cite{SerfatySerra},
as well as a summary in Appendix \ref{sec:obstacle}.

We need 
to find the droplet $D_s$ associated with the potential $U_s(\bx)=U(\bx) + \frac{s}{\beta} f(\bx)$. It is provided by the solution
to the {\it obstacle problem}, see Prop. ~3.3 in \cite{LebleSerfaty} (same notation except that their $V=2U$) 
and Thm.~1.1 in \cite{SerfatySerra}. In these works the normal speed of the boundary of the domain is obtained.
Let us denote by $\partial D_s$ the boundary of $D_s$.
As $s$ increases infinitesimally by $ds$ the new boundary is defined by the collection of points
\be  \label{newboundary}
{ {\partial D}_{s+ds}} {\equiv} 
\{ {\bf x} +  v_s({\bf x}) \, {\bf n}_s({\bf x}) \, ds  \, \mid \,   {\bf x} \in \partial D_s \} 
\ee

where the normal velocity $v_s({\bf x})$ is given by 
\be \label{motion}
v_s({\bf x}) = \frac{1}{\beta \Delta U_s({\bf x})} ( [ \vec \nabla f^{D_s} ] \cdot {\bf n}_s({\bf x}) ) \, .
\ee 
Here ${\bf n}_s({\bf x})$ denotes the unit normal to $\partial D_s$
pointing outward from the droplet $D_s$, i.e. toward the exterior domain
$D_s^c$. As above, $f^{D_s}$ denotes the function equal to $f$ inside $D_s$,
harmonic in $D_s^c$, continuous across $\partial D_s$, and satisfying
the zero-flux condition at infinity. Again, the notation
\be
[ \vec \nabla g ]({\bf x}) \cdot {\bf n}_s({\bf x})
=
\vec \nabla g({\bf x} + 0^+ {\bf n}_s({\bf x}))\cdot {\bf n}_s({\bf x})
-
\vec \nabla g({\bf x} - 0^+ {\bf n}_s({\bf x}))\cdot {\bf n}_s({\bf x})
\ee 
denotes the jump of the normal derivative of a function $g$ across the boundary,
with the convention exterior minus interior. We see that here $g({\bf x})=f^{D_s}(\bx)$, hence in order to compute  
the normal velocity $v_s({\bf x})$ one needs to be able to find the harmonic extension
of the function $f({\bf x})$ with respect to the domain $D_s$. In practice, this will be possible by applying this formula 
perturbatively in $s$ starting from the unperturbed droplet $D_0$. 
\\

{\bf Remark}. In order to get an intuition on the origin of the formula \eqref{motion},
we can verify that under its dynamics the total mass is conserved. Indeed one has, using \eqref{motion}
\bea \label{massconservation}
&&  \partial_s \int_{D_s}  d^d{\bf x} \, \rho_s({\bf x}) = \partial_s 
 \int_{D_s} \frac{d^d {\bf x}}{\Omega_d} \Delta (U({\bf x}) + \frac{s}{\beta} f({\bf x}) ) \\
&&  = \frac{1}{\beta \Omega_d}  \int_{D_s} d^d {\bf x}  \Delta  f({\bf x}) 
+ \frac{1}{\Omega_d}  \int_{\partial D_s} d\sigma  \, v_s({\bf x})  \, \Delta (U({\bf x}) + \frac{s}{\beta} f({\bf x}) ) \nn \\
&& = \frac{1}{\beta \Omega_d }  \int_{D_s} d^d {\bf x}  \Delta  f({\bf x}) 
+ \frac{1}{ \beta \Omega_d }  \int_{\partial D_s^+} d\sigma  (  \vec \nabla f^{D_s}  \cdot {\bf n}_s({\bf x}) )  
- \frac{1}{ \beta \Omega_d }  \int_{\partial D_s^-} d\sigma  (  \vec \nabla f  \cdot {\bf n}_s({\bf x}) ) = 0 \nn 
\eea  
where $d\sigma$ is the $d-1$-dimensional area element 
and we use the notation $\int_{\partial D_s^\pm} d\sigma  (  \vec \nabla g  \cdot {\bf n}_s({\bf x}) )
= \int_{\partial D_s} d\sigma\,
\partial_n^\pm g$. To arrive at the last line of \eqref{massconservation} we have used the 
two integration by parts formulae
for the interior and exterior, respectively
\bea 
&& \int_{D_s} d^d {\bf x} \,  \Delta  f({\bf x}) = \int_{\partial D_s^-} d\sigma  \, (  \vec \nabla f  \cdot {\bf n}_s({\bf x}) ) \\
&& \int_{\partial D_s^+} d\sigma  (  \vec \nabla f^{D_s}  \cdot {\bf n}_s({\bf x}) ) = 
\lim_{R\to+\infty}
\int_{S_R}d\sigma\,
\vec\nabla f^{D_s}\cdot{\bf n}_R
-
\int_{D_s^c}d^d{\bf x}\,\Delta f^{D_s}
=
0 .
\eea 
where the boundary term vanishes at infinity. Indeed, in the definition of the harmonic extension $f^{D_s}$ we have imposed, in addition to $\Delta f^{D_s}=0$
outside of $D_s$, 
the zero flux condition 
\be
\lim_{R\to+\infty}
\int_{S_R} d\sigma\,
\vec \nabla f^{D_s}({\bf x})\cdot{\bf n}_R({\bf x})
=
0,
\ee
where $S_R$ is a sphere of radius $R$ and ${\bf n}_R$ is its outward normal.
For $d>2$, this excludes the monopole term proportional to $r^{2-d}$: at infinity $f^{D_s}$ must be a constant plus terms decaying faster than $r^{2-d}$. In $d=2$, the corresponding condition is the absence of a logarithmic term, so that $f^{D_s}$ remains bounded at infinity.

We see that the Laplacian of the shifted potential exactly cancels going from
the second to the third line in \eqref{massconservation}, allowing for
the integration by parts leading to mass conservation. This is only
possible because of the specific form of the denominator 
in the obstacle formula \eqref{motion}. 
Consequently,
there is a conservation law of the form $\partial_s \rho_s(x) = - \vec \nabla \cdot {\bf j}({\bf x})$
such that the normal flux of the current ${\bf j}({\bf x}) \cdot {\bf n}_s({\bf x}) = \rho_s({\bf x}) v_s({\bf x})  $ through the
boundary compensates the excess equilibrium charge created in the bulk by the change in the external potential.
This is closely related to the formulation in terms of a transport map between the original
and perturbed equilibrium measures used in \cite{LebleSerfaty}, together with the
free-boundary motion estimates of \cite{SerfatySerra}.
\\

{\bf Simplification of the starting point}. One can simplify the calculation of the cumulants by taking another derivative 
with respect to $s$ of \eqref{starting} and obtain an explicit expression.
For the integral below, we assume that $f$ is compactly supported,
or more generally that the boundary term at infinity vanishes. 
Indeed one can show that
\be \label{starting2}
\frac{1}{N^2} \partial_s^2 \chi(s,N)= \frac{1}{ \beta \Omega_d} \int_{\mathbb{R}^d} d^d \bx \, 
\vec{\nabla} f({\bf x})   \cdot
\vec{\nabla} f^{D_s}( {\bf x}) 
= \frac{1}{ \beta \Omega_d} \int_{\mathbb{R}^d} d^d \bx \, 
\vec \nabla f^{D_s}({\bf x})  \cdot \vec \nabla f^{D_s}( {\bf x}) 
\ee 
This formula will be the starting point of our calculation of the cumulants.
\\

{\it Proof}. We use similar manipulations as in \eqref{massconservation}. One 
writes  
\bea 
 &&  \partial_s \int_{D_s}  d^d{\bf x} \, \rho_s({\bf x}) f({\bf x}) = \partial_s 
 \int_{D_s} \frac{d^d {\bf x}}{\Omega_d} \Delta (U({\bf x}) + \frac{s}{\beta} f({\bf x}) ) f({\bf x}) \\
&&  = \frac{1}{\beta \Omega_d}  \int_{D_s} d^d {\bf x} f({\bf x})  \Delta  f({\bf x})  
+ \frac{1}{\Omega_d}  \int_{\partial D_s} d\sigma  \, v_s({\bf x})  \, \Delta (U({\bf x}) + \frac{s}{\beta} f({\bf x}) ) f({\bf x}) \nn \\
&& = \frac{1}{ \beta \Omega_d }  \int_{D_s} d^d {\bf x}  f({\bf x}) \, \Delta  f({\bf x})  
+ \frac{1}{\beta \Omega_d }  \int_{\partial D_s^+} d\sigma   \,  \vec \nabla f^{D_s}  \cdot {\bf n}_s({\bf x})   \, f({\bf x})
- \frac{1}{\beta \Omega_d }  \int_{\partial D_s^-} d\sigma \,   \vec \nabla f  \cdot {\bf n}_s({\bf x}) \,  f({\bf x}) \nn
\eea  
We now use that, by integration by parts inside the domain
\be 
\int_{D_s} d^d {\bf x} \,  f({\bf x}) \Delta  f({\bf x}) = \int_{\partial D_s^-} d\sigma  \, (  \vec \nabla f  \cdot {\bf n}_s({\bf x}) ) f({\bf x})
- \int_{D_s} d^d {\bf x} (\vec \nabla f({\bf x}) )^2
\ee 
This leads to
\be 
\partial_s \int_{D_s}  d^d{\bf x} \, \rho_s({\bf x}) f({\bf x}) = 
- \frac{1}{ \beta \Omega_d } \int_{D_s} d^d {\bf x} (\vec \nabla f({\bf x}) )^2 
+ \frac{1}{\beta \Omega_d }  \int_{\partial D_s^+} d\sigma  \,   \vec \nabla f^{D_s}({\bf x})  \cdot {\bf n}_s({\bf x}) \,   f({\bf x})
\ee 
We now use that, by integration by parts outside the domain
\be
\begin{aligned}
\int_{\partial D_s^+} d\sigma\,
\left(\vec \nabla f^{D_s}({\bf x}) \cdot{\bf n}_s({\bf x}) \right) f({\bf x})
&=
\lim_{R\to+\infty}
\int_{S_R}d\sigma\,
f({\bf x})\,\vec\nabla f^{D_s}({\bf x})\cdot{\bf n}_R({\bf x})
\\
&\quad
-
\int_{D_s^c} d^d{\bf x}\,
f({\bf x})\,\Delta f^{D_s}({\bf x})
-
\int_{D_s^c} d^d{\bf x}\,
\vec\nabla f({\bf x})\cdot\vec\nabla f^{D_s}({\bf x}) .
\end{aligned}
\ee
Then, under the decay/zero-flux assumptions, the term at infinity vanishes, and since $\Delta f^{D_s}({\bf x})=0$ outside $D_s$, we obtain
\bea  
&& \partial_s \int_{D_s}  d^d{\bf x} \, \rho_s({\bf x}) f({\bf x}) = - \frac{1}{ \beta \Omega_d} 
( \int_{D_s} d^d {\bf x} (\vec \nabla f({\bf x}) )^2  
+  \int_{D_s^c} d^d {\bf x} \vec \nabla f({\bf x}) \cdot  \vec \nabla f^{D_s}({\bf x}) ) \\
&& = - \frac{1}{\beta \Omega_d}  \int_{\mathbb{R}^d } d^d {\bf x} \vec \nabla f({\bf x}) \cdot \vec  \nabla f^{D_s}({\bf x}) \\
&& = - \frac{1}{ \beta \Omega_d}  \int_{\mathbb{R}^d } d^d {\bf x} \vec \nabla f^{D_s}({\bf x}) \cdot \vec   \nabla f^{D_s}({\bf x}) 
\eea  
The last line is obtained by noting that $\vec  \nabla f^{D_s}({\bf x})=\vec  \nabla f({\bf x})$ for ${\bf x} \in D_s$, 
and that 
\be \int_{D_s^c} d^d{\bf x}\,
\left(
\vec\nabla f({\bf x})-\vec\nabla f^{D_s}({\bf x})
\right)\cdot\vec\nabla f^{D_s}({\bf x})
=0
\ee
This identity is obtained by integration by parts, 
using the fact that 
$f^{D_s}({\bf x})=f({\bf x})$ for ${\bf x} \in \partial D_s$ (since
by definition it is continuous) and that 
$\Delta f^{D_s}({\bf x})=0$ for ${\bf x} \in D_s^c$.
The boundary term at infinity vanishes,
under the decay assumptions on $f$ and the zero-flux/decay condition
defining $f^{D_s}${\red .}
\\

{{\bf Non-connected droplet}. The geometric shape-derivative calculation can formally be extended to
disconnected droplets, provided their boundaries remain smooth and their
topology does not change under the perturbation. However, this does not
necessarily imply the validity of the resulting cumulant asymptotics:
in the planar determinantal case, fluctuations of the particle numbers
between disconnected components have been shown to produce additional
oscillatory discrete Gaussian contributions, which can contribute at
order one, even to higher cumulants
\cite{AmeurCronvall2026,AmeurCharlierCronvall2025}.
Therefore, we do not claim that our cumulant
formulas extend unchanged to general disconnected droplets.}
\\

{\bf Special tools in $d=2$: the Riemann map and Julia's formula}. In $d=2$, when the droplet has a single smooth boundary component, one can
parametrize $\partial D_s$ using the exterior Riemann map. Representing
${\bf x}=(x,y)$ by the complex number $z=x+\ii y$, let $\psi_s$ be the
conformal map from the exterior of the droplet onto the exterior of the unit
disk,
\be
\psi_s: D_s^c \to \mathbb D^c,
\qquad
\mathbb D^c=\{w\in\mathbb C:\ |w|>1\}.
\label{uniformizingmap}
\ee
We shall use standard facts from conformal mapping theory, in particular the
exterior Riemann map and its boundary correspondence for smooth Jordan curves
\cite{Nehari1952,Pommerenke1992}.
We choose the normalization at infinity
\be
\psi_s(\infty)=\infty,
\qquad
\psi_s(z)=a_s z+O(1)\quad \text{as }z\to\infty,
\qquad
a_s>0.
\label{normalizationpsi}
\ee
This fixes the map uniquely. With this normalization, $\partial D_s$ is mapped
onto the unit circle $\partial\mathbb D$.

Since conformal maps preserve harmonicity in two dimensions, this map is useful
for computing the exterior harmonic extension $f^{D_s}$. Let
\be
{\sf g}_s(\theta)
=
f\!\left(\psi_s^{-1}(e^{\ii\theta})\right),
\qquad 0\leq\theta<2\pi,
\ee
be the boundary data transported to the unit circle, and define its Fourier
coefficients by
\be
{\sf g}_s(\theta)
=
\sum_{k\in\mathbb Z}\hat g_k(s)e^{\ii k\theta},
\qquad
\hat g_k(s)
=
\int_0^{2\pi}\frac{d\theta}{2\pi}\,
{\sf g}_s(\theta)e^{-\ii k\theta}.
\label{defgs}
\ee
The exterior harmonic extension of these boundary data to $\mathbb D^c$ is
then
\be
g_s^{\mathbb D}(w)
=
\sum_{k\geq0}\hat g_k(s)\,\bar w^{-k}
+
\sum_{k\leq -1}\hat g_k(s)\,w^k,
\qquad |w|>1.
\label{Poissong}
\ee
This is the Poisson extension in the exterior disk, with the boundedness
condition at infinity. Pulling back by $\psi_s$, one obtains the exterior part
of $f^{D_s}$:
\be
f^{D_s}(z)
=
g_s^{\mathbb D}(\psi_s(z)),
\qquad z\in D_s^c.
\label{mapfg}
\ee
Inside the droplet, one has simply $f^{D_s}(z)=f(z)$.

\begin{rmk*}
In most of our applications, $f$ is a real function of $z$ and $\bar z$. The
notation
$
f\!\left(\psi_s^{-1}(w)\right)
$
is then a shorthand for
$
f\!\left(\psi_s^{-1}(w),\overline{\psi_s^{-1}(w)}\right).
$
Equivalently, if one regards the second argument as a function of $\bar w$,
this may be written as $
f\!\left(\psi_s^{-1}(w),\overline{\psi_s^{-1}}(\bar w)\right)$.
The boundary function ${\sf g}_s$ is real-valued, and therefore
\be
\hat g_{-k}(s)=\hat g_k(s)^*.
\ee
Consequently \(g_s^{\mathbb D}\), and hence \(f^{D_s}\) on \(D_s^c\), is real.
\end{rmk*}

Next, Julia's variational formula \cite{Julia1922}, closely related to the
Hadamard-Schiffer theory of domain variations \cite{Schiffer1946,SchifferSpencer1954},
describes how the exterior Riemann map changes under an infinitesimal normal
motion of the boundary. With the normalization
\eqref{normalizationpsi}, and with the normal velocity $v_s$ defined with
respect to the outward normal from $D_s$ to $D_s^c$, it reads
\be
\partial_s\log\psi_s(z)
=
\frac{1}{2\pi}
\oint_{\partial D_s}
\frac{\psi_s(t)+\psi_s(z)}{\psi_s(t)-\psi_s(z)}
|\psi_s'(t)|^2\,v_s(t)\,d\ell(t),
\qquad z\in D_s^c.
\label{julia}
\ee
Here \(\partial_s\) is taken at fixed exterior point \(z\), \(d\ell(t)\) is the
arclength element on \(\partial D_s\), and \(\psi_s'(t)\) denotes the boundary
limit of the complex derivative of \(\psi_s\). The velocity \(v_s(t)\) is a
real signed normal velocity; positive \(v_s\) means that the droplet moves
outward. Julia's variational formula, in a form close to the one used here, can be found for instance in \cite[Eq.~(15.24)]{BergmanSchiffer1953}.

\begin{rmk*}
Julia's formula is a shape-derivative formula for a prescribed boundary
velocity. In the applications below, this velocity is itself determined by the
Coulomb-gas obstacle problem and depends on \(U_s\) and \(f^{D_s}\). Thus, when
differentiating quantities depending on \(s\), one must keep track both of the
explicit \(s\)-dependence of \(U_s=U+s f/\beta\) and of the induced motion of
the boundary.
\end{rmk*}

In our case, using the obstacle expression \eqref{motion} for \(v_s\),
Julia's formula can be rewritten in complex notation as
\bea \label{julia2}
\partial_s\log\psi_s(z)
=
\frac{1}{\pi}
\oint_{\partial D_s}
\frac{\psi_s(t)+\psi_s(z)}
{\psi_s(t)-\psi_s(z)}
|\psi_s'(t)|^2
\frac{1}{\beta\Delta U_s(t)}
\,{\rm Im}\!\left(dt\,[\partial_z f^{D_s}](t)\right).
\eea
Here \(\partial D_s\) is oriented counterclockwise and $\partial_z=\frac12\left(\partial_x-i\partial_y\right)$ is evaluated on the contour. We used that, along \(\partial D_s\),
$[\partial_n f^{D_s}]\,d\ell
=
2\,{\rm Im}\!\left(dt\,[\partial_z f^{D_s}](t)\right)$,
where the jump is exterior minus interior and \({\bf n}\) is the outward normal
from \(D_s\) to \(D_s^c\).

Finally one can obtain an "inverse Julia" formula for the evolution of the inverse Riemann map, see Appendix \ref{app:inverseJulia}. 
A complementary Green-function formulation is recalled in Appendix \ref{app:greenhadamard}.

\section{Derivation of the general formulas} \label{sec:derivation-general}

\subsection{Third derivative and third cumulant} \label{sec:third-derivative}

We start from the finite-\(s\) second derivative formula
\be
\frac{1}{N^2}\partial_s^2\chi(s,N)
=
\frac{1}{\beta\Omega_d}E(s),
\qquad
E(s)=
\int_{\mathbb R^d}d^d{\bf x}\,
|\vec \nabla f^{D_s}({\bf x})|^2 . \label{defEs}
\ee
We will first derive a formula for the finite-\(s\) third derivative, i.e. compute 
\be \label{der3chis}
\frac{1}{N^2}\partial_s^3\chi(s,N)
=
\frac{1}{\beta\Omega_d}E'(s).
\ee

Before starting, let us fix some notations. For a function defined on both sides of $\partial D_s$, we use the notation
\be
g_s^\pm({\bf x})
=
\lim_{\epsilon\downarrow0}
g_s({\bf x}\pm\epsilon{\bf n}_s({\bf x})) ,
\qquad {\bf x}\in\partial D_s ,
\ee
where ${\bf n}_s$ is the outward normal to $D_s$. Thus $-$ denotes the
limit from the interior of $D_s$, while $+$ denotes the limit from the exterior.

Let \(X_s({\bf y})\in \partial D_s\) be the position of the boundary point labelled by
\({\bf y}\in\partial D_0\).
The boundary motion is described by 
\be
\frac{dX_s({\bf y})}{ds}
=
v_s(X_s({\bf y}))\,{\bf n}_s(X_s({\bf y})),
\qquad
X_0({\bf y})={\bf y},
\ee
where ${\bf n}_s({\bf x})$ is the outward unit normal.

We shall repeatedly use the following elementary shape-derivative formula, valid
for a smooth family of domains $D_s$ and for a smooth function $g_s({\bf x})$ in $D_s$
\be \label{derivativeboundary}
\frac{d}{ds}\int_{D_s} d^d{\bf x}\,g_s({\bf x})
=
\int_{D_s} d^d{\bf x}\,\partial_s g_s({\bf x})
+
\int_{\partial D_s}d\sigma_s\,g_s^-({\bf x})\,v_s({\bf x}) \;,
\ee
where $d\sigma_s$ is the area element of $\partial D_s$, where the subscript $s$ emphasizes that it is $s$-dependent. The last term involves $g_s^-$ 
since, during an infinitesimal change $s\to s+ds$, the moving boundary
adds a thin layer of thickness $v_s({\bf x})ds$ along $\partial D_s$
(or subtracts it if $v_s({\bf x})<0$).
The contribution of this layer is therefore
$ds\int_{\partial D_s}d\sigma_s\,g_s^- v_s$. The value $g_s^-$ appears because,
independently of the sign of $v_s$, the function $g_s$ is integrated over
the interior of the domain $D_s$.
The first term comes
from the explicit $s$-dependence of $g_s$ in the bulk.

To lighten the notations below, when no ambiguity is possible, we use the convention
\be
\int_D g
:=
\int_D d^d{\bf x}\,g({\bf x}),
\qquad
\int_{\partial D} h
:=
\int_{\partial D}d\sigma\,h({\bf x}) ,
\qquad
\int_{\partial D_s} h
:=
\int_{\partial D_s}d\sigma_s\,h({\bf x})
.
\ee
When the domain is $\partial D_s$, limits from the interior and exterior are
denoted by the superscripts $-$ and $+$, respectively.

Let us split $E(s)$ in \eqref{defEs} into interior and exterior parts,
\be
E(s) = E_{\rm int}(s) + E_{\rm ext}(s)
=
\int_{D_s}|\vec \nabla f|^2
+
\int_{D_s^c}|\vec \nabla f_{\rm ext}^{D_s}|^2 ,
\ee
where \(f_{\rm ext}^{D_s}\) is harmonic in \(D_s^c\), equals \(f\) on
\(\partial D_s\), and satisfies the condition at infinity. 
The subscript "${\rm ext}$" emphasizes that it is evaluated
on the complementary domain.

Since \(f\) itself
does not depend on \(s\), the derivative of the interior part, computed using \eqref{derivativeboundary}, comes only from
the moving boundary:
\be
\frac{d}{ds} E_{\rm int}(s) = \frac{d}{ds}\int_{D_s}|\vec \nabla f|^2
=
\int_{\partial D_s}\,
|\vec \nabla f^-|^2 v_s .
\ee

To differentiate the exterior part of $E(s)$ we use the analog of 
\eqref{derivativeboundary} for an integral over the exterior domain, which is 
\be
\frac{d}{ds}\int_{D_s^c} g_s
=
\int_{D_s^c}\partial_s g_s
-
\int_{\partial D_s} g_s^+ v_s .
\ee
The minus sign comes from the fact that, when $D_s$ expands outward, the
exterior domain $D_s^c$ loses a thin layer. Applying this rule we find
\be
\frac{d}{ds} E_{\rm ext}(s) =  
\frac{d}{ds}\int_{D_s^c}|\vec \nabla f_{\rm ext}^{D_s}|^2
=
-\int_{\partial D_s} \,
\left|\left(\vec\nabla f_{\rm ext}^{D_s}\right)^+\right|^2 v_s
+
2\int_{D_s^c} \,
\vec \nabla f_{\rm ext}^{D_s}\cdot
\vec \nabla \partial_s f_{\rm ext}^{D_s}.
\ee

Using integration by parts in the exterior domain and \(\Delta f_{\rm ext}^{D_s}=0\),
\be
2\int_{D_s^c}\vec \nabla f_{\rm ext}^{D_s}\cdot\vec \nabla \partial_s f_{\rm ext}^{D_s}
=
-2\int_{\partial D_s}
\partial_{n_s}^+ f^{D_s}\,
\left(\partial_s f_{\rm ext}^{D_s}\right)^+  \;,
\ee
where the boundary term at infinity
$2\lim_{R\to+\infty}
\int_{S_R}d\sigma\,
\partial_s f_{\rm ext}^{D_s}\,
\vec\nabla f_{\rm ext}^{D_s}\cdot{\bf n}_R$
vanishes from the zero flux/decay assumption.
The minus sign appears because the outward normal of the exterior domain on
\(\partial D_s\) is \(-{\bf n}_s\).

Now we need to evaluate $\partial_s f_{\rm ext}^{D_s}$ on the boundary.
To this aim, recall that the boundary condition for the exterior harmonic extension is
\be
f_{\rm ext}^{D_s}(X_s({\bf y}))
=
f(X_s({\bf y})).
\ee
Differentiating this boundary condition with respect to $s$, at fixed boundary
label ${\bf y}$, gives
\be
\left(\partial_s f_{\rm ext}^{D_s}\right)^+
+
v_s\,\partial_{n_s}^+ f^{D_s}
=
v_s\,\partial_{n_s}^- f^{D_s}.
\ee
Indeed, the left-hand side contains both the explicit variation of the exterior
harmonic extension with $s$ and the displacement of the boundary point, whereas
the right-hand side contains only the displacement of the fixed function $f$.
Therefore
\be \label{derivativefDs}
\left(\partial_s f_{\rm ext}^{D_s}\right)^+
=
-v_s
\left(
\partial_{n_s}^+ f^{D_s}
-
\partial_{n_s}^- f^{D_s}
\right)
=
-v_s[\partial_n f^{D_s}]
\qquad \text{on } \partial D_s .
\ee

Substituting this boundary identity into the exterior contribution, and adding
the interior contribution, gives
\be
E'(s)
=
\int_{\partial D_s}
\left[
\left|\left(\vec \nabla f^{D_s}\right)^-\right|^2
-
\left|\left(\vec \nabla f^{D_s}\right)^+\right|^2
-
2\partial_{n_s}^+ f^{D_s}
\left(
\partial_{n_s}^- f^{D_s}
-
\partial_{n_s}^+ f^{D_s}
\right)
\right]v_s .
\ee
Here the superscripts $+$ and $-$ denote respectively the boundary limits
taken from the exterior and from the interior of $D_s$. In particular
$\left(\vec \nabla f^{D_s}\right)^+$ is the gradient of the exterior harmonic
extension, while $\left(\vec \nabla f^{D_s}\right)^-=
\left(\vec \nabla f \right)^-$, i.e. it equals the gradient of the
original function $f$ inside $D_s$.

On the boundary, the two limits of $f^{D_s}$ coincide. Hence their tangential
gradients are equal:
\be
\vec \tau = (\vec \nabla_{\partial D_s} f^{D_s})^-
=
(\vec \nabla_{\partial D_s} f^{D_s})^+ .
\ee
Writing $a^\pm=\partial_{n_s}^\pm f^{D_s}$,
one has
\be
\left|\left(\vec \nabla f^{D_s}\right)^-\right|^2
=
|\vec\tau|^2+(a^-)^2,
\qquad
\left|\left(\vec \nabla f^{D_s}\right)^+\right|^2
=
|\vec\tau|^2+(a^+)^2 .
\ee
Therefore the expression in brackets becomes
\be
|\vec\tau|^2+(a^-)^2
-
|\vec\tau|^2-(a^+)^2
-
2a^+(a^- -a^+)
=
(a^+-a^-)^2 .
\ee
Thus
\be
E'(s)
=
\int_{\partial D_s}\,
\left(
\partial_{n_s}^+ f^{D_s}
-
\partial_{n_s}^- f^{D_s}
\right)^2
v_s .
\ee
Equivalently, with the jump notation,
\be
E'(s)
=
\int_{\partial D_s}\,
\left(
[\vec \nabla f^{D_s}]\cdot{\bf n}_s
\right)^2
v_s .
\ee
Inserting $v_s$ from the obstacle formula, which reads
\be
v_s({\bf x})
=
\frac{
[\vec \nabla f^{D_s}]({\bf x})\cdot{\bf n}_s({\bf x})
}{
\beta\,\Delta U_s({\bf x})
},
\qquad {\bf x}\in\partial D_s,
\ee
and restoring the more explicit notations, we obtain
\be
E'(s)
=
\frac{1}{\beta}
\int_{\partial D_s}d\sigma_s\,
\frac{
\left(
[\vec \nabla f^{D_s}]({\bf x})\cdot{\bf n}_s({\bf x})
\right)^3
}{
\Delta U_s({\bf x})
}.
\ee
Finally, inserting this into \eqref{der3chis} gives the finite-\(s\) third derivative formula
\be \label{3derivativechis}
\frac{1}{N^2}\partial_s^3\chi(s,N)
=
\frac{1}{\beta^2\Omega_d}
\int_{\partial D_s} d\sigma_s\,
\frac{
\left(
[\vec \nabla f^{D_s}]({\bf x})\cdot{\bf n}_s({\bf x})
\right)^3
}{
\Delta U_s({\bf x})
}
.
\ee

Evaluating this expression at \(s=0\), and using
that $\langle{\mathcal L}_N^3\rangle_c
=
-\frac{1}{N^3}\partial_s^3\chi(0,N)$
we obtain the following general formula for the third cumulant 
\be \label{cum3general}
\langle{\mathcal L}_N^3\rangle_c
\simeq
-\frac{1}{N}
\frac{1}{\beta^2\Omega_d}
\int_{\partial D_0}dS\,
\frac{
\left(
[\vec \nabla f^{D_0}]({\bf x})\cdot{\bf n}({\bf x})
\right)^3
}{
\Delta U({\bf x})
}.
\ee
This is  valid in any space dimension for a smooth 
enough droplet $D_0$.

\subsection{Fourth derivative and fourth cumulant} \label{sec:fourth-derivative}

We now differentiate the finite-$s$ formula for the third derivative obtained
above in \eqref{3derivativechis}. To simplify the notation, we set
\be
K_s({\bf x})
=
[\vec \nabla f^{D_s}]({\bf x})\cdot{\bf n}_s({\bf x})
=
\partial_{n_s}^+ f^{D_s}({\bf x})
-
\partial_{n_s}^- f^{D_s}({\bf x}),
\qquad
A_s({\bf x})=\Delta U_s({\bf x}) ,
\label{defKsAs}
\ee
for ${\bf x}\in\partial D_s$. With this notation, the third derivative reads
\be
\frac{1}{N^2}\partial_s^3\chi(s,N)
=
\frac{1}{\beta^2\Omega_d}
\int_{\partial D_s}
\frac{K_s^3}{A_s}.
\label{chis3KsAs}
\ee
Now we want to evaluate the fourth derivative
\be
\frac{1}{N^2}\partial_s^4\chi(s,N)
=
\frac{1}{\beta^2\Omega_d}
\frac{d}{ds}
\int_{\partial D_s}
\frac{K_s^3}{A_s}.
\label{chis4surface}
\ee

We shall use the following shape-derivative formula for surface integrals.
Let \(X_s({\bf y})\), with \({\bf y}\in\partial D_0\), be a parametrization of
\(\partial D_s\), and let \(J_s({\bf y})\) be the corresponding surface
Jacobian, so that
\be
d\sigma_s(X_s({\bf y}))=J_s({\bf y})\,d\sigma_0({\bf y}) .
\ee
We also denote by
\be
M_s({\bf x})
=
\vec\nabla_{\partial D_s}\cdot{\bf n}_s({\bf x}),
\qquad
{\bf x}\in\partial D_s,
\ee
the mean curvature of \(\partial D_s\), with the outward-normal convention.
With this convention \(M_s=(d-1)/R\) for a sphere of radius \(R\).
Then
\be
\int_{\partial D_s}F_s
=
\int_{\partial D_0}d\sigma_0({\bf y})\,
J_s({\bf y})F_s(X_s({\bf y})) .
\ee
Differentiating at fixed label \({\bf y}\) gives
\be
\frac{d}{ds}\int_{\partial D_s}F_s
=
\int_{\partial D_0}d\sigma_0({\bf y})\,
J_s({\bf y})
\left[
\frac{d}{ds}F_s(X_s({\bf y}))
+
M_s(X_s({\bf y}))v_s(X_s({\bf y}))
F_s(X_s({\bf y}))
\right],
\label{surfacederivativeparam}
\ee
where we used, for the normal motion considered here,
(see Appendix \ref{app:jacobian} for a derivation)
\be
\frac{d}{ds}J_s({\bf y})
=
M_s(X_s({\bf y}))v_s(X_s({\bf y}))J_s({\bf y}) .
\label{surfacejacobian}
\ee
Equivalently, after changing variables back to
\({\bf x}=X_s({\bf y})\), this can be written as
\be
\frac{d}{ds}\int_{\partial D_s}F_s
=
\int_{\partial D_s}
\left[
\left.\frac{d}{ds}F_s(X_s({\bf y}))\right|_{{\bf y}=X_s^{-1}({\bf x})}
+
M_s({\bf x})v_s({\bf x})F_s({\bf x})
\right].
\label{surfacederivative}
\ee

For later use, it is convenient to introduce a compact notation for the first
term in \eqref{surfacederivative}. If \(B_s\) is a quantity defined on
\(\partial D_s\), we write
\be
\dot B_s^\partial({\bf x})
:=
\left.
\frac{d}{ds}B_s(X_s({\bf y}))
\right|_{{\bf y}=X_s^{-1}({\bf x})},
\qquad {\bf x}\in\partial D_s .
\label{defboundarydot}
\ee
Here the dot denotes differentiation at fixed boundary label \({\bf y}\),
followed by the change of variables back to the current boundary point
\({\bf x}\in\partial D_s\). The superscript \(\partial\) is only a reminder
that this is a derivative along the moving boundary; it is unrelated to the
superscripts \(+\) and \(-\), which denote exterior and interior boundary
limits.

In particular, if \(B_s\) is the restriction to \(\partial D_s\) of a smooth
bulk function, one has
\be
\dot B_s^\partial
=
\partial_s B_s
+
v_s\partial_{n_s}B_s
\qquad \text{on } \partial D_s .
\label{boundarydotchainrule}
\ee
This follows from the chain rule and the equation of motion
\[
\frac{dX_s({\bf y})}{ds}
=
v_s(X_s({\bf y})){\bf n}_s(X_s({\bf y})) .
\]
Here and below
\(\partial_{n_s}B_s=\vec\nabla B_s\cdot{\bf n}_s\), and all quantities are
evaluated on \(\partial D_s\).

With this notation, \eqref{surfacederivative} becomes
\be
\frac{d}{ds}\int_{\partial D_s}F_s
=
\int_{\partial D_s}
\left[
\dot F_s^\partial
+
M_s v_s F_s
\right].
\label{surfacederivativecompact}
\ee

Applying \eqref{surfacederivativecompact} to
\[
F_s=\frac{K_s^3}{A_s},
\]
we get
\be
\begin{aligned}
\frac{d}{ds}
\int_{\partial D_s}
\frac{K_s^3}{A_s}
=
\int_{\partial D_s}
\bigg[
3\frac{K_s^2}{A_s}\dot K_s^\partial
-
\frac{K_s^3}{A_s^2}\dot A_s^\partial
+
M_s v_s\frac{K_s^3}{A_s}
\bigg].
\end{aligned}
\label{surfaceFderivative}
\ee
We now compute the two boundary derivatives which appear in this formula.

First, since \(A_s=\Delta U_s=\Delta U+s\Delta f/\beta\) is a smooth bulk
function, the chain rule \eqref{boundarydotchainrule} gives
\be
\dot A_s^\partial
=
\partial_s A_s
+
v_s\partial_{n_s}A_s
=
\frac{1}{\beta}\Delta f
+
v_s\partial_{n_s}A_s
\qquad \text{on } \partial D_s .
\label{boundaryderivativeAs}
\ee

It remains to compute \(\dot K_s^\partial\). The exterior part of
\(f^{D_s}\), denoted by \(f_{\rm ext}^{D_s}\), is harmonic in \(D_s^c\) and
satisfies \(f_{\rm ext}^{D_s}=f\) on \(\partial D_s\). 
Applying formula \eqref{derivativefDs} we obtain 
\be
\left(\partial_s f_{\rm ext}^{D_s}\right)^+
=
-v_sK_s
\qquad \text{on } \partial D_s .
\label{boundarydus}
\ee
Using the obstacle formula
\be
v_s=\frac{K_s}{\beta A_s},
\label{vsKsAs}
\ee
this becomes
\be
\left(\partial_s f_{\rm ext}^{D_s}\right)^+
=
-\frac{1}{\beta}\frac{K_s^2}{A_s}
\qquad \text{on } \partial D_s .
\label{boundarydusKs}
\ee

We denote by \(\Lambda_s^+\) the exterior Dirichlet-to-Neumann operator on
\(\partial D_s\). Namely, for a boundary function \(h\), let \(w_h\) be the
solution of
\be
\Delta w_h=0\quad \text{in }D_s^c,
\qquad
w_h=h \quad \text{on } \partial D_s  ,
\ee
with the same condition at infinity as for \(f^{D_s}\). We then define
\be
\Lambda_s^+ h
=
\partial_{n_s}^+ w_h .
\label{defLambdas}
\ee

We now apply this definition to the boundary function
\be
h_s
=
-\frac{1}{\beta}\frac{K_s^2}{A_s}.
\ee
Indeed, \eqref{boundarydusKs} says precisely that
\be
\left(\partial_s f_{\rm ext}^{D_s}\right)^+
=
h_s
\qquad \text{on } \partial D_s .
\ee
Since \(\partial_s f_{\rm ext}^{D_s}\) is harmonic in \(D_s^c\) and satisfies
the same condition at infinity, uniqueness of the exterior
Dirichlet problem gives
\be
\partial_s f_{\rm ext}^{D_s}=w_{h_s}
\qquad \text{in }D_s^c .
\ee
Therefore
\be
\partial_{n_s}^+ (\partial_s f_{\rm ext}^{D_s})
=
\Lambda_s^+ h_s
=
-\frac{1}{\beta}
\Lambda_s^+\left(\frac{K_s^2}{A_s}\right).
\label{normaldus}
\ee

We now differentiate
\[
K_s
=
\partial_{n_s}^+ f_{\rm ext}^{D_s}
-
\partial_{n_s}^- f^{D_s}
\]
along the moving boundary. Although \(f^{D_s}\) is not smooth across
\(\partial D_s\), its two restrictions are smooth up to the boundary on their
respective sides: \(f_{\rm ext}^{D_s}\) in \(D_s^c\), and \(f\) in \(D_s\).
Thus the chain rule may be applied separately to the two one-sided limits.
For the exterior term, differentiating
\[
\partial_{n_s}^+ f_{\rm ext}^{D_s}
=
{\bf n}_s\cdot\left(\vec\nabla f_{\rm ext}^{D_s}\right)^+
\]
along the moving boundary gives
\be
\left(\partial_{n_s}^+ f_{\rm ext}^{D_s}\right)^{\dot\partial}
=
\dot{\bf n}_s^\partial\cdot
\left(\vec\nabla f_{\rm ext}^{D_s}\right)^+
+
\partial_{n_s}^+\left(\partial_s f_{\rm ext}^{D_s}\right)
+
v_s
\left(\partial_{n_s}^2 f_{\rm ext}^{D_s}\right)^+ .
\ee
Similarly, since the interior function is the fixed function \(f\), one has
\be
\left(\partial_{n_s}^- f^{D_s}\right)^{\dot\partial}
=
\dot{\bf n}_s^\partial\cdot
\left(\vec\nabla f^{D_s}\right)^-
+
v_s
\left(\partial_{n_s}^2 f^{D_s}\right)^- .
\ee
Subtracting the two identities gives
\be
\begin{aligned}
\dot K_s^\partial
&=
\dot{\bf n}_s^\partial\cdot
\left[
\left(\vec\nabla f_{\rm ext}^{D_s}\right)^+
-
\left(\vec\nabla f^{D_s}\right)^-
\right]
+
\partial_{n_s}^+\left(\partial_s f_{\rm ext}^{D_s}\right)
\\
&\quad
+
v_s
\left[
\left(\partial_{n_s}^2 f_{\rm ext}^{D_s}\right)^+
-
\left(\partial_{n_s}^2 f^{D_s}\right)^-
\right].
\end{aligned}
\ee
The first term vanishes. Indeed, since \({\bf n}_s\cdot{\bf n}_s=1\), one has
\[
\dot{\bf n}_s^\partial\cdot{\bf n}_s=0,
\]
so \(\dot{\bf n}_s^\partial\) is tangential to \(\partial D_s\). Moreover the
boundary values of \(f_{\rm ext}^{D_s}\) and \(f\) coincide, hence their
tangential gradients are equal. Therefore
\be
\left(\vec\nabla f_{\rm ext}^{D_s}\right)^+
-
\left(\vec\nabla f^{D_s}\right)^-
=
K_s{\bf n}_s
\qquad \text{on } \partial D_s,
\ee
and the term containing \(\dot{\bf n}_s^\partial\) is zero. Thus
\be
\dot K_s^\partial
=
\partial_{n_s}^+\left(\partial_s f_{\rm ext}^{D_s}\right)
+
v_s
\left[
\left(\partial_{n_s}^2 f_{\rm ext}^{D_s}\right)^+
-
\left(\partial_{n_s}^2 f^{D_s}\right)^-
\right].
\label{boundaryderivativeKsfirst}
\ee

To evaluate the last bracket, we use the decomposition of the Laplacian {on} the boundary,
\be
\Delta g
=
\partial_{n_s}^2 g
+
M_s\partial_{n_s}g
+
\Delta_{\partial D_s}g ,
\label{laplaciandecomp}
\ee
where \(\Delta_{\partial D_s}\) is the Laplace-Beltrami operator on
\(\partial D_s\), and
\(M_s=\vec\nabla_{\partial D_s}\cdot{\bf n}_s\) is the mean curvature with the
outward-normal convention. This formula is used on both sides of the boundary,
with the same normal \({\bf n}_s\), pointing outward from \(D_s\).

Applying \eqref{laplaciandecomp} to the exterior harmonic function
\(f_{\rm ext}^{D_s}\), and taking the boundary limit from the exterior, gives
\be
0
=
\left(\Delta f_{\rm ext}^{D_s}\right)^+
=
\left(\partial_{n_s}^2 f_{\rm ext}^{D_s}\right)^+
+
M_s\partial_{n_s}^+ f^{D_s}
+
\Delta_{\partial D_s}\left(f^{D_s}\right)^+ .
\ee
On the other hand, applying the same decomposition to the interior function
\(f^{D_s}=f\), and taking the boundary limit from the interior, gives
\be
\Delta f
=
\left(\partial_{n_s}^2 f^{D_s}\right)^-
+
M_s\partial_{n_s}^- f^{D_s}
+
\Delta_{\partial D_s}\left(f^{D_s}\right)^- .
\ee
The boundary values of the two functions coincide,
\[
\left(f^{D_s}\right)^+
=
\left(f^{D_s}\right)^-
=
f
\qquad \text{on } \partial D_s ,
\]
and therefore their surface Laplacians also coincide:
\be
\Delta_{\partial D_s}\left(f^{D_s}\right)^+
=
\Delta_{\partial D_s}\left(f^{D_s}\right)^- .
\ee
Subtracting the interior identity from the exterior identity, the
Laplace-Beltrami terms cancel and we obtain
\be
-\Delta f
=
\left[
\left(\partial_{n_s}^2 f_{\rm ext}^{D_s}\right)^+
-
\left(\partial_{n_s}^2 f^{D_s}\right)^-
\right]
+
M_s
\left(
\partial_{n_s}^+ f^{D_s}
-
\partial_{n_s}^- f^{D_s}
\right).
\ee
Using the definition $K_s=\partial_{n_s}^+ f^{D_s}-\partial_{n_s}^- f^{D_s}$,
this gives
\be
\left(\partial_{n_s}^2 f_{\rm ext}^{D_s}\right)^+
-
\left(\partial_{n_s}^2 f^{D_s}\right)^-
=
-M_sK_s-\Delta f
\qquad \text{on } \partial D_s .
\label{normal2jump}
\ee

Combining \eqref{normaldus}, \eqref{boundaryderivativeKsfirst}, and
\eqref{normal2jump}, we obtain
\be
\dot K_s^\partial
=
-\frac{1}{\beta}
\Lambda_s^+\left(\frac{K_s^2}{A_s}\right)
-
v_s\left(M_sK_s+\Delta f\right).
\label{boundaryderivativeKs}
\ee
Using again \eqref{vsKsAs}, this can be written as
\be
\dot K_s^\partial
=
-\frac{1}{\beta}
\Lambda_s^+\left(\frac{K_s^2}{A_s}\right)
-
\frac{K_s}{\beta A_s}
\left(M_sK_s+\Delta f\right).
\label{boundaryderivativeKs2}
\ee

Substituting \eqref{boundaryderivativeAs} and
\eqref{boundaryderivativeKs2} into \eqref{surfaceFderivative}, and using
\eqref{vsKsAs} in the surface-measure term, gives
\bea
\frac{d}{ds}
\int_{\partial D_s}
\frac{K_s^3}{A_s}
&=&
\frac{1}{\beta}
\int_{\partial D_s}
\bigg[
-3\frac{K_s^2}{A_s}
\Lambda_s^+\left(\frac{K_s^2}{A_s}\right)
-3M_s\frac{K_s^4}{A_s^2}
-3\frac{K_s^3\Delta f}{A_s^2}
\nn\\
&&\hspace{1.0cm}
-\frac{K_s^3\Delta f}{A_s^2}
-\frac{K_s^4\partial_{n_s}A_s}{A_s^3}
+
M_s\frac{K_s^4}{A_s^2}
\bigg].
\label{surfaceFderivativeexpanded}
\eea
Collecting the terms proportional to \(M_s\) and to \(\Delta f\), this becomes
\be
\frac{d}{ds}
\int_{\partial D_s}
\frac{K_s^3}{A_s}
=
\frac{1}{\beta}
\int_{\partial D_s}
\left[
-3\frac{K_s^2}{A_s}
\Lambda_s^+\left(\frac{K_s^2}{A_s}\right)
-4\frac{K_s^3\Delta f}{A_s^2}
-2M_s\frac{K_s^4}{A_s^2}
-\frac{K_s^4\partial_{n_s}A_s}{A_s^3}
\right].
\label{surfaceFderivativefinal}
\ee

Inserting \eqref{surfaceFderivativefinal} into \eqref{chis4surface}, we find
the finite-\(s\) fourth derivative formula
\be
\frac{1}{N^2}\partial_s^4\chi(s,N)
=
\frac{1}{\beta^3\Omega_d}
\int_{\partial D_s}
\left[
-3\frac{K_s^2}{A_s}
\Lambda_s^+\left(\frac{K_s^2}{A_s}\right)
-4\frac{K_s^3\Delta f}{A_s^2}
-2M_s\frac{K_s^4}{A_s^2}
-\frac{K_s^4\partial_{n_s}A_s}{A_s^3}
\right].
\label{chis4finite}
\ee

Finally, evaluating at \(s=0\), using
\be
\langle{\mathcal L}_N^4\rangle_c
=
\frac{1}{N^4}\partial_s^4\chi(0,N),
\ee
and writing
\be
K({\bf x})
=
[\vec\nabla f^{D_0}]({\bf x})\cdot{\bf n}({\bf x}),
\qquad
{\bf x}\in\partial D_0,
\label{defKzero}
\ee
we obtain the formula for the fourth cumulant
\be
\begin{aligned}
\langle{\mathcal L}_N^4\rangle_c
&\simeq
-\frac{1}{\beta^3\Omega_dN^2}
\int_{\partial D_0}dS\,
\bigg[
3\frac{K^2}{\Delta U}
\Lambda^+\left(\frac{K^2}{\Delta U}\right)
+
4\frac{K^3\Delta f}{(\Delta U)^2}
\\
&\hspace{3.0cm}
+
2M\frac{K^4}{(\Delta U)^2}
+
\frac{K^4\partial_n\Delta U}{(\Delta U)^3}
\bigg].
\end{aligned}
\label{cum4general}
\ee
Here all quantities in the integrand are evaluated on \(\partial D_0\),
\({\bf n}\) is the outward unit normal, \(M=\vec\nabla_{\partial D_0}\cdot{\bf n}\),
\(\partial_n\Delta U=\vec\nabla(\Delta U)\cdot{\bf n}\), and \(\Lambda^+\) is
the exterior Dirichlet-to-Neumann operator associated with \(D_0\).

\subsection{Specialization to simple geometries} \label{sec:special-geometries}

We now explain how the general formulas \eqref{cum3general} and
\eqref{cum4general} reduce to the explicit expressions stated in the
introduction.

\subsubsection{Disk in $d=2$}

Let \(D_0\) be the disk of radius \(R_0\), with outward normal
\({\bf n}={\bf e}_r\). Then \(dS=R_0d\phi\), \(\Omega_2=2\pi\), and
\be
\Delta U
=
U''(R_0)+\frac{1}{R_0}U'(R_0)
=
U''(R_0)+\frac{1}{R_0^2},
\ee
where we used \(R_0U'(R_0)=1\). If
\({\sf f}(\phi)=f(R_0,\phi)\), the exterior harmonic extension is
\be
f_{\rm ext}^{D_0}(r,\phi)
=
\hat{\sf f}_0
+
\sum_{n\neq0}
\hat{\sf f}_n
\left(\frac{R_0}{r}\right)^{|n|}
e^{in\phi},
\qquad r>R_0 .
\ee
Hence
\be
\partial_r^+ f^{D_0}
=
-\frac{1}{R_0}H\partial_\phi{\sf f},
\qquad
\partial_r^- f^{D_0}
=
\partial_r f(r,\phi)|_{r=R_0}.
\ee
Thus, with the notation used in the introduction,
\be
K(\phi)
=
[\partial_n f^{D_0}]
=
-\left(
\partial_r f(r,\phi)|_{r=R_0}
+
\frac{1}{R_0}H\partial_\phi{\sf f}(\phi)
\right)
=
-J(\phi),
\ee
where
\be
J(\phi)
=
\partial_r f(r,\phi)|_{r=R_0}
+
\frac{1}{R_0}H\partial_\phi{\sf f}(\phi).
\ee
Substituting \(K=-J\) in \eqref{cum3general} gives
\be
\langle{\mathcal L}_N^3\rangle_c
\simeq
\frac{R_0}{2\pi N\beta^2\Delta U}
\int_0^{2\pi}d\phi\,J(\phi)^3.
\ee
which recovers \eqref{cum3circleintro}.

For the fourth cumulant, we also need
\be
\Lambda^+h
=
-\frac{1}{R_0}H\partial_\phi h,
\qquad
M=\frac{1}{R_0},
\qquad
\partial_n\Delta U
=
U'''(R_0)+\frac{U''(R_0)}{R_0}
-\frac{1}{R_0^3}.
\ee
Note that \(R_0U'(R_0)=1\) can be used only after differentiating.
Using \(K=-J\) and the fact that \(\Delta U\) is constant on the boundary,
the Dirichlet-to-Neumann term becomes
\be
3\frac{K^2}{\Delta U}
\Lambda^+\left(\frac{K^2}{\Delta U}\right)
=
-\frac{3}{R_0(\Delta U)^2}
J^2H\partial_\phi(J^2).
\ee
The term cubic in \(K\) gives
\be
4\frac{K^3\Delta f}{(\Delta U)^2}
=
-4\frac{J^3\Delta f}{(\Delta U)^2}.
\ee
Finally, the curvature and normal-derivative terms combine as
\be
2M\frac{K^4}{(\Delta U)^2}
+
\frac{K^4\partial_n\Delta U}{(\Delta U)^3}
=
\frac{J^4}{(\Delta U)^2}
\left[
\frac{2}{R_0}
+
\frac{
U'''(R_0)+R_0^{-1}U''(R_0)-R_0^{-3}
}{\Delta U}
\right].
\ee
After multiplying by the global factor
\(-R_0/(2\pi\beta^3N^2)\) from \eqref{cum4general}, this gives
\be
\begin{aligned}
\langle{\mathcal L}_N^4\rangle_c
&\simeq
\frac{1}{2\pi\beta^3N^2(\Delta U)^2}
\int_0^{2\pi}d\phi\,
\bigg[
3J(\phi)^2H\partial_\phi\!\left(J(\phi)^2\right)
+
4R_0J(\phi)^3\Delta f(R_0,\phi)
\\
&\hspace{3.0cm}
-
\frac{
R_0U'''(R_0)+3U''(R_0)+R_0^{-2}
}{\Delta U}
J(\phi)^4
\bigg],
\end{aligned}
\ee
which is \eqref{4cumdiskintro}.

\subsubsection{Half-space in $d=2$}

Let \(D_0=\{x<0\}\), with boundary \(x=0\) and outward normal
\({\bf n}={\bf e}_x\). Then \(dS=dy\), \(\Omega_2=2\pi\), \(M=0\), and
\be
\Delta U=\tilde U''(0),
\qquad
\partial_n\Delta U=\tilde U'''(0).
\ee
Writing \({\sf f}(y)=f(0,y)\), the exterior harmonic extension in \(x>0\)
has Fourier representation
\be
f_{\rm ext}^{D_0}(x,y)
=
\int_{\mathbb R}\frac{dq}{2\pi}\,
\hat{\sf f}_q e^{-|q|x}e^{iqy}.
\ee
Therefore
\be
\partial_x^+f^{D_0}
=
-H\partial_y{\sf f},
\qquad
\partial_x^-f^{D_0}
=
\partial_x f(x,y)|_{x=0}.
\ee
Thus
\be
K(y)
=
[\partial_n f^{D_0}]
=
-\left[
\partial_x f(x,y)|_{x=0}
+
H\partial_y{\sf f}(y)
\right]
=
-g(y),
\ee
where
\be
g(y)
=
\partial_x f(x,y)|_{x=0}
+
H\partial_y{\sf f}(y).
\ee
Substituting \(K=-g\) in \eqref{cum3general} gives immediately
\be
\langle{\mathcal L}_N^3\rangle_c
\simeq
\frac{1}{2\pi N\beta^2\tilde U''(0)}
\int_{\mathbb R}dy\,g(y)^3,
\ee
which is \eqref{cum3lineintro}.

For the fourth cumulant, the exterior Dirichlet-to-Neumann operator is
\be
\Lambda^+h
=
-H\partial_y h.
\ee
Since \(M=0\), the curvature term drops out. Moreover, \(\Delta U=\tilde
U''(0)\) is constant along the boundary. Therefore
\be
3\frac{K^2}{\Delta U}
\Lambda^+\left(\frac{K^2}{\Delta U}\right)
=
-\frac{3}{\tilde U''(0)^2}
g^2H\partial_y(g^2),
\ee
\be
4\frac{K^3\Delta f}{(\Delta U)^2}
=
-\frac{4g^3(\Delta f)(0,y)}{\tilde U''(0)^2},
\ee
and
\be
\frac{K^4\partial_n\Delta U}{(\Delta U)^3}
=
\frac{g^4\tilde U'''(0)}{\tilde U''(0)^3}.
\ee
Substitution into \eqref{cum4general} gives
\be
\langle{\mathcal L}_N^4\rangle_c
\simeq
\frac{1}{2\pi\beta^3N^2\tilde U''(0)^2}
\int_{-\infty}^{+\infty}dy\,
\left[
3g^2H\partial_y(g^2)
+
4g^3(\Delta f)(0,y)
-
\frac{\tilde U'''(0)}{\tilde U''(0)}g^4
\right],
\ee
which is \eqref{cum4lineintro}. An independent Fourier check of the half-plane formula is given in Appendix
\ref{app:halfspacefouriercheck}.

\subsubsection{Sphere in $d=3$}

Let \(D_0\) be the ball of radius \(R_0\), with outward normal
\({\bf n}={\bf e}_r\). Then \(dS=R_0^2d\Omega\), \(\Omega_3=4\pi\), and
\be
\Delta U
=
U''(R_0)+\frac{2}{R_0}U'(R_0)
=
U''(R_0)+\frac{2}{R_0^3},
\ee
where we used \(R_0^2U'(R_0)=1\). Using the spherical-harmonic expansion
\eqref{defflm}, the exterior harmonic extension is
\be \label{harmonicextension3D}
f_{\rm ext}^{D_0}(r,\theta,\phi)
=
f_{00}Y_{00}
+
\sum_{\ell\geq1}\sum_{m=-\ell}^{\ell}
f_{\ell m}
\left(\frac{R_0}{r}\right)^{\ell+1}
Y_{\ell m}(\theta,\phi).
\ee
Therefore
\be
\partial_r^+f^{D_0}
=
-\frac{1}{R_0}
\sum_{\ell\geq1}\sum_{m=-\ell}^{\ell}
(\ell+1)f_{\ell m}Y_{\ell m}(\theta,\phi),
\qquad
\partial_r^-f^{D_0}
=
\partial_r f(r,\theta,\phi)|_{r=R_0}.
\ee
Thus
\be
K(\theta,\phi)
=
[\partial_n f^{D_0}]
=
-J(\theta,\phi),
\ee
where
\be
J(\theta,\phi)
=
\partial_r f(r,\theta,\phi)|_{r=R_0}
+
\frac{1}{R_0}
\sum_{\ell\geq1}\sum_{m=-\ell}^{\ell}
(\ell+1)f_{\ell m}Y_{\ell m}(\theta,\phi).
\ee
Substituting \(K=-J\) into \eqref{cum3general} gives
\be
\langle{\mathcal L}_N^3\rangle_c
\simeq
\frac{R_0^2}{4\pi N\beta^2\Delta U}
\int_{S^2}d\Omega\,J(\theta,\phi)^3,
\ee
which is \eqref{cum3sphere}.

For the fourth cumulant, the exterior Dirichlet-to-Neumann operator is diagonal
in spherical harmonics (as can be {seen} from its definition by taking $\partial_r|_{r=R_0}$ in 
\eqref{harmonicextension3D}):
\be
\Lambda^+Y_{\ell m}
=
-\frac{\ell+1}{R_0}Y_{\ell m},
\qquad \ell\geq1,
\qquad
\Lambda^+Y_{00}=0.
\ee
Writing
\be
J(\theta,\phi)^2
=
\sum_{\ell=0}^{\infty}\sum_{m=-\ell}^{\ell}
q_{\ell m}Y_{\ell m}(\theta,\phi),
\ee
one obtains
\be
\int_{S^2}d\Omega\,
J^2\Lambda^+(J^2)
=
-\frac{1}{R_0}
\sum_{\ell\geq1}\sum_{m=-\ell}^{\ell}
(\ell+1)|q_{\ell m}|^2 .
\ee
The remaining geometric ingredients are
\be
M=\frac{2}{R_0},
\qquad
\partial_n\Delta U
=
U'''(R_0)
+
\frac{2}{R_0}U''(R_0)
-
\frac{2}{R_0^4}.
\ee
Using \(K=-J\) in \eqref{cum4general}, the Dirichlet-to-Neumann term gives
\be
\frac{3R_0}{4\pi\beta^3N^2(\Delta U)^2}
\sum_{\ell\geq1}\sum_{m=-\ell}^{\ell}
(\ell+1)|q_{\ell m}|^2,
\ee
and the cubic term gives
\be
\frac{1}{4\pi\beta^3N^2(\Delta U)^2}
\int_{S^2}d\Omega\,
4R_0^2J^3(\Delta f)_{r=R_0}.
\ee
Finally, the curvature and normal-derivative terms combine into
\be
-\frac{1}{4\pi\beta^3N^2(\Delta U)^2}
\int_{S^2}d\Omega\,
\frac{
R_0^2U'''(R_0)
+
6R_0U''(R_0)
+
6R_0^{-2}
}{\Delta U}
J^4.
\ee
Thus the specialization of \eqref{cum4general} to the sphere gives
\be
\begin{aligned}
\langle{\mathcal L}_N^4\rangle_c
&\simeq
\frac{1}{4\pi\beta^3N^2(\Delta U)^2}
\bigg[
3R_0
\sum_{\ell\geq1}\sum_{m=-\ell}^{\ell}
(\ell+1)|q_{\ell m}|^2
\\
&\hspace{1.0cm}
+
\int_{S^2}d\Omega\,
\left(
4R_0^2J^3(\Delta f)_{r=R_0}
-
\frac{
R_0^2U'''(R_0)
+
6R_0U''(R_0)
+
6R_0^{-2}
}{\Delta U}
J^4
\right)
\bigg].
\end{aligned}
\ee
This agrees with \eqref{4cumsphereintro}. 
\\

{\bf Remark}. 
The coefficients $q_{\ell m}$ can be expressed explicitly in terms of the boundary data.
Let
\be
\partial_r f(R_0,\theta,\phi)
=
\sum_{\ell=0}^{\infty}\sum_{m=-\ell}^{\ell}
a_{\ell m}Y_{\ell m}(\theta,\phi).
\ee
Then
\be
J(\theta,\phi)
=
\sum_{\ell=0}^{\infty}\sum_{m=-\ell}^{\ell}
j_{\ell m}Y_{\ell m}(\theta,\phi),
\ee
with
\be
j_{\ell m}
=
a_{\ell m}
+
\frac{\ell+1}{R_0}f_{\ell m}{\bf 1}_{\ell\geq1}.
\ee
Since $q_{\ell m}$ are the spherical-harmonic coefficients of $J^2$, one has
\be
q_{\ell m}
=
\sum_{\ell_1,m_1}
\sum_{\ell_2,m_2}
j_{\ell_1m_1}j_{\ell_2m_2}
{\mathcal G}^{\ell m}_{\ell_1m_1,\ell_2m_2},
\ee
where the Gaunt coefficient is
\be
{\mathcal G}^{\ell m}_{\ell_1m_1,\ell_2m_2}
=
\int_{S^2}d\Omega\,
Y_{\ell m}^*
Y_{\ell_1m_1}
Y_{\ell_2m_2}.
\ee
Equivalently, in terms of Wigner $3j$ symbols,
\be
{\mathcal G}^{\ell m}_{\ell_1m_1,\ell_2m_2}
=
(-1)^m
\left[
\frac{(2\ell+1)(2\ell_1+1)(2\ell_2+1)}{4\pi}
\right]^{1/2}
\begin{pmatrix}
\ell & \ell_1 & \ell_2\\
0 & 0 & 0
\end{pmatrix}
\begin{pmatrix}
\ell & \ell_1 & \ell_2\\
-m & m_1 & m_2
\end{pmatrix}.
\ee
Thus, explicitly,
\be
q_{\ell m}
=
\sum_{\ell_1,m_1}
\sum_{\ell_2,m_2}
\left(
a_{\ell_1m_1}
+
\frac{\ell_1+1}{R_0}f_{\ell_1m_1}{\bf 1}_{\ell_1\geq1}
\right)
\left(
a_{\ell_2m_2}
+
\frac{\ell_2+1}{R_0}f_{\ell_2m_2}{\bf 1}_{\ell_2\geq1}
\right)
{\mathcal G}^{\ell m}_{\ell_1m_1,\ell_2m_2}.
\ee


\section{Calculation of third cumulant in $d=2$ using conformal tools}
\label{sec:conformal-tools}

In this section we use complex analysis tools, such as the Riemann maps and Julia's formula, to derive independently the formula for the third cumulant. 

\subsection{Simplified starting formula and second cumulant} \label{sec:conformal-starting} 

We now give an alternative derivation of the starting formula in \(d=2\),
using the exterior Riemann map. This derivation is specific to simply
connected droplets with smooth boundary, but it is useful for the direct
complex-analytic calculation of the third cumulant below.

Starting from \eqref{starting2}, we write, in \(d=2\),
\be
\frac{1}{N^2}\partial_s^2\chi(s,N)
=
\frac{1}{2\pi\beta}E(s),
\qquad
E(s)=A(s)+B(s),
\ee
where
\be
A(s)
=
\int_{D_s} d^2{\bf x}\,
\vec\nabla f\cdot\vec\nabla f,
\qquad
B(s)
=
\int_{D_s^c} d^2{\bf x}\,
\vec\nabla f^{D_s}\cdot\vec\nabla f^{D_s}.
\ee
Here \(f^{D_s}=f\) in \(D_s\), while in \(D_s^c\) it is the exterior harmonic
extension of the boundary value of \(f\).

It is useful to introduce a bilinear version of these two terms. For two smooth functions \(h\) and \(f\), define
\be
A_s(h,f)
=
\int_{D_s} d^2{\bf x}\,
\vec\nabla h\cdot\vec\nabla f,
\qquad
B_s(h,f)
=
\int_{D_s^c} d^2{\bf x}\,
\vec\nabla h^{D_s}\cdot\vec\nabla f^{D_s}.
\label{defABbilinear}
\ee
Thus \(A(s)=A_s(f,f)\) and \(B(s)=B_s(f,f)\).

We first rewrite the exterior contribution \(B_s(h,f)\). Let
\(\psi_s:D_s^c\to\mathbb D^c\) be the exterior Riemann map introduced in
\eqref{uniformizingmap}, where $\mathbb D^c$ is the exterior of the unit disk, and write
\be
w=\psi_s(z),
\qquad
z=\psi_s^{-1}(w).
\ee
We define the functions on \(\mathbb D^c\)
\be
{\mathfrak f}_s(w)
=
f^{D_s}(\psi_s^{-1}(w)),
\qquad
{\mathfrak h}_s(w)
=
h^{D_s}(\psi_s^{-1}(w)).
\ee
They are harmonic in \(\mathbb D^c\). Their boundary values on the unit circle
are
\be
{\sf f}_s(\phi)
=
f(\psi_s^{-1}(e^{i\phi})),
\qquad
{\sf h}_s(\phi)
=
h(\psi_s^{-1}(e^{i\phi})).
\ee
Equivalently, \({\mathfrak f}_s\) is the exterior-disk harmonic extension
\(g_s^{\mathbb D}\) in \eqref{Poissong}, when the boundary datum is \(f\).

We claim that
\be
B_s(h,f)
=
\int_{\mathbb D^c} d^2w\,
\vec\nabla {\mathfrak h}_s\cdot\vec\nabla {\mathfrak f}_s .
\label{confinvB}
\ee
This is the conformal invariance of the Dirichlet integral in two dimensions.
Indeed, write
\[
\psi_s(z)=\psi_s^{(1)}(x,y)+i\psi_s^{(2)}(x,y),
\qquad z=x+iy,
\]
and denote by
\[
{\rm Jac}\,\psi_s(z)
=
\begin{pmatrix}
\partial_x \psi_s^{(1)} & \partial_y \psi_s^{(1)}\\
\partial_x \psi_s^{(2)} & \partial_y \psi_s^{(2)}
\end{pmatrix}
\]
the real Jacobian matrix of the map
\[
(x,y)\mapsto
\big(\psi_s^{(1)}(x,y),\psi_s^{(2)}(x,y)\big).
\]
Since \(\psi_s\) is holomorphic, its real and imaginary parts satisfy the
Cauchy-Riemann equations
\be
\partial_x \psi_s^{(1)}
=
\partial_y \psi_s^{(2)},
\qquad
\partial_y \psi_s^{(1)}
=
-\partial_x \psi_s^{(2)}.
\ee
Therefore
\be
{\rm Jac}\,\psi_s(z)
=
\begin{pmatrix}
\partial_x \psi_s^{(1)} & \partial_y \psi_s^{(1)}\\
-\partial_y \psi_s^{(1)} & \partial_x \psi_s^{(1)}
\end{pmatrix}.
\ee
It follows that
\be
{\rm Jac}\,\psi_s(z)\,
{\rm Jac}\,\psi_s(z)^T
=
\left[
\left(\partial_x \psi_s^{(1)}\right)^2
+
\left(\partial_y \psi_s^{(1)}\right)^2
\right]{\rm Id}
=
|\psi_s'(z)|^2{\rm Id}.
\ee
Also,
\be
d^2w
=
|\det {\rm Jac}\,\psi_s(z)|\,d^2z
=
|\psi_s'(z)|^2d^2z .
\ee
Moreover, since
\[
h^{D_s}(z)
=
{\mathfrak h}_s(\psi_s(z)),
\qquad
f^{D_s}(z)
=
{\mathfrak f}_s(\psi_s(z)),
\]
the chain rule gives
\be
\vec\nabla_z h^{D_s}(z)
=
{\rm Jac}\,\psi_s(z)^T
\vec\nabla_w{\mathfrak h}_s(w),
\qquad
\vec\nabla_z f^{D_s}(z)
=
{\rm Jac}\,\psi_s(z)^T
\vec\nabla_w{\mathfrak f}_s(w).
\ee
Therefore
\be
\vec\nabla_z h^{D_s}\cdot\vec\nabla_z f^{D_s}
=
|\psi_s'(z)|^2
\vec\nabla_w{\mathfrak h}_s\cdot\vec\nabla_w{\mathfrak f}_s .
\ee
Substituting this identity into \(B_s(h,f)\), and using
\(d^2w=|\psi_s'(z)|^2d^2z\), gives \eqref{confinvB}.

We now evaluate the right-hand side of \eqref{confinvB} as a boundary integral
on the unit circle. Let \(a(\phi)\) be a boundary function on
\(\partial\mathbb D\), with Fourier series
\be
a(\phi)
=
\sum_{k\in\mathbb Z}\hat a_k e^{ik\phi},
\qquad
\hat a_k
=
\int_0^{2\pi}\frac{d\phi}{2\pi}\,
a(\phi)e^{-ik\phi}.
\ee
Its bounded exterior harmonic extension is
\be
a^{\mathbb D}(r,\phi)
=
\hat a_0
+
\sum_{k\geq1}\hat a_k r^{-k}e^{ik\phi}
+
\sum_{k\leq -1}\hat a_k r^{k}e^{ik\phi},
\qquad r>1.
\ee
Therefore, for \(r=1\),
\be
\partial_r a^{\mathbb D}|_{r=1}
=
-\sum_{k\in\mathbb Z}|k|\hat a_k e^{ik\phi}
=
-H\partial_\phi a,
\label{normaldiskhilbert}
\ee
where the Hilbert transform is defined by
\be
\widehat{Hg}_k
=
-i\,{\rm sign}(k)\,\hat g_k.
\ee
Since the outward normal of the exterior domain \(\mathbb D^c\) on the inner
boundary \(r=1\) is \(-{\bf e}_r\), Green's formula gives
\be
\int_{\mathbb D^c} d^2w\,
\vec\nabla a^{\mathbb D}\cdot\vec\nabla b^{\mathbb D}
=
-\int_0^{2\pi}d\phi\,a(\phi)\partial_r b^{\mathbb D}(1,\phi)
=
\int_0^{2\pi}d\phi\,a(\phi)H\partial_\phi b(\phi).
\label{diskdirichletboundary}
\ee

Applying this with \(a={\sf h}_s\) and \(b={\sf f}_s\), we obtain from
\eqref{confinvB}
\be
B_s(h,f)
=
\int_0^{2\pi}d\phi\,
{\sf h}_s(\phi)H\partial_\phi{\sf f}_s(\phi).
\label{Bbilinearcircle}
\ee
Equivalently, let us introduce the normalized scalar product on the unit circle,
\be
\langle a,b\rangle
=
\int_0^{2\pi}\frac{d\phi}{2\pi}\,
a(\phi)b(\phi)
=
\sum_{k\in\mathbb Z}\hat a_{-k}\hat b_k,
\label{circlebracket}
\ee
and for later use below, the shorthand
\be
\langle a\rangle
:=
\langle 1,a\rangle
=
\langle a,1\rangle
=
\int_0^{2\pi}\frac{d\phi}{2\pi}\,a(\phi)
=
\hat a_0 .
\label{circleaverage}
\ee
Then we can write
\be
B_s(h,f)
=
2\pi\langle{\sf h}_s,H\partial_\phi{\sf f}_s\rangle.
\ee
In particular, taking \(h=f\),
\be
B(s)
=
2\pi\langle{\sf f}_s,H\partial_\phi{\sf f}_s\rangle
=
2\pi
\sum_{k\in\mathbb Z}|k|\hat f_k(s)\hat f_{-k}(s),
\label{Bg}
\ee
where \(\hat f_k(s)\) are the Fourier coefficients of the boundary function
\({\sf f}_s\).

Note that a more direct perturbative construction of the exterior harmonic extension near
the disk, based on the multipole expansion in the exterior of a perturbed
circle, is given in Appendix \ref{app:perturbativeharmonicdisk}.

\subsection{Calculation of the third cumulant for the disk} \label{sec:conformal-disk}

We now specialize to the case where \(D_0=\mathbb D\) is the unit disk and $U$ is radial. Thus
\(\psi_0(z)=z\). The general-radius formula is recovered by rescaling, and we
compare with \eqref{cum3circleintro} at the end.

The third cumulant is obtained from 
\be
\langle{\mathcal L}_N^3\rangle_c 
=
-\frac{1}{N^3}\partial_s^3\chi(0,N)
=
-\frac{1}{2\pi N\beta}E'(0).
\label{cum3fromEprime}
\ee
where we used \eqref{der3chis}. We compute \(E'(0)\) by differentiating the interior and exterior Dirichlet
forms.

Let \(v(t)\), \(t\in\partial\mathbb D\), be the outward normal velocity of the
boundary at \(s=0\). We write the infinitesimal deformation of the exterior
Riemann map as
\be
\psi_s(z)
=
z\exp\left(s\eta(z)\right)+O(s^2),
\qquad |z|>1.
\label{psieta}
\ee
Julia's formula \eqref{julia}, evaluated at \(s=0\), gives
\be
\eta(z)
=
\oint_{\partial\mathbb D}
\frac{dt}{2\pi i\,t}
\frac{t+z}{t-z}\,v(t),
\qquad |z|>1.
\label{etaJulia}
\ee
Writing
\be
v(t)=\sum_{k\in\mathbb Z}\hat v_k t^k,
\qquad |t|=1,
\ee
and using, for \(|z|>1\),
\be
\frac{t+z}{t-z}
=
-1-2\sum_{p\geq1}\left(\frac{t}{z}\right)^p,
\ee
we obtain
\be
\eta(z)
=
-\hat v_0
-
2\sum_{p\geq1}\hat v_{-p}z^{-p}.
\ee

Taking the boundary limit \(z=e^{i\phi}\), and using that \(v\) is real, this gives
\be
\eta(e^{i\phi})
=
-v(e^{i\phi})+iH v(e^{i\phi}).
\label{etaBoundary}
\ee
In particular, \({\rm Re}\,\eta=-v\), as it should: an outward displacement of
the unit circle corresponds to a decrease of the leading coefficient
$c_s=\lim_{z\to\infty}\psi_s(z)/z$, and hence an increase of the logarithmic
capacity $c_s^{-1}$.

We now differentiate the transported boundary datum
\be
{\sf f}_s(\phi)
=
f\!\left(\psi_s^{-1}(e^{i\phi})\right).
\ee
Using \eqref{psieta}, the inverse map satisfies
\be
\psi_s^{-1}(w)
=
w\exp\left(-s\eta(w)\right)+O(s^2).
\ee
Taking \(w=e^{i\phi}\) and using \eqref{etaBoundary}, 
we obtain
\be
\psi_s^{-1}(e^{i\phi})
=
e^{i\phi}
\exp\left[
s v(e^{i\phi})
-
is\,H v(e^{i\phi})
\right]
+O(s^2).
\ee
Thus, in polar coordinates, the point
\(\psi_s^{-1}(e^{i\phi})\) has, to first order in \(s\),
\be
r_s(\phi)
=
1+s\,v(e^{i\phi})+O(s^2),
\qquad
\phi_s
=
\phi-s\,H v(e^{i\phi})+O(s^2).
\ee
Therefore
\be
{\sf f}_s(\phi)
=
f(r_s(\phi),\phi_s)
=
f(1,\phi)
+
s\left[
v(e^{i\phi})\partial_r f(1,\phi)
-
H v(e^{i\phi})\partial_\phi{\sf f}(\phi)
\right]
+O(s^2).
\ee
Hence
\be
\left.\partial_s{\sf f}_s(\phi)\right|_{s=0}
=
v(e^{i\phi})\partial_r f(1,\phi)
-
H v(e^{i\phi})\partial_\phi{\sf f}(\phi).
\label{fsdot}
\ee
and similarly
\be
\left.\partial_s{\sf h}_s(\phi)\right|_{s=0}
=
v(\phi)\partial_r h(1,\phi)
-
(Hv)(\phi)\partial_\phi{\sf h}(\phi).
\label{hsdot}
\ee

We now differentiate the interior term of \(E(s)\). We use the elementary
Hadamard formula for a moving domain,
\be
\frac{d}{ds}\int_{D_s}F({\bf x})\,d^2{\bf x}
=
\int_{\partial D_s}F({\bf x})v_s({\bf x})\,d\ell({\bf x}),
\ee
valid when \(F\) has no explicit \(s\)-dependence. Applying it to
\(F=\vec\nabla h\cdot\vec\nabla f\), and using \(D_0=\mathbb D\), gives
\be
\left.\frac{d}{ds}A_s(h,f)\right|_{s=0}
=
\int_{\partial\mathbb D}
v\,\vec\nabla h\cdot\vec\nabla f\,d\ell.
\ee
On the unit circle, \(d\ell=d\phi\) and
\[
\vec\nabla h\cdot\vec\nabla f
=
\partial_r h\,\partial_r f
+
\partial_\phi{\sf h}\,\partial_\phi{\sf f}.
\]
Therefore
\be
\left.\frac{d}{ds}A_s(h,f)\right|_{s=0}
=
2\pi
\left\langle
v\left(
\partial_r h\,\partial_r f
+
\partial_\phi{\sf h}\,\partial_\phi{\sf f}
\right)
\right\rangle.
\label{Aprimebilinear}
\ee
All functions {on} the right-hand side are evaluated at \(r=1\).

For the exterior term, we differentiate
$B_s(h,f)
=
2\pi\langle{\sf h}_s,H\partial_\phi{\sf f}_s\rangle$.
Let
\be
\dot{\sf f}
:=
\left.\partial_s{\sf f}_s\right|_{s=0},
\qquad
\dot{\sf h}
:=
\left.\partial_s{\sf h}_s\right|_{s=0}.
\ee
Then
\be
\left.\frac{d}{ds}B_s(h,f)\right|_{s=0}
=
2\pi
\left[
\langle \dot{\sf h},H\partial_\phi{\sf f}\rangle
+
\langle {\sf h},H\partial_\phi\dot{\sf f}\rangle
\right].
\ee
Using that \(H\partial_\phi\) is self-adjoint on the circle,
\be
\langle a,H\partial_\phi b\rangle
=
\langle H\partial_\phi a,b\rangle,
\label{Hpartialselfadjoint}
\ee
we rewrite the second term as
\be
\langle {\sf h},H\partial_\phi\dot{\sf f}\rangle
=
\langle H\partial_\phi{\sf h},\dot{\sf f}\rangle.
\ee
Therefore
\be
\left.\frac{d}{ds}B_s(h,f)\right|_{s=0}
=
2\pi
\left[
\langle \dot{\sf h},H\partial_\phi{\sf f}\rangle
+
\langle H\partial_\phi{\sf h},\dot{\sf f}\rangle
\right].
\ee
Now inserting \eqref{fsdot} and its analogue
\be
\dot{\sf h}
=
v\partial_r h
-
(Hv)\partial_\phi{\sf h} \quad , \quad \dot{\sf f}
=
v\partial_r f
-
(Hv)\partial_\phi{\sf f} ,
\ee
we obtain
\be
\begin{aligned}
\left.\frac{d}{ds}B_s(h,f)\right|_{s=0}
=
2\pi\bigg[
&
\left\langle
v\partial_r h\,H\partial_\phi{\sf f}
\right\rangle
-
\left\langle
(Hv)\partial_\phi{\sf h}\,H\partial_\phi{\sf f}
\right\rangle
\\
&+
\left\langle
H\partial_\phi{\sf h}\,
v\partial_r f
\right\rangle
-
\left\langle
H\partial_\phi{\sf h}\,
(Hv)\partial_\phi{\sf f}
\right\rangle
\bigg].
\end{aligned}
\label{Bprimebilinear}
\ee

The terms containing \(Hv\) can be rewritten so as to factor out \(v\). We use
the antisymmetry of \(H\),
\[
\langle Hv,a\rangle=-\langle v,Ha\rangle,
\]
and the Hilbert-transform identity
\be
H(aHb+bHa)
=
(Ha)(Hb)-ab,
\label{Hilbertidentity}
\ee
valid when \(a\) and \(b\) have zero average. In the present application
\(a=\partial_\phi{\sf h}\) and \(b=\partial_\phi{\sf f}\), so this condition is
satisfied. A proof is given in Appendix \ref{app:Hilbertidentity}.
Applying \eqref{Hilbertidentity}
with \(a=\partial_\phi{\sf h}\) and \(b=\partial_\phi{\sf f}\), the sum of
\eqref{Aprimebilinear} and \eqref{Bprimebilinear} simplifies to
\be
\left.\frac{d}{ds}\left(A_s(h,f)+B_s(h,f)\right)\right|_{s=0}
=
2\pi
\left\langle
v\,
Lh\,Lf
\right\rangle,
\label{Eprimebilinear}
\ee
where we introduced the boundary operator
\be
Lf
=
\partial_r f(1,\phi)
+
H\partial_\phi{\sf f}(\phi),
\qquad
{\sf f}(\phi)=f(1,\phi).
\label{defLcircle}
\ee

Taking \(h=f\), we get
\be
E'(0)
=
2\pi
\left\langle
v\,(Lf)^2
\right\rangle.
\label{Eprimecircle}
\ee
It remains to insert the normal velocity associated with the Coulomb-gas
deformation \(U_s=U+s f/\beta\). With our jump convention,
\be
K
=
[\partial_n f^{D_0}]
=
\partial_r^+ f^{D_0}
-
\partial_r^- f^{D_0}.
\ee
For the unit disk,
\be
\partial_r^+f^{D_0}
=
-H\partial_\phi{\sf f},
\qquad
\partial_r^-f^{D_0}
=
\partial_r f(1,\phi),
\ee
hence
\be
K=-Lf.
\label{KminusL}
\ee
The normal velocity is
\be
v
=
\frac{K}{\beta\Delta U}
=
-\frac{Lf}{\beta\Delta U},
\label{velocitycircle}
\ee
where \(\Delta U=U''(1)+U'(1)\) is constant on the boundary. Therefore
\be
E'(0)
=
-\frac{2\pi}{\beta\Delta U}
\left\langle
(Lf)^3
\right\rangle.
\label{Eprimefinalcircle}
\ee
Using \eqref{cum3fromEprime}, we finally obtain
\be
\langle{\mathcal L}_N^3\rangle_c
\simeq
\frac{1}{N\beta^2\Delta U}
\left\langle
\left(
\partial_r f(1,\phi)
+
H\partial_\phi{\sf f}(\phi)
\right)^3
\right\rangle.
\label{cum3unitdiskcomplex}
\ee
This is the unit-radius version of \eqref{cum3circleintro}. For a disk of
radius \(R_0\), one has instead
\[
Lf
=
\partial_r f(R_0,\phi)
+
\frac{1}{R_0}H\partial_\phi{\sf f}(\phi),
\]
and \eqref{cum3unitdiskcomplex} becomes \eqref{cum3circleintro}.

\subsection{Extension of the conformal derivation to a general simply connected droplet} \label{sec:conformal-general-droplet}

The disk calculation above extends naturally to any simply connected droplet
with smooth boundary. We describe the modification, since it also clarifies the
geometric meaning of the Hilbert transform appearing in the disk formula.

Let \(\psi_0:D_0^c\to\mathbb D^c\) be the exterior Riemann map, normalized as
in \eqref{normalizationpsi}, and let
\[
\varphi_0=\psi_0^{-1}
\]
be its inverse. We parametrize the boundary by
\be
t(\theta)=\varphi_0(e^{i\theta}),
\qquad
0\leq\theta<2\pi .
\label{generalconformalboundaryparam}
\ee
We also define the boundary metric factor
\be
m(\theta)=|\varphi_0'(e^{i\theta})|.
\label{generalconformalm}
\ee
Then the arclength element on \(\partial D_0\) is
\be
d\ell=m(\theta)d\theta.
\label{generalconformalmeasure}
\ee
For a function \(f\), we denote its boundary value pulled back to the unit
circle by
\be
\tilde f(\theta)
=
f(t(\theta))
=
f(\varphi_0(e^{i\theta})).
\label{generalconformalftilde}
\ee

The exterior harmonic extension of \(f\) in \(D_0^c\) is obtained by pulling
back the exterior harmonic extension of \(\tilde f\) in \(\mathbb D^c\). Thus,
if \(\tilde f^{\mathbb D}\) denotes the bounded harmonic function in
\(\mathbb D^c\) with boundary value \(\tilde f\), then
\be
f^{D_0}(z)
=
\tilde f^{\mathbb D}(\psi_0(z)),
\qquad z\in D_0^c.
\label{generalconformalextension}
\ee
On the unit circle one has
\be
\partial_\rho \tilde f^{\mathbb D}\big|_{\rho=1^+}
=
-H\partial_\theta\tilde f,
\label{generalconformalunitdtn}
\ee
where \(w=\rho e^{i\theta}\). Since the conformal map rescales lengths by
\(m(\theta)\) on the boundary, the physical exterior normal derivative is
\be
\partial_n^+ f^{D_0}(t(\theta))
=
\frac{1}{m(\theta)}
\partial_\rho \tilde f^{\mathbb D}(e^{i\theta})
=
-\frac{1}{m(\theta)}H\partial_\theta\tilde f(\theta).
\label{generalconformaldtn}
\ee
This is the conformal form of the exterior Dirichlet-to-Neumann operator.

It is therefore natural to introduce
\be
L_{D_0}f(t(\theta))
=
\partial_n f(t(\theta))
+
\frac{1}{m(\theta)}H\partial_\theta\tilde f(\theta).
\label{generalconformalL}
\ee
With the jump convention used throughout the paper, exterior minus interior,
we have
\be
K(t(\theta))
=
[\partial_n f^{D_0}](t(\theta))
=
\partial_n^+ f^{D_0}(t(\theta))
-
\partial_n f(t(\theta))
=
-L_{D_0}f(t(\theta)).
\label{generalconformalK}
\ee

Let the boundary move with physical normal velocity \(v(t)\). In the
\(w\)-plane this corresponds to a radial velocity
\[
\tilde v(\theta)=\frac{v(t(\theta))}{m(\theta)}.
\]
Indeed, a radial displacement \(d\rho\) of the point \(e^{i\theta}\) in the
\(w\)-plane is mapped by \(\varphi_0\) to a physical normal displacement
\(m(\theta)d\rho\) of the boundary point \(t(\theta)\). Thus the disk
calculation applies in the \(w\)-plane with the replacement
\[
v\longrightarrow \tilde v=\frac{v}{m}.
\]

Repeating the disk computation in the \(\theta\)-coordinate gives
\be
E'(0)
=
\int_0^{2\pi}d\theta\,
m(\theta)\,
v(t(\theta))\,
\left(L_{D_0}f(t(\theta))\right)^2.
\label{generalconformalEprimeTheta}
\ee
Using \(d\ell=m(\theta)d\theta\), this is simply
\be
E'(0)
=
\int_{\partial D_0}d\ell\,
v\,
\left(L_{D_0}f\right)^2.
\label{generalconformalEprime}
\ee
Finally, the obstacle formula gives
\be
v
=
\frac{K}{\beta\Delta U}
=
-\frac{L_{D_0}f}{\beta\Delta U}.
\label{generalconformalvelocity}
\ee
Substituting into \eqref{generalconformalEprime}, we obtain
\be
E'(0)
=
\frac{1}{\beta}
\int_{\partial D_0}d\ell\,
\frac{K^3}{\Delta U}.
\label{generalconformalEprimeK}
\ee
Since
\[
\langle{\mathcal L}_N^3\rangle_c
=
-\frac{1}{2\pi N\beta}E'(0),
\]
this gives
\be
\langle{\mathcal L}_N^3\rangle_c
\simeq
-\frac{1}{2\pi N\beta^2}
\int_{\partial D_0}d\ell\,
\frac{K^3}{\Delta U},
\label{generalconformalcum3}
\ee
which is the \(d=2\) case of the general formula \eqref{cum3general}.
\\

\begingroup
\hypersetup{linkcolor=black}

\addtocontents{toc}{\protect\setcounter{tocdepth}{1}}
\section[Limit inversion in the 2d determinantal case]{Limit inversion in the 2d determinantal case}
\label{app:analyticU}\label{sec:rigorous}

In this section we work with $d=2$, $\beta=2$, and $U,f$ real analytic near the droplet. To justify the limit interchange in \eqref{starting}, we use a uniform expansion of the tilted mean, from the bulk Bergman expansion \cite{Berman2009},  and the orthogonal-polynomial asymptotics of \cite{HedenmalmWennman2021} close to the boundary.

\subsection{Assumptions and statement}
\label{anapp:assumptions}
With the normalization of \eqref{eq:pdf}--\eqref{energy}, we have
\begin{equation}
 \dd\mathbb P_U=\frac1{Z_{N,U}}\prod_{i<j}|z_i-z_j|^2
 e^{-2N\sum_iU(z_i)}\prod_i\dd^2z_i,
 \qquad \mathcal L_N(f)=\sum_{i=1}^N f(z_i),
 \label{anapp:law}
\end{equation}
where $\dd^2z=\dd x\,\dd y$ and $\Delta=\partial_x^2+\partial_y^2$. We assume that:
\begin{enumerate}[(i)]
\item $U\in C^\infty(\mathbb C;\mathbb R)$, $f\in C_c^\infty(\mathbb C;\mathbb R)$, and $U(z)\ge(1+\eta)\log|z|$ for large $|z|$, for some $\eta>0$.
\item The droplet is the closure of a bounded simply connected domain $D$ with real-analytic Jordan boundary. On a neighborhood $V$ of $\overline D$, the functions $U,f$ are real analytic and $a:=\Delta U>0$.
\item There is no exterior contact set: the equilibrium measure and potential satisfy
\begin{equation}
 \dd\nu=\frac{a}{2\pi}\mathbf1_D\dd^2z,\qquad \nu(\mathbb C)=1,\qquad
 \widehat U(z)=\ell+\int\log|z-w|\,\dd\nu(w),
 \label{anapp:equilibrium}
\end{equation}
for some constant $\ell\in\mathbb R$, with $\widehat U=U$ on $\overline D$ and $\widehat U<U$ on $\mathbb C\setminus\overline D$.
\end{enumerate}
We choose $V$ bounded so that $a\ge c>0$ on $\overline V$. We only require analyticity on $V$, which is compatible with $f\in C_c^\infty$.

We denote by $n$ the outward normal to $D$ and by $\Lambda^+$ the exterior Dirichlet-to-Neumann operator, with boundedness at infinity.  We recall that $f^D=f$ on $\overline D$ and that $f^D$ is harmonic and bounded in $\mathbb C\setminus\overline D$.  We write
\begin{equation}
 J=\partial_n f-\Lambda^+(f|_{\partial D})=-[\partial_n f^D],
 \qquad M=\operatorname{div}_{\partial D}n.
 \label{anapp:boundary-data}
\end{equation}

In the following main result of this section,  the notation $X_N=\OO_{U,f}(Y_N)$ means that the implicit multiplicative constants only depend on $U$ and $f$: $|X_N|\leq C(U,f)|Y_N|$ for all $N$.

\begin{proposition}\label{anapp:prop}
Under assumptions $(i)$--$(iii)$, we have
\begin{align}
 \langle\mathcal L_N(f)^3\rangle_c
 &=\frac1{8\pi N}\int_{\partial D}\frac{J^3}{a}\,\dd S
   +\OO_{U,f}(N^{-2}),\label{anapp:c3}\\
 \langle\mathcal L_N(f)^4\rangle_c
 &=\frac1{16\pi N^2}\int_{\partial D}
 \left[\frac{4(\Delta f)J^3}{a^2}
 -\left(\frac{2M}{a^2}+\frac{\partial_n a}{a^3}\right)J^4
 -3\frac{J^2}{a}\Lambda^+\!\left(\frac{J^2}{a}\right)\right]\dd S
 +\OO_{U,f}(N^{-3}).\label{anapp:c4}
\end{align}
\end{proposition}

We first describe the proof. For $U_s=U+sf/2$, we denote the equilibrium domain by $D_s$ and write
\begin{equation}
 M_N(s)=\langle\mathcal L_N(f)\rangle_{U_s},\qquad
 \mu(s)=\frac1{2\pi}\int_{D_s}f\Delta U_s\,\dd^2z.
 \label{anapp:mean}
\end{equation}
From Lemma~\ref{anapp:mean-expansion}, we have an expansion of $M_N$ with smooth coefficients and a uniform $\OO(N^{-32})$ remainder. We combine the derivative bound \eqref{anapp:crude-derivatives} with the finite-difference estimate \eqref{anapp:finite-difference} to obtain $M_N^{(r)}(0)=N\mu^{(r)}(0)+\OO(1)$, $r=2,3$. By the tilt identity \eqref{anapp:exact-tilt}, it then remains to compute $\mu''(0)$ and $\mu'''(0)$.

\subsection{Equilibrium droplets}
\label{anapp:geometry-section}
We write $a_s=\Delta U_s$. { Let $\nu_{s,\tau}$ be the minimizer of $$\int U_s\,\dd\nu-\frac12\iint\log|z-w|\,\dd\nu(z)\dd\nu(w)$$ among positive measures of mass $\tau$. We use $\tau=j/N$ for degree-$j$ orthogonal polynomials in Lemma~\ref{anapp:HW}.}

For a smooth Jordan curve $\Gamma$,  we call a collar a neighborhood of the form $\{\zeta+tn_\Gamma(\zeta):\zeta\in\Gamma,\ |t|<\varepsilon\}$, where $n_\Gamma$ is the outward unit normal and $\varepsilon>0$ is small enough that these coordinates are unique. Its inner part corresponds to $-\varepsilon<t<0$.

\begin{lemma}\label{anapp:analytic-family}
There exist $s_0,\delta>0$,  with $\delta<\min\{1,\eta\}$,  such that, for $|s|\le s_0$ and $|\tau-1|\le\delta$, we have
\begin{equation}
 \dd\nu_{s,\tau}=\frac{a_s}{2\pi}\mathbf1_{D_{s,\tau}}\dd^2z,
 \qquad D_{0,1}=D.
 \label{anapp:mass-measure}
\end{equation}
The boundaries are real-analytic Jordan curves.  The exterior conformal maps $\Phi_{s,\tau}:\{|w|>1\}\to\mathbb C\setminus\overline{D_{s,\tau}}$, normalized by
\[
 \Phi_{s,\tau}(w)=c_{s,\tau}w+\OO(1),\qquad c_{s,\tau}>0,
\]
extend injectively to a common domain $\{|w|>\rho\}$, $\rho<1$. Moreover, on smaller common collars they depend real analytically on $(s,\tau)$, and { $|\partial_w\Phi_{s,\tau}|$ is uniformly bounded away from zero}.

Writing $D_s=D_{s,1}$, $\phi_{s,\tau}=\Phi_{s,\tau}^{-1}$ and $J_{s,\tau}=\partial_n f-\Lambda^+_{s,\tau}f$, we have the normal velocities
\begin{equation}
 v_s=-\frac{J_{s,\tau}}{2a_s},\qquad
 v_\tau=\frac{|\phi_{s,\tau}'|}{a_s}>0
 \quad\hbox{on }\partial D_{s,\tau}.
 \label{anapp:velocities}
\end{equation}
In particular, $D_{s,\tau}$ increases with $\tau$, and we have
\begin{equation}
 \partial_\tau\nu_{s,\tau}=\omega_{s,\tau}^{\infty},\qquad
 \dd\omega_{s,\tau}^{\infty}=\frac{|\phi_{s,\tau}'|}{2\pi}\,\dd S.
 \label{anapp:mass-derivative}
\end{equation}
{ The obstacle inequality is $\widehat U_{s,\tau}\le U_s$, where $\widehat U_{s,\tau}(z)=\ell_{s,\tau}+\int\log|z-w|\,\dd\nu_{s,\tau}(w)$ equals $U_s$ on $\overline{D_{s,\tau}}$. It is strict outside $\overline{D_{s,\tau}}$, uniformly on compact sets separated from these domains.}
\end{lemma}

\begin{proof}
We first construct the domains and measures in \eqref{anapp:mass-measure}, and then verify equilibrium. For an exterior map $\Phi$ near $\Phi_{0,1}$, we set $Z(\theta)=\Phi(e^{i\theta})$, $X=-i\partial_\theta Z$, and
\[
 H(e^{ik\theta})=-i\sgn(k)e^{ik\theta},\qquad
 \mathscr S h=h_0+2\sum_{k<0}h_ke^{ik\theta}=h-iHh
 \quad(h\hbox{ real}).
\]
By \eqref{inversejuliaK}, the prescribed normal velocities in \eqref{anapp:velocities} correspond to
\begin{align}
 \partial_s Z&=\mathscr F_s(s,Z):=-X\mathscr S\!\left(
 \frac{\nabla f(Z)\cdot X+H\partial_\theta(f\circ Z)}{2a_s(Z)|X|^2}\right),\notag\\
 \partial_\tau Z&=\mathscr F_\tau(s,Z):=X\mathscr S\!\left(\frac1{a_s(Z)|X|^2}\right).
 \label{anapp:analytic-evolution}
\end{align}
Here the scalar product is in $\mathbb R^2$. On the scale $\|h\|_b=\sum_k(1+|k|)|h_k|e^{b|k|}$,  a Cauchy estimate gives
\begin{equation}
 \|\mathscr F_q(s,Z)-\mathscr F_q(s,\widetilde Z)\|_{b'}
 \le\frac C{b-b'}\|Z-\widetilde Z\|_b,\qquad
 \|\mathscr F_q(s,Z)\|_{b'}\le\frac C{b-b'},
 \label{anapp:CK-estimate}
\end{equation}
for $q=s,\tau$ and $0<b'<b<b_0$, on a small ball about the initial analytic curve. The factor $(b-b')^{-1}$ accounts for one tangential derivative. The denominators stay away from zero and the vector fields admit analytic complexifications, so \eqref{anapp:CK-estimate} allows us to apply the Cauchy--Kowalevski theorem on Banach scales \cite{Nishida1977}, first in $s$ at $\tau=1$, then in $\tau$ for each $s$. For small $s_0,\delta$, the maps remain univalent on $\{|w|>\rho\}$, with $|\Phi'|\ge c>0$, and depend analytically on $(s,\tau)$ on smaller collars.

We write $P\nu(z)=\int\log|z-w|\,\dd\nu(w)$. Reynolds' formula differentiates integrals over moving domains: $\partial_q\int_{D_q}h_q\,\dd^2z=\int_{D_q}\partial_qh_q\,\dd^2z+\int_{\partial D_q}h_qv_q\,\dd S$. Along the first flow, it gives, with $J_s=J_{s,1}$,
\begin{equation}
 \partial_s\nu_{s,1}
 =\frac1{4\pi}\left((\Delta f)\mathbf1_{D_s}\dd^2z-J_s\dd S\right)
 =\frac1{4\pi}\Delta f^{D_s},
 \label{anapp:measure-derivative}
\end{equation}
distributionally. Thus the mass remains one and, writing $m_s=f^{D_s}(\infty)$, we have
\[
 \partial_sP\nu_{s,1}=\tfrac12(f^{D_s}-m_s),\qquad
 \partial_s\ell_{s,1}=\tfrac12m_s,\qquad \ell_{0,1}=\ell.
\]
Along the second flow, Reynolds' formula gives \eqref{anapp:mass-derivative}, hence $\nu_{s,\tau}(\mathbb C)=\tau$. By the mean-value property under $\Phi_{s,\tau}$, we have $P\omega_{s,\tau}^{\infty}=\log c_{s,\tau}$ inside $D_{s,\tau}$. We choose $\partial_\tau\ell_{s,\tau}=-\log c_{s,\tau}$. On a fixed disk $B\subset D_{s,\tau}$ common to all parameters, the function
\[
 q_{s,\tau}=U_s-\ell_{s,\tau}-P\nu_{s,\tau}
\]
is constant first along $s$ at $\tau=1$, then along $\tau$. It therefore vanishes on $B$, and then throughout $D_{s,\tau}$ by harmonic unique continuation.

The logarithmic potential of the bounded density is $C^1$ across the boundary. Since $q_{s,\tau}=0$ inside, this gives $q_{s,\tau}=\partial_nq_{s,\tau}=0$ on the boundary. As $\Delta q_{s,\tau}=a_s$ outside, $\partial_n^2q_{s,\tau}=a_s$ on the exterior side, hence
\begin{equation}
 q_{s,\tau}(\zeta+tn)=\tfrac12a_s(\zeta)t^2+\OO(t^3),
 \qquad \zeta\in\partial D_{s,\tau},\quad t\downarrow0,
 \label{anapp:quadratic-gap}
\end{equation}
uniformly in the parameters. The exterior gap follows near the boundary from $a_s\ge c>0$, on separated compact sets from $(iii)$ and continuity, and at infinity from $(i)$ and $\delta<\eta$. Thus the measures satisfy the equilibrium variational conditions. For every $\tau$, the bounded exterior harmonic function $\partial_s\widehat U_{s,\tau}$ has boundary value $f/2$, so it equals $f^{D_{s,\tau}}/2$. Differentiating $\partial_nq_{s,\tau}=0$ gives $J_{s,\tau}/2+a_sv_s=0$.
\end{proof}

\subsection{Mean expansion}
\label{anapp:expansions-section}
We denote by $p_{j,N,s}$ the degree-$j$ orthonormal polynomial for $e^{-2NU_s}\dd^2z$, with positive leading coefficient, and write
\[
 \|h\|_{N,s}^2=\int_{\mathbb C}|h|^2e^{-2NU_s}\dd^2z.
\]
We denote by $\mathcal Q_{s,\tau}$ the bounded holomorphic function in the exterior of $D_{s,\tau}$ with $\Re\mathcal Q_{s,\tau}=U_s$ on the boundary and $\Im\mathcal Q_{s,\tau}(\infty)=0$.
Since $U_s\circ\Phi_{s,\tau}$ is real analytic on $|w|=1$, its Fourier coefficients decay exponentially. The exterior holomorphic Fourier series with this real part therefore defines $\mathcal Q_{s,\tau}\circ\Phi_{s,\tau}$ across the circle. After shrinking the collar, both $\mathcal Q_{s,\tau}$ and $\phi_{s,\tau}$ extend holomorphically across $\partial D_{s,\tau}$, with $\phi_{s,\tau}\ne0$. We may thus define the harmonic function
\begin{equation}
 \breve U_{s,\tau}=\tau\log|\phi_{s,\tau}|+\Re\mathcal Q_{s,\tau}
 \label{anapp:harmonic-continuation}
\end{equation}
on the exterior and an inner collar. On the exterior, $\breve U_{s,\tau}-\widehat U_{s,\tau}$ is harmonic, bounded at infinity and zero on the boundary, hence vanishes by uniqueness for the exterior Dirichlet problem. Thus $\breve U_{s,\tau}$ is the harmonic continuation of $\widehat U_{s,\tau}|_{\mathbb C\setminus\overline{D_{s,\tau}}}$. Unlike $\widehat U_{s,\tau}$, this continuation need not equal $U_s$ inside the droplet.

\begin{lemma}\label{anapp:HW}
For every fixed integer $L\ge0$, after  possibly decreasing $s_0,\delta$ from Lemma~\ref{anapp:analytic-family}, there exist bounded holomorphic functions $B_{\ell,s,\tau}$ on the exterior and a common inner collar, and smooth cutoffs $\chi_{s,\tau}$ equal to one near the boundary and in the exterior, such that, for $\tau=j/N$,
\begin{align}
 F_{j,N,s}^{(L)}(z)
 &=N^{1/4}\sqrt{\phi_{s,\tau}'(z)}\,
 \phi_{s,\tau}(z)^j e^{N\mathcal Q_{s,\tau}(z)}
 \sum_{\ell=0}^L N^{-\ell}B_{\ell,s,\tau}(z),\label{anapp:HW-form}\\
 \|p_{j,N,s}-\chi_{s,\tau}F_{j,N,s}^{(L)}\|_{N,s}
 &\le C_LN^{-L-1}.\label{anapp:HW-error}
\end{align}
(The square root in \eqref{anapp:HW-form} is chosen positive at infinity.) We extend $\chi_{s,\tau}F^{(L)}$ by zero where $F^{(L)}$ is undefined. Every fixed number of $(s,\tau)$ derivatives of $B_{\ell,s,\tau}\circ\Phi_{s,\tau}$ is uniformly bounded on a smaller collar. Moreover, we have
\begin{equation}
 |B_{0,s,\tau}|^2=\frac{\sqrt{a_s}}{2\pi^{3/2}}
 \quad\hbox{on }\partial D_{s,\tau}.
 \label{anapp:HW-leading}
\end{equation}
\end{lemma}

\begin{proof}
We apply \cite[Theorem~3.1.2]{HedenmalmWennman2021} and check that its estimates are uniform in $(s,\tau)$. We write
\[
 R_{s,\tau}=(U_s-\breve U_{s,\tau})\circ\Phi_{s,\tau}.
\]
From \eqref{anapp:quadratic-gap}, we have
\begin{equation}
 R_{s,\tau}(w)=(1-|w|^2)^2R_{0,s,\tau}(w),\qquad
 R_{0,s,\tau}(e^{i\theta})=
 \frac18a_s(\Phi_{s,\tau}(e^{i\theta}))|\Phi_{s,\tau}'(e^{i\theta})|^2>0,
 \label{anapp:uniform-weight}
\end{equation}
where $R_{0,s,\tau}$ extends real analytically across the circle. We denote by $R^{\rm pol}$ the polarization, with $\bar w$ replaced by an independent variable. From Lemma~\ref{anapp:analytic-family} and the exterior gaps, we have, on a fixed complex neighborhood $\mathcal W$ of the diagonal annulus,
\begin{gather*}
 \sup_{s,\tau}\left(\|R_{s,\tau}^{\rm pol}\|_{\infty,\mathcal W}
 +\|R_{0,s,\tau}^{\rm pol}\|_{\infty,\mathcal W}
 +\|(R_{0,s,\tau}^{\rm pol})^{-1}\|_{\infty,\mathcal W}\right)\le C,\\
 \inf_{s,\tau,\,\rho<|w|<\rho^{-1}}R_{0,s,\tau}(w)\ge c,\qquad
 \inf_{s,\tau,\,|w|\ge\rho^{-1}}\frac{R_{s,\tau}(w)}{\log|w|}\ge c.
\end{gather*}
We therefore have a uniform family in the sense of \cite[Definition~6.1.1]{HedenmalmWennman2021}. We choose $\chi_{s,\tau}\circ\Phi_{s,\tau}$ from a fixed radial cutoff whose transition lies in a smaller inner collar. On $\supp\bar\partial\chi_{s,\tau}$, we have a uniform positive lower bound for the phase and $a_s\ge c$. The $\bar\partial$ estimate in Section~3.6 of that paper is then exponentially small uniformly in $(s,\tau)$. Combining it with Proposition~6.2.1 and the projection argument in Theorem~3.1.2, we obtain \eqref{anapp:HW-error}.

 In the recursion of \cite[Section~4.2]{HedenmalmWennman2021} for $B_{\ell,s,\tau}\circ\Phi_{s,\tau}$, the Fourier projections $P$ act on the fixed circle, so $[\partial_s,P]=[\partial_\tau,P]=0$. Differentiating this recursion and using Lemma~\ref{anapp:analytic-family} and Cauchy estimates on smaller collars, we obtain by induction in $\ell$,
\[
 \sup_{s,\tau}\|\partial_s^r\partial_\tau^t
 (B_{\ell,s,\tau}\circ\Phi_{s,\tau})\|_\infty\le C_{\ell,r,t}.
\]
Finally, we obtain \eqref{anapp:HW-leading} from \cite[Theorem~1.5.4]{HedenmalmWennman2021}, converting its area measure $\dd^2z/\pi$ and Laplacian $\Delta/4$ to our conventions.
\end{proof}

\begin{lemma}\label{anapp:mean-expansion}
For every fixed integer $A\ge1$, we have, on a sufficiently small fixed interval $|s|\le s_0$,
\begin{equation}
 M_N(s)=N\mu(s)+\sum_{\ell=0}^{A-1}N^{-\ell}b_\ell(s)+E_{N,A}(s),
 \qquad \|E_{N,A}\|_\infty\le C_AN^{-A}.
 \label{anapp:mean-expansion-eq}
\end{equation}
The coefficients $b_\ell$ are independent of $N$ and have bounded derivatives of every fixed order.
\end{lemma}

\begin{proof}
As usual, writing the Vandermonde in \eqref{anapp:law} in the orthonormal basis $(p_{j,N,s})_{j<N}$ and integrating out $N-1$ variables, we obtain the one-point intensity
\[
 \rho_{N,s}^{(1)}(z)=e^{-2NU_s(z)}\sum_{j=0}^{N-1}|p_{j,N,s}(z)|^2,
 \qquad \int_{\mathbb C}\rho_{N,s}^{(1)}\,\dd^2z=N.
\]
Since $M_N(s)=\int_{\mathbb C}f\rho_{N,s}^{(1)}\,\dd^2z$, we have
\begin{equation}
 M_N(s)=\sum_{j=0}^{N-1}\int f|p_{j,N,s}|^2e^{-2NU_s}\dd^2z.
 \label{anapp:mean-trace}
\end{equation}
We choose a fixed smooth partition $f=g_{\rm b}+g_{\rm e}+g_{\rm o}$ into bulk, edge and exterior parts. Their supports lie, respectively, in a compact subset of every $D_s$, in a thin collar of $\partial D$, and in a compact subset of every $\mathbb C\setminus\overline{D_s}$. By $v_\tau>0$, we can choose $\tau_-<1$, then the collar and $s_0$ small enough that
\begin{equation}
 \operatorname{dist}(\supp g_{\rm e},\overline{D_{s,\tau}})\ge c>0
 \quad\hbox{for $\tau$ near $\tau_-$}.
 \label{anapp:lower-separation}
\end{equation}
We make these choices within the parameter and holomorphy regions of Lemma~\ref{anapp:HW}, at the required precision.

For the bulk part, we have by \cite[Theorem~3.9]{Berman2009}
\begin{equation}
 \sum_{j<N}\int g_{\rm b}|p_{j,N,s}|^2e^{-2NU_s}\dd^2z
 =\frac N{2\pi}\int_{D_s}g_{\rm b}a_s\dd^2z
 +\sum_{\ell=0}^{A-1}N^{-\ell}b_{{\rm b},\ell}(s)+\OO(N^{-A}).
 \label{anapp:bulk}
\end{equation}
Here we use that $\supp g_{\rm b}$ stays away from $\partial D_s$, $a_s\ge c>0$, and all fixed derivatives of $U_s$ are uniformly bounded there. The coefficients depend smoothly on finitely many derivatives of $U_s$, so we also have $\|b_{{\rm b},\ell}\|_{C^r}\le C_{\ell,r}$.

We next estimate the exterior part and the low-degree edge terms.   By  \cite[Proposition~2.2.2]{HedenmalmWennman2021}, with $Q=U_s$ and $m=N$, any polynomial with $\deg p\le N\tau_0$ and $\|p\|_{N,s}=1$ satisfies, uniformly in $s$ and $\tau_0$ near $1$,
\begin{equation}
 |p(z)|e^{-NU_s(z)}\le C\sqrt N\,
 e^{-N(U_s-\widehat U_{s,\tau_0})(z)},\qquad z\notin D_{s,\tau_0}.
 \label{anapp:polynomial-bound}
\end{equation}
With $\tau_0=1$ and $\tau_0=\tau_-$, respectively, we have
\[
 \sum_{j<N}\left|\int g_{\rm o}|p_{j,N,s}|^2e^{-2NU_s}\dd^2z\right|
 +\sum_{j<N\tau_-}\left|\int g_{\rm e}|p_{j,N,s}|^2e^{-2NU_s}\dd^2z\right|
 \le CN^2e^{-cN}.
\]

For the remaining edge terms, we write $\tau=j/N$ and $F=F_{j,N,s}^{(L)}$. By Cauchy--Schwarz, we have
\begin{equation}
 \left|\int g_{\rm e}(|p_{j,N,s}|^2-|\chi_{s,\tau}F|^2)e^{-2NU_s}\dd^2z\right|
 \le\|g_{\rm e}\|_\infty\|p_{j,N,s}-\chi_{s,\tau}F\|_{N,s}
 (1+\|\chi_{s,\tau}F\|_{N,s})\le C_LN^{-L-1}.
 \label{anapp:L2-to-mean}
\end{equation}
With the change of variables $z=\Phi_{s,\tau}(w)$, we may write the approximating integral as
\begin{equation}
 \sqrt N\int g_{\rm e}(\Phi(w))\chi(\Phi(w))^2|\Phi'(w)|
 e^{-2NR_{s,\tau}(w)}
 \left|\sum_{\ell=0}^L N^{-\ell}B_{\ell,s,\tau}(\Phi(w))\right|^2\dd^2w.
 \label{anapp:scalar-integral}
\end{equation}
We have an exponentially small contribution outside a fixed collar of $|w|=1$. Inside this collar, we use the coordinates
\begin{equation}
 w=re^{i\theta},\qquad u=\sgn(r-1)\sqrt{2R_{s,\tau}(re^{i\theta})}.
 \label{anapp:gaussian-coordinate}
\end{equation}
By \eqref{anapp:uniform-weight}, we have uniform derivative bounds for this change of coordinates and its inverse.  Write $r=r(s,\tau,u,\theta)$ for the inverse and $w=re^{i\theta}$. Expanding the squared sum in \eqref{anapp:scalar-integral} and using $\dd^2w=r\partial_u r\,\dd u\,\dd\theta$, we obtain the coefficients
\[
\begin{aligned}
 A_\ell(s,\tau,u,\theta)
 ={}&\xi(u)g_{\rm e}(\Phi(w))\chi(\Phi(w))^2|\Phi'(w)|\,r\partial_u r\\
 &\times\sum_{\substack{p+q=\ell\\0\le p,q\le L}}
 B_{p,s,\tau}(\Phi(w))\overline{B_{q,s,\tau}(\Phi(w))},
 \qquad 0\le\ell\le2L.
\end{aligned}
\]
Here $\xi\in C_c^\infty(\mathbb R)$ is fixed, supported in the common $u$-interval and equal to one near zero. The integral is therefore $\sqrt N\sum_{\ell=0}^{2L}N^{-\ell}\int_0^{2\pi}\int_{\mathbb R}e^{-Nu^2}A_\ell\,\dd u\,\dd\theta$, up to an exponentially small error. Taylor's formula gives
\begin{equation}
 \sqrt N\int e^{-Nu^2}A_\ell(s,\tau,u,\theta)\dd u
 =\sqrt\pi\sum_{q=0}^K\frac{N^{-q}}{4^q q!}
 \partial_u^{2q}A_\ell(s,\tau,0,\theta)+\OO(N^{-K-1}),
 \label{anapp:Gaussian-expansion}
\end{equation}
uniformly in the parameters.   Taking $L,K$ sufficiently large, we obtain
\begin{equation}
 \int g_{\rm e}|p_{j,N,s}|^2e^{-2NU_s}\dd^2z
 =\sum_{\ell=0}^{A}N^{-\ell}C_\ell(s,j/N)+\OO(N^{-A-1}).
 \label{anapp:degree-expansion}
\end{equation}
We have uniform bounds on every fixed number of $(s,\tau)$ derivatives of the coefficients. By \eqref{anapp:lower-separation}, these coefficients vanish near $\tau_-$. From \eqref{anapp:HW-leading} and \eqref{anapp:Gaussian-expansion}, we have
\begin{equation}
 C_0(s,\tau)=\frac1{2\pi}\int_0^{2\pi}
 g_{\rm e}(\Phi_{s,\tau}(e^{i\theta}))\dd\theta
 =\int g_{\rm e}\,\dd\omega_{s,\tau}^{\infty}.
 \label{anapp:leading-degree}
\end{equation}
We extend $C_\ell$ by zero below $\tau_-$. By Euler-Maclaurin, with Bernoulli numbers $\mathfrak b_{2k}$, we have
\begin{equation}
 \sum_{j=0}^{N-1}C_\ell(s,j/N)
 =N\int_{\tau_-}^1C_\ell(s,\tau)\dd\tau-\frac12C_\ell(s,1)+\sum_{k=1}^K\frac{\mathfrak b_{2k}}{(2k)!}N^{1-2k}
 \partial_\tau^{2k-1}C_\ell(s,1)+\OO(N^{1-2K}).
 \label{anapp:EM}
\end{equation}
We choose $2K>A+1$ and sum \eqref{anapp:degree-expansion}, with total error $\OO(N^{-A})$. Finally, by \eqref{anapp:mass-derivative}, we have
\[
 \int_{\tau_-}^1C_0(s,\tau)\dd\tau
 =\int g_{\rm e}\,\dd(\nu_{s,1}-\nu_{s,\tau_-})
 =\frac1{2\pi}\int_{D_s}g_{\rm e}a_s\dd^2z.
\]
Combining this with \eqref{anapp:bulk}, we obtain \eqref{anapp:mean-expansion-eq} and the coefficient bounds.
\end{proof}

\subsection{Derivatives of the tilted mean}
\label{anapp:differentiation-section}
We now compare $M_N^{(r)}(0)$ with $N\mu^{(r)}(0)$. The same argument applies to any fixed $r\ge2$, with the expansion order and finite-difference step chosen in terms of $r$. We restrict to $r=2,3$, for which we have computed the corresponding shape derivatives.

\begin{lemma}\label{anapp:derivatives}
For $r=2,3$, we have
\begin{equation}
 M_N^{(r)}(0)=N\mu^{(r)}(0)+\OO_{U,f}(1).
 \label{anapp:derivative-estimate}
\end{equation}
Consequently,
\begin{equation}
 \langle\mathcal L_N(f)^3\rangle_c=N^{-1}\mu''(0)+\OO(N^{-2}),\qquad
 \langle\mathcal L_N(f)^4\rangle_c=-N^{-2}\mu'''(0)+\OO(N^{-3}).
 \label{anapp:cumulant-reduction}
\end{equation}
\end{lemma}
\begin{proof}
Since $|\mathcal L_N(f)|\le N\|f\|_\infty$, the moment--cumulant formula gives
\[
 |\langle\mathcal L_N(f)^q\rangle_{c,U_s}|
 \le C_qN^q\|f\|_\infty^q,\qquad q\ge2.
\]
Differentiating the finite-$N$ expectation, we therefore have
\begin{equation}
 M_N^{(r)}(s)=(-N)^r\langle\mathcal L_N(f)^{r+1}\rangle_{c,U_s},\qquad
 \|M_N^{(r)}\|_\infty\le C_rN^{2r+1}\quad(1\le r\le4).
 \label{anapp:crude-derivatives}
\end{equation}
We take $A=32$ and abbreviate $E_N=E_{N,A}$ in Lemma~\ref{anapp:mean-expansion}. From the preceding bounds, we have
\[
 \|E_N\|_\infty\le CN^{-32},\qquad
 \|E_N^{(r)}\|_\infty\le C_rN^{2r+1}\quad(1\le r\le4).
\]
Writing $\Delta_hE(x)=E(x+h)-E(x)$, we have by the fundamental theorem of calculus
\[
 h^{-r}(\Delta_h^rE)(0)=\int_{[0,1]^r}E^{(r)}(h(t_1+\cdots+t_r))\dd t_1\cdots\dd t_r.
\]
For $r=2,3$, take $h=N^{-10}$ and $N$ large enough that $3h\le s_0$. Then $[0,rh]\subset[-s_0,s_0]$, so
\begin{equation}
 |E_N^{(r)}(0)|
 \le 2^rh^{-r}\|E_N\|_\infty+rh\|E_N^{(r+1)}\|_\infty
 \le C_r\bigl(N^{-32+10r}+N^{2r-7}\bigr)=\OO(N^{-1}).
 \label{anapp:finite-difference}
\end{equation}
Using the above estimate in \eqref{anapp:mean-expansion-eq}, we obtain \eqref{anapp:derivative-estimate}.  Finally, the tilt gives
\begin{equation}
 \chi_N(s):=\log\langle e^{-Ns\mathcal L_N(f)}\rangle_U,\qquad
 \chi_N'(s)=-NM_N(s),\qquad
 \langle\mathcal L_N(f)^m\rangle_c
 =(-1)^{m-1}N^{1-m}M_N^{(m-1)}(0).
 \label{anapp:exact-tilt}
\end{equation}
Combining these identities with \eqref{anapp:derivative-estimate}, we obtain \eqref{anapp:cumulant-reduction}.
\end{proof}

\subsection{Shape derivatives}
\label{anapp:shape-section}
\begin{proof}[Proof of Proposition~\ref{anapp:prop}]
From \eqref{anapp:measure-derivative} and integration by parts, we have
\begin{equation}
 \mu'(s)=-\frac1{4\pi}\int_{\mathbb C}|\nabla f^{D_s}|^2\dd^2z.
 \label{anapp:mu-energy}
\end{equation}
Differentiating this energy as in \eqref{3derivativechis}, we obtain
\begin{equation}
 \mu''(s)=\frac1{8\pi}\int_{\partial D_s}\frac{J_s^3}{a_s}\,\dd S.
 \label{anapp:mu-second}
\end{equation}
Applying \eqref{surfaceFderivativefinal} from Section~\ref{sec:fourth-derivative}, with $\beta=2$, $K_s=-J_s$ and $A_s=a_s$, gives $\mu'''(0)$ from \eqref{anapp:mu-second}. Substituting in \eqref{anapp:cumulant-reduction}, we obtain \eqref{anapp:c3}--\eqref{anapp:c4}.
\end{proof}

\setcounter{thm}{\value{proposition}}
\begin{rem}[General $d$ and $\beta$]\label{anapp:general-beta}
Let $d\ge2$, $\beta>0$, $U_s=U+sf/\beta$, and let $\nu_s$ be the probability equilibrium measure. We assume that the droplet deformation is sufficiently smooth for the shape derivatives in Section~\ref{sec:derivation-general}, and write
\begin{equation}
 \mathcal E_d(U_s)=\int U_s\,\dd\nu_s
 +\frac12\iint V_d(|x-y|)\,\dd\nu_s(x)\dd\nu_s(y),\qquad
 R_{N,d}(s)=\log Z_{N,U_s}+\beta N^2\mathcal E_d(U_s).
 \label{anapp:general-d-remainder}
\end{equation}
Using $\partial_s\mathcal E_d(U_s)=\beta^{-1}\int f\,\dd\nu_s$ and the exact tilt identity, we have, for $m=3,4$,
\begin{equation}
 \langle\mathcal L_N(f)^m\rangle_c
 =(-1)^{m+1}\beta N^{2-m}
   \partial_s^m\mathcal E_d(U_s)\big|_{s=0}
 +(-1)^mN^{-m}R_{N,d}^{(m)}(0).
 \label{anapp:general-d-cumulants}
\end{equation}
We would therefore obtain the leading cumulant formulas from $R_{N,d}^{(3)}(0),R_{N,d}^{(4)}(0)=o(N^2)$. With the stronger bounds $\OO(N)$, we would have errors $\OO(N^{-2})$ and $\OO(N^{-3})$, respectively. 

We can state the missing estimate more precisely for $d=2$. For fixed $\beta>0$, we use $U_s=U+sf/\beta$; we still have \eqref{anapp:exact-tilt}, and the shape calculation uses $v_s=-J_s/(\beta a_s)$. In our normalization, we have from \cite[Theorem~1.1]{BBNY19} (with their $V=2U_s$ and inverse temperature $\beta/2$) and \cite[Proposition~2.13]{LebleSerfaty}
\begin{equation}
 \log Z_{N,U_s}=-\beta N^2\mathcal E(U_s)
 +\frac\beta4N\log N+c_\beta N
 +N\left(\frac\beta4-1\right)\operatorname{Ent}(\nu_s)+r_N(s),
 \label{anapp:free-energy}
\end{equation}
where
\[
 \mathcal E(U_s)=\int U_s\,\dd\nu_s
 -\frac12\iint\log|z-w|\,\dd\nu_s(z)\dd\nu_s(w),\qquad
 \operatorname{Ent}(\nu_s)=\int_{D_s}\frac{a_s}{2\pi}\log\frac{a_s}{2\pi}\,\dd^2z.
\]
Here we have $r_N=o(N)$, and from the first reference $r_N=\OO(N^{1-\kappa})$ for any fixed $0<\kappa<1/24$.  See also \cite[Theorem~2]{SerfatyAnyd} for a power-law error in a free-energy expansion using the thermal equilibrium measure. The coefficient of $N\log N$ is independent of $s$, so this term cancels in $\log Z_{N,U_s}-\log Z_{N,U}$. 

Evaluating the cumulants as in \eqref{anapp:general-d-cumulants} therefore requires estimates on derivatives of $r_N$. Using $\partial_s\mathcal E(U_s)=\mu(s)/\beta$ and bounded derivatives of the entropy, we see that the condition
\begin{equation}
 |r_N^{(3)}(0)|+|r_N^{(4)}(0)|=\OO(N)
 \label{anapp:general-beta-condition}
\end{equation}
would imply \eqref{anapp:cumulant-reduction}. We would then have \eqref{anapp:c3}--\eqref{anapp:c4}, with $1/(8\pi)$ and $1/(16\pi)$ replaced by $1/(2\pi\beta^2)$ and $1/(2\pi\beta^3)$, respectively. The bound on $r_N$ itself does not control the derivatives in \eqref{anapp:general-beta-condition}. 
\end{rem}

\addtocontents{toc}{\protect\setcounter{tocdepth}{2}}
\endgroup

\section{Conclusion}
\label{sec:conclusion}

In this work we derived explicit formulas for the leading third and fourth
cumulants of smooth linear statistics in Coulomb gases beyond rotational
invariance. The formulas hold for general smooth droplets in dimension
\(d\geq2\), under the assumption that the perturbed droplets remain smooth and
undergo no topological transition, provided the required limit interchange
is justified. Section~\ref{sec:rigorous} proves this for $d=2$, $\beta=2$
under its analytic assumptions.  Remark~\ref{anapp:general-beta} gives a
sufficient condition for general $d$ and $\beta$. They are expressed intrinsically on the
boundary of the equilibrium droplet, in terms of the jump of the normal
derivative of the harmonic extension of the test function, the exterior
Dirichlet-to-Neumann operator, the mean curvature, and normal derivatives of
the background density. In \(d=2\), we also recovered the same structure using
conformal maps, Julia's variational formula, and the Hilbert transform. The
general formulas reduce to the previously known radial results and pass
nontrivial checks from translated elliptic droplets and spheroidal
deformations.

Several natural questions remain open. First, it would be important to extend
the analysis to singular geometries, such as droplets with corners, cusps,
hard edges, or topology changes such as holes \cite{Estienne_corner,Johansson_corner}. In such cases the boundary operators
appearing here become singular, and the scaling of the cumulants may change.
Second, one would like to approach nonsmooth test functions, in particular
indicator functions. This would connect the present smooth-statistics formulas
to full counting statistics and to the microscopic boundary layers that appear
near the edge of the droplet. Third, the tilted potential
\(U_s=U+s f/\beta\) suggests a geometric interpretation of the analytic
structure of the cumulant generating function: singularities in \(s\) should
correspond to phase transitions of the associated obstacle problem, such as
loss of smoothness, nucleation of holes, or changes in topology. Finally, it
would be interesting to extend the method to higher cumulants and to other
long-range interacting systems, such as Riesz gases, where the role of the
harmonic extension should be replaced by suitable nonlocal boundary operators.\\

{\bf Acknowledgments.} The authors thank Sylvia Serfaty for useful discussions {and Sung-Soo Byun for a careful reading of the manuscript}.
PB acknowledges support from NSF grant DMS-2348202.
GS and PLD acknowledge support from ANR Grant No. ANR23-CE30-0020-01 EDIPS.
     This research was supported in part by grant NSF PHY-2309135 to the Kavli Institute for Theoretical Physics (KITP). The authors acknowledge the use of generative AI to assist in derivation and manuscript writing.  All mathematical ideas and proof strategies originated from the authors, and all the scientific content is their responsibility.

\appendix

\section{Exact {large $N$} cumulants for a translated elliptic deformation of the disk}
\label{app:elliptictranslationcheck}

We now give a check involving an elliptic deformation
combined with a translation. It leads 
to an exact formula for the leading large-$N$ cumulants, which we compare with our general formulas. 

We take
\be
U(z)=\frac{|z|^2}{2},
\qquad
f(z)=\operatorname{Re}z^2+a\,x+b\,y
=
x^2-y^2+a\,x+b\,y,
\qquad
\beta=2.
\label{elliptictranslationUf}
\ee
Thus
\be
U_s
=
U+\frac{s}{\beta}f
=
\frac12|z|^2+\frac{s}{2}\left(x^2-y^2+a\,x+b\,y\right).
\label{elliptictranslationUs}
\ee
We assume $|s|<1$, so that the tilted quadratic potential is confining.
For this deformation, the leading large-\(N\) cumulant generating function is
\be
\frac{1}{N^2}\chi(s,N)
\simeq
-\frac12\log(1-s^2)
+
\frac{a^2}{4}\frac{s^2}{1+s}
+
\frac{b^2}{4}\frac{s^2}{1-s}.
\label{elliptictranslationchi}
\ee
Consequently, for \(q\geq2\),
\be
\langle{\mathcal  L}_N^q\rangle_c
\simeq
N^{2-q}
\left[
(q-1)!\,{\bf 1}_{q\ {\rm even}}
+
\frac{q!}{4}\left(a^2+(-1)^q b^2\right)
\right].
\label{elliptictranslationcumulantsgeneral}
\ee
In particular,
\be
\langle{\mathcal  L}_N^2\rangle_c
\simeq
1+\frac{a^2+b^2}{2},
\qquad
\langle{\mathcal  L}_N^3\rangle_c
\simeq
\frac{3}{2N}(a^2-b^2),
\qquad
\langle{\mathcal  L}_N^4\rangle_c
\simeq
\frac{6}{N^2}(1+a^2+b^2).
\label{elliptictranslationcumulants234}
\ee
Note that the case \(a=b=0\) reduces to a pure elliptic deformation.

Let us derive \eqref{elliptictranslationchi}. Completing the square in
\eqref{elliptictranslationUs}, one can write \(U_s\), up to an additive
constant, as a translated elliptic potential,
\be
U_s
=
\frac12
\left[
(1+s)(x-x_0)^2
+
(1-s)(y-y_0)^2
\right]
-\frac12\left[(1+s)x_0^2+(1-s)y_0^2\right],
\ee
where
\be
x_0=-\frac{as}{2(1+s)},
\qquad
y_0=-\frac{bs}{2(1-s)}.
\label{elliptictranslationcenter}
\ee
Equivalently, this is the translated elliptic droplet of
\cite{ByunPlanar23}, in the \(c=0\) elliptic case, with parameter
\(\tau=-s\). Hence the droplet is the ellipse
\be
D_s
=
\left\{
\frac{1+s}{1-s}(x-x_0)^2
+
\frac{1-s}{1+s}(y-y_0)^2
\leq1
\right\}.
\label{elliptictranslationdroplet}
\ee
Since \(f\) is harmonic, \(\Delta f=0\), and hence \(\Delta U_s=2\). Therefore
the equilibrium density is constant, \(\rho_s=1/\pi\), and the starting formula
gives
\be
-\frac{1}{N^2}\partial_s\chi(s,N)
=
\frac{1}{\pi}
\int_{D_s}d^2{\bf x}\,f({\bf x}).
\label{elliptictranslationchiderivative}
\ee

We evaluate the integral by shifting to the center of the ellipse:
\(X=x-x_0\), \(Y=y-y_0\). The semiaxes are
\[
A_s=\sqrt{\frac{1-s}{1+s}},
\qquad
B_s=\sqrt{\frac{1+s}{1-s}},
\qquad
A_sB_s=1.
\]
By symmetry, the integrals of \(X\) and \(Y\) vanish, and
\be
\frac{1}{\pi}
\int_{D_s}d^2{\bf x}\,(x^2-y^2)
=
-\frac{s}{1-s^2}
+
x_0^2-y_0^2.
\ee
Adding the linear terms gives
\be
\frac{1}{\pi}
\int_{D_s}d^2{\bf x}\,f({\bf x})
=
-\frac{s}{1-s^2}
-
\frac{a^2s(s+2)}{4(1+s)^2}
+
\frac{b^2s(s-2)}{4(1-s)^2}.
\label{elliptictranslationintegral}
\ee
Combining \eqref{elliptictranslationchiderivative} and
\eqref{elliptictranslationintegral}, we find
\be
\frac{1}{N^2}\partial_s\chi(s,N)
=
\frac{s}{1-s^2}
+
\frac{a^2s(s+2)}{4(1+s)^2}
-
\frac{b^2s(s-2)}{4(1-s)^2}.
\ee
Integrating in \(s\), with \(\chi(0,N)=0\), gives
\eqref{elliptictranslationchi}.

We now compare with the disk formulas in the main text. First, the variance
formula gives
\be
\langle{\mathcal  L}_N^2\rangle_c
\simeq
\frac{1}{2\pi\beta}
\left[
\int_{|z|\leq1}d^2{\bf x}\,|\vec\nabla f|^2
+
2\pi\sum_{n\in\mathbb Z}|n|\hat f_n\hat f_{-n}
\right].
\ee
For
\[
f(r,\phi)=r^2\cos(2\phi)+a r\cos\phi+b r\sin\phi,
\]
one has
\be
\frac{1}{2\pi\beta}
\int_{|z|\leq1}d^2{\bf x}\,|\vec\nabla f|^2
=
\frac12+\frac{a^2+b^2}{4},
\ee
and
\be
\frac{1}{\beta}
\sum_{n\in\mathbb Z}|n|\hat f_n\hat f_{-n}
=
\frac12+\frac{a^2+b^2}{4}.
\ee
Thus
\be
\langle{\mathcal  L}_N^2\rangle_c
\simeq
1+\frac{a^2+b^2}{2},
\ee
in agreement with \eqref{elliptictranslationcumulants234}.

Next, the third cumulant formula \eqref{cum3circleintro} gives the same result
as \eqref{elliptictranslationcumulants234}. Indeed, here
\[
R_0=1,
\qquad
\Delta U=2,
\qquad
\beta=2,
\]
and
\be
J(\phi)
=
\partial_r f|_{r=1}+H\partial_\phi{\sf f}
=
4\cos(2\phi)+2a\cos\phi+2b\sin\phi.
\label{elliptictranslationJ}
\ee
Therefore
\be
\langle{\mathcal  L}_N^3\rangle_c
\simeq
\frac{1}{8N}\left\langle J^3\right\rangle
=
\frac{3}{2N}(a^2-b^2),
\ee
again in agreement with \eqref{elliptictranslationcumulants234}.

Finally, the fourth cumulant formula \eqref{4cumdiskintro} gives a nontrivial
check. Since \(U(r)=r^2/2\), one has
\[
R_0=1,
\qquad
\Delta U=2,
\qquad
U''(1)=1,
\qquad
U'''(1)=0,
\]
and, since \(f\) is harmonic, \(\Delta f=0\). Hence
\eqref{4cumdiskintro} reduces to
\be
\langle{\mathcal  L}_N^4\rangle_c
\simeq
\frac{1}{2\pi\beta^3N^2(\Delta U)^2}
\int_0^{2\pi}d\phi\,
\left[
3J^2H\partial_\phi(J^2)
-
2J^4
\right],
\ee
with \(J\) given in \eqref{elliptictranslationJ}. A direct Fourier calculation
gives
\be
\int_0^{2\pi}d\phi\,
\left[
3J^2H\partial_\phi(J^2)
-
2J^4
\right]
=
384\pi(1+a^2+b^2).
\ee
Using \(\beta=2\) and \(\Delta U=2\), we obtain
\be
\langle{\mathcal  L}_N^4\rangle_c
\simeq
\frac{6}{N^2}(1+a^2+b^2),
\ee
which agrees with \eqref{elliptictranslationcumulants234}.

\section{Exact {large $N$} cumulants from a spheroidal deformation of the ball}
\label{app:spheroidcheck}

We give a three-dimensional check based on the classical fact that the
Newtonian potential of a homogeneous ellipsoid is quadratic inside the
ellipsoid. We consider \(d=3\), and take
\be
U(\bx)=\frac{|\bx|^2}{2},
\qquad
\bx=(x_1,x_2,x_3),
\ee
together with the harmonic quadratic perturbation
\be
f(\bx)
=
x_3^2-\frac{x_1^2+x_2^2}{2}.
\label{spheroidtestfunction}
\ee
Equivalently, in spherical coordinates,
\be
f(\bx)=r^2P_2(\cos\theta),
\qquad
P_2(\mu)=\frac12(3\mu^2-1),
\qquad
\mu=\cos\theta.
\ee
We write
\be
t=\frac{s}{\beta}.
\ee
Then
\be
U_s
=
U+\frac{s}{\beta}f
=
\frac12(1-t)(x_1^2+x_2^2)
+
\frac12(1+2t)x_3^2.
\label{spheroidUs}
\ee
The confining range is $-\tfrac12<t<1$.

For this deformation, the leading large-\(N\) cumulant generating function is
given parametrically by
\be
\frac{1}{N^2}\chi(s,N)
\simeq
-\frac{\beta}{5}
\int_0^{s/\beta}d\tau\,
\left[
\kappa(\tau)^{4/3}
-
\kappa(\tau)^{-2/3}
\right].
\label{spheroidchi}
\ee
Here \(\kappa=\kappa(t)\) is the axis ratio of the spheroid,
\be
\kappa=\frac{c}{a},
\ee
where \(a\) is the equatorial semiaxis and \(c\) the polar semiaxis. It is
defined implicitly by
\be
1+2t
=
3N_3(\kappa),
\label{spheroidktrelation}
\ee
where \(N_3(\kappa)\) is the depolarization factor along the symmetry axis of a
spheroid. Equivalently,
\be
t=\frac{3N_3(\kappa)-1}{2}.
\ee
The spherical point is \(\kappa=1\), for which \(N_3(1)=1/3\), hence \(t=0\).

Expanding \eqref{spheroidchi} near \(s=0\), one finds
\be
\frac{1}{N^2}\chi(s,N)
\simeq
\frac{s^2}{2\beta}
-
\frac{25s^3}{63\beta^2}
+
\frac{1250s^4}{1764\beta^3}
+
O(s^5).
\label{spheroidchiexpansion}
\ee
Therefore
\be
\langle{\mathcal L}_N^2\rangle_c
\simeq
\frac{1}{\beta},
\qquad
\langle{\mathcal L}_N^3\rangle_c
\simeq
\frac{50}{21N\beta^2},
\qquad
\langle{\mathcal L}_N^4\rangle_c
\simeq
\frac{2500}{147N^2\beta^3}.
\label{spheroidcumulants234}
\ee

Let us derive \eqref{spheroidchi}. Since \(f\) is harmonic, one has
\[
\Delta f=0,
\]
and therefore
\[
\Delta U_s=3.
\]
Thus the equilibrium density is constant,
\be
\rho_s=\frac{\Delta U_s}{4\pi}=\frac{3}{4\pi}.
\ee
We look for a spheroidal droplet
\be
D_s
=
\left\{
\frac{x_1^2+x_2^2}{a^2}
+
\frac{x_3^2}{c^2}
\leq1
\right\}.
\label{spheroiddroplet}
\ee
The total mass is one, hence
\be
\frac{3}{4\pi}\,\frac{4\pi}{3}a^2c=1,
\qquad\text{so}\qquad
a^2c=1.
\label{spheroidvolumeconstraint}
\ee
Introducing the axis ratio
\[
\kappa=\frac{c}{a},
\]
the constraint \eqref{spheroidvolumeconstraint} gives
\be
a=\kappa^{-1/3},
\qquad
c=\kappa^{2/3}.
\label{spheroidaxes}
\ee

We now use the classical ellipsoid theorem. If \(\Phi_s\) denotes the
Newtonian potential generated by the uniform probability measure on \(D_s\),
then inside \(D_s\)
\be
\Phi_s(\bx)
=
C_s
-
\frac12
\left[
3N_1(\kappa)(x_1^2+x_2^2)
+
3N_3(\kappa)x_3^2
\right],
\label{spheroidinsidepotential}
\ee
where
\be
N_1(\kappa)=N_2(\kappa)=\frac{1-N_3(\kappa)}{2},
\qquad
2N_1(\kappa)+N_3(\kappa)=1.
\ee
The factor \(3\) in \eqref{spheroidinsidepotential} comes from our density
\(\rho_s=3/(4\pi)\).

For a spheroid, the depolarization factor along the symmetry axis can be
written as
\be
N_3(\kappa)
=
\frac{\kappa}{2}
\int_0^\infty
\frac{d\lambda}
{(1+\lambda)(\kappa^2+\lambda)^{3/2}}.
\label{spheroidN3integral}
\ee
Equivalently, for the prolate case \(\kappa>1\),
\be
N_3(\kappa)
=
\frac{1-e^2}{2e^3}
\left[
\log\left(\frac{1+e}{1-e}\right)-2e
\right],
\qquad
e=\sqrt{1-\kappa^{-2}},
\ee
whereas for the oblate case \(0<\kappa<1\),
\be
N_3(\kappa)
=
\frac{1+\eta^2}{\eta^3}
\left(
\eta-\arctan\eta
\right),
\qquad
\eta=\sqrt{\kappa^{-2}-1}.
\ee

The equilibrium condition is that \(U_s+\Phi_s\) is constant inside the
droplet. Comparing the coefficients of \(x_1^2+x_2^2\) and \(x_3^2\) in
\eqref{spheroidUs} and \eqref{spheroidinsidepotential}, we obtain
\be
1-t=3N_1(\kappa),
\qquad
1+2t=3N_3(\kappa).
\ee
These two equations are equivalent because
\(2N_1+N_3=1\). Hence \(\kappa\) is determined by
\eqref{spheroidktrelation}. Positive \(t\) corresponds to a stronger
confinement in the \(x_3\)-direction, and hence to an oblate spheroid
\(\kappa<1\); negative \(t\) corresponds to a prolate spheroid \(\kappa>1\).

We can now compute the derivative of \(\chi\). The starting formula gives
\be
-\frac{1}{N^2}\partial_s\chi(s,N)
=
\int_{D_s}d^3\bx\,\rho_s(\bx)f(\bx).
\label{spheroidchiderivative}
\ee
For a uniform ellipsoid,
\be
\frac{3}{4\pi}
\int_{D_s}d^3\bx\,x_1^2
=
\frac{a^2}{5},
\qquad
\frac{3}{4\pi}
\int_{D_s}d^3\bx\,x_2^2
=
\frac{a^2}{5},
\qquad
\frac{3}{4\pi}
\int_{D_s}d^3\bx\,x_3^2
=
\frac{c^2}{5}.
\ee
Therefore
\be
\int_{D_s}d^3\bx\,\rho_s(\bx)f(\bx)
=
\frac{c^2-a^2}{5}.
\ee
Using \eqref{spheroidaxes}, this becomes
\be
\int_{D_s}d^3\bx\,\rho_s(\bx)f(\bx)
=
\frac15
\left[
\kappa^{4/3}
-
\kappa^{-2/3}
\right].
\label{spheroidmeanf}
\ee
Combining \eqref{spheroidchiderivative} and \eqref{spheroidmeanf}, and using
\(t=s/\beta\), gives \eqref{spheroidchi}.

To obtain the cumulants explicitly, expand \(N_3(\kappa)\) around
\(\kappa=1\). Writing \(\kappa=1+\epsilon\), one has
\be
N_3(1+\epsilon)
=
\frac13
-
\frac{4}{15}\epsilon
+
\frac{6}{35}\epsilon^2
-
\frac{32}{315}\epsilon^3
+
O(\epsilon^4).
\ee
Hence
\be
t
=
-\frac{2}{5}\epsilon
+
\frac{9}{35}\epsilon^2
-
\frac{16}{105}\epsilon^3
+
O(\epsilon^4).
\ee
Inverting this relation and substituting in \eqref{spheroidchi} gives the
expansion \eqref{spheroidchiexpansion}, and hence the cumulants
\eqref{spheroidcumulants234}.

We finally compare with the main formulas for the spherical droplet. At
\(s=0\), the droplet is the unit ball. On the unit sphere,
\be
f|_{r=1}
=
P_2(\cos\theta),
\qquad
\partial_r f|_{r=1}
=
2P_2(\cos\theta).
\ee
Since \(P_2\) is a spherical harmonic of degree \(\ell=2\), its exterior
harmonic extension has exterior normal derivative
\[
\partial_r^+f^{D_0}
=
-3P_2(\cos\theta).
\]
Thus, with the notation of the introduction,
\be
J
=
\partial_r f|_{r=1}
-
\partial_r^+f^{D_0}
=
5P_2(\cos\theta).
\label{spheroidJ}
\ee

The variance formula gives
\be
\langle{\mathcal L}_N^2\rangle_c
\simeq
\frac{1}{4\pi\beta}
\int_{\mathbb R^3}d^3\bx\,|\vec\nabla f^{D_0}|^2.
\ee
The interior contribution is
\[
2\int_{S^2}d\Omega\,P_2(\cos\theta)^2,
\]
and the exterior contribution is
\[
3\int_{S^2}d\Omega\,P_2(\cos\theta)^2.
\]
Since
\[
\int_{S^2}d\Omega\,P_2(\cos\theta)^2=\frac{4\pi}{5},
\]
the total energy is \(4\pi\), and therefore
\be
\langle{\mathcal L}_N^2\rangle_c
\simeq
\frac{1}{\beta},
\ee
in agreement with \eqref{spheroidcumulants234}.

For the third cumulant, using \eqref{cum3sphere}, \(R_0=1\), and
\(\Delta U=3\), we obtain
\be
\langle{\mathcal L}_N^3\rangle_c
\simeq
\frac{1}{12\pi N\beta^2}
\int_{S^2}d\Omega\,J^3.
\ee
Using \(J=5P_2\) and
\[
\int_{S^2}d\Omega\,P_2(\cos\theta)^3=\frac{8\pi}{35},
\]
this gives
\be
\langle{\mathcal L}_N^3\rangle_c
\simeq
\frac{50}{21N\beta^2},
\ee
again in agreement with \eqref{spheroidcumulants234}.

For the fourth cumulant, we use the spherical formula \eqref{4cumsphereintro}.
Here \(U''(1)=1\), \(U'''(1)=0\), \(\Delta U=3\), and \(\Delta f=0\). Thus the
curvature and normal-derivative terms give the coefficient
\[
\frac{1^2U'''(1)+6U''(1)+6}{\Delta U}=4.
\]
Next,
\be
J^2
=
25P_2(\cos\theta)^2
=
5P_0(\cos\theta)
+
\frac{50}{7}P_2(\cos\theta)
+
\frac{90}{7}P_4(\cos\theta).
\ee
Converting this expansion to orthonormal spherical harmonics, one finds
\be
\sum_{\ell\geq1}\sum_{m=-\ell}^{\ell}
(\ell+1)|q_{\ell m}|^2
=
\frac{24000\pi}{49},
\ee
where \(J^2=\sum_{\ell,m}q_{\ell m}Y_{\ell m}\). Moreover
\be
\int_{S^2}d\Omega\,J^4
=
\frac{1500\pi}{7}.
\ee
Therefore \eqref{4cumsphereintro} gives
\be
\langle{\mathcal L}_N^4\rangle_c
\simeq
\frac{1}{4\pi\beta^3N^2(\Delta U)^2}
\left[
3\frac{24000\pi}{49}
-
4\frac{1500\pi}{7}
\right]
=
\frac{2500}{147N^2\beta^3},
\ee
which agrees with \eqref{spheroidcumulants234}.

\section{The case $d=1$} \label{app:1d}

In $d=1$, the leading Coulomb-gas action can be written explicitly.
Let $U_s=U+s f/\beta$ and let $D_s=(a_s,b_s)$, where
\be
U_s'(a_s)=-1,
\qquad
U_s'(b_s)=1 .
\ee
Then
\be
\rho_s(x)=\frac12 U_s''(x),
\qquad a_s<x<b_s,
\ee
and the leading contribution to the cumulant generating function is
\be
\frac{1}{N^2}\chi(s,N)
\simeq
-\beta\left({\mathcal E}[U_s]-{\mathcal E}[U]\right),
\ee
where
\be
{\mathcal E}[U_s]
=
\frac12\left(U_s(a_s)+U_s(b_s)\right)
-
\frac14\int_{a_s}^{b_s}dx\,
\left[(U_s'(x))^2+1\right].
\ee
We now extract the first few cumulants from this explicit formula. Differentiating
the endpoint equations gives
\be
\frac{da_s}{ds}
=
-\frac{f'(a_s)}{\beta U_s''(a_s)},
\qquad
\frac{db_s}{ds}
=
-\frac{f'(b_s)}{\beta U_s''(b_s)} .
\ee
From the expression above for ${\mathcal E}[U_s]$, or equivalently from
\[
-\frac{1}{N^2}\partial_s\chi(s,N)
=
\frac12\int_{a_s}^{b_s}dx\,U_s''(x)f(x),
\]
one obtains
\be
\frac{1}{N^2}\partial_s^2\chi(s,N)
=
\frac{1}{2\beta}
\int_{a_s}^{b_s}dx\, [f'(x)]^2 .
\ee
Differentiating once more yields
\be
\frac{1}{N^2}\partial_s^3\chi(s,N)
=
\frac{1}{2\beta^2}
\left[
\frac{f'(a_s)^3}{U_s''(a_s)}
-
\frac{f'(b_s)^3}{U_s''(b_s)}
\right].
\ee
A further differentiation gives
\be
\begin{aligned}
\frac{1}{N^2}\partial_s^4\chi(s,N)
=
\frac{1}{2\beta^3}
\bigg[
&
-\frac{4f'(a_s)^3f''(a_s)}{U_s''(a_s)^2}
+
\frac{f'(a_s)^4U_s'''(a_s)}{U_s''(a_s)^3}
\\
&+
\frac{4f'(b_s)^3f''(b_s)}{U_s''(b_s)^2}
-
\frac{f'(b_s)^4U_s'''(b_s)}{U_s''(b_s)^3}
\bigg].
\end{aligned}
\ee
Thus, setting $a=a_0$, $b=b_0$, and using
$\langle{\mathcal L}_N^q\rangle_c=N^{-q}(-\partial_s)^q\chi(0,N)$, we find
\be
\langle{\mathcal L}_N^2\rangle_c
\simeq
\frac{1}{2\beta}
\int_a^b dx\, [f'(x)]^2 ,
\ee
\be
\langle{\mathcal L}_N^3\rangle_c
\simeq
-\frac{1}{2N\beta^2}
\left[
\frac{f'(a)^3}{U''(a)}
-
\frac{f'(b)^3}{U''(b)}
\right],
\ee
and
\be
\begin{aligned}
\langle{\mathcal L}_N^4\rangle_c
\simeq
\frac{1}{2\beta^3N^2}
\bigg[
&
-\frac{4f'(a)^3f''(a)}{U''(a)^2}
+
\frac{f'(a)^4U'''(a)}{U''(a)^3}
\\
&+
\frac{4f'(b)^3f''(b)}{U''(b)^2}
-
\frac{f'(b)^4U'''(b)}{U''(b)^3}
\bigg].
\end{aligned}
\ee

These formulas are consistent with the general formulas derived in the main
text. In $d=1$, one has $\Omega_1=2$, and the boundary of the droplet
$D_0=(a,b)$ consists of two points. The outward normal is
\be
n(a)=-1,
\qquad
n(b)=1 .
\ee
The exterior harmonic extension is constant on each exterior interval,
so that the exterior Dirichlet energy vanishes and
\be
\int_{\mathbb R}dx\,\left|(f^{D_0})'(x)\right|^2
=
\int_a^b dx\, [f'(x)]^2 .
\ee
The general variance formula therefore gives precisely
\be
\langle{\mathcal L}_N^2\rangle_c
\simeq
\frac{1}{2\beta}
\int_a^b dx\, [f'(x)]^2 .
\ee

For the third cumulant, the normal jumps are
\be
[\partial_n f^{D_0}](a)=f'(a),
\qquad
[\partial_n f^{D_0}](b)=-f'(b).
\ee
Replacing the boundary integral by a sum over the two endpoints, the general
third cumulant formula gives
\be
\langle{\mathcal L}_N^3\rangle_c
\simeq
-\frac{1}{2N\beta^2}
\left[
\frac{f'(a)^3}{U''(a)}
-
\frac{f'(b)^3}{U''(b)}
\right],
\ee
in agreement with the direct calculation above.

Similarly, the fourth cumulant formula reduces correctly. In one dimension
there is no tangential part and no curvature contribution, while the exterior
Dirichlet-to-Neumann contribution vanishes. Moreover
\be
\partial_n U''(a)=-U'''(a),
\qquad
\partial_n U''(b)=U'''(b),
\ee
and $\Delta f=f''$. Hence the general fourth cumulant formula gives
\be
\begin{aligned}
\langle{\mathcal L}_N^4\rangle_c
\simeq
-\frac{1}{2\beta^3N^2}
\sum_{x=a,b}
\left[
4\frac{[\partial_n f^{D_0}](x)^3 f''(x)}{U''(x)^2}
+
\frac{[\partial_n f^{D_0}](x)^4\partial_n U''(x)}{U''(x)^3}
\right],
\end{aligned}
\ee
which is exactly the endpoint formula written above.

Let us finally relate this appendix to previous works. The variance for a
general smooth linear statistics function in the one-dimensional jellium model
was obtained in \cite{FlackMajumdarSchehrJelliumVariance}. Related observables,
such as truncated linear statistics, gap probabilities and full counting
statistics, were studied in
\cite{FlackMajumdarSchehrTruncatedOCP,FlackMajumdarSchehrFCSOCP}.
The rotationally invariant formulas of \cite{OurCoulomb} include the
$d=1$ case, but then rotational invariance means that the test function is even,
$f(x)=f(|x|)$, and the droplet is the symmetric interval $[-R,R]$.
Higher cumulants for the one-dimensional trapped Riesz gas, including the
Coulomb case as the value $k=-1$, were obtained in
\cite{LeDoussalSchehrRieszCumulants}, in particular for monomial test functions
in the long-range regime.
\\

\section{Connection with the obstacle problem}
\label{sec:obstacle}

The equilibrium problem for the Coulomb gas is closely related to the classical
obstacle problem. In its simplest form, and with our sign convention for the
Laplacian, the obstacle problem asks for a function \(u\) lying above a
prescribed obstacle \(h\), i.e. satisfying \(u\geq h\), with \(\Delta u=0\)
away from the coincidence set and \(\Delta u\leq0\) in the distributional
sense. Equivalently, it asks for the minimizer of a Dirichlet energy under the
constraint \(u\geq h\), with suitable boundary or growth conditions at infinity
\cite{SaffTotik1997,KinderlehrerStampacchia1980,Friedman1982,CaffarelliSalsa2005}.

The solution has two regions. In the non-coincidence region one has
\[
u>h,
\qquad
\Delta u=0,
\]
whereas in the coincidence region one has
\[
u=h.
\]
The boundary between these two regions is the free boundary.

In the Coulomb gas, let
\[
u({\bf x})
=
\int_D d^d{\bf y}\,\rho({\bf y})V_d(|{\bf x}-{\bf y}|)
\]
be the Coulomb potential generated by the equilibrium density, with
\[
\Delta V_d(|{\bf x}|)
=
-\Omega_d\delta({\bf x}).
\]
For instance, in \(d=2\),
\[
V_2(r)=-\log r,
\qquad
\Omega_2=2\pi .
\]
The Euler--Lagrange conditions for the equilibrium measure state that, for
some constant \(\lambda\),
\[
u({\bf x})+U({\bf x})=-\lambda
\qquad
{\bf x}\in D,
\]
and
\[
u({\bf x})+U({\bf x})\geq -\lambda
\qquad
{\bf x}\in \mathbb R^d .
\]
Thus \(u\) is constrained to lie above the obstacle
\[
h({\bf x})=-\lambda-U({\bf x}),
\]
and the droplet \(D\) is the coincidence set
\[
D=\{{\bf x}:u({\bf x})=h({\bf x})\}.
\]
The complement \(D^c\) is the non-coincidence region, where the potential is
harmonic:
\[
\Delta u=0
\qquad
{\bf x}\in D^c.
\]
Inside the droplet, however, \(u=h=-\lambda-U\), and therefore
\[
\Delta u=-\Delta U.
\]
On the other hand, by the definition of the Coulomb potential,
\[
\Delta u
=
-\Omega_d\rho
\qquad
{\bf x}\in D.
\]
Hence
\[
\rho({\bf x})
=
\frac{\Delta U({\bf x})}{\Omega_d},
\qquad
{\bf x}\in D.
\]
This is the usual obstacle-problem interpretation of the equilibrium measure:
the density is obtained by applying the Laplacian to the obstacle in the
coincidence set, and the droplet boundary is the corresponding free boundary.

For the perturbed potential
\[
U_s=U+\frac{s}{\beta}f,
\]
the obstacle becomes
\[
h_s({\bf x})=-\lambda_s-U_s({\bf x}),
\]
and the coincidence set becomes the perturbed droplet \(D_s\). Thus changing
\(s\) moves the free boundary. The shape derivatives used in the main text are
the infinitesimal form of this free-boundary motion, under the assumption that
the droplet remains smooth and that no topological transition occurs.

The velocity of the free boundary under smooth perturbations of the obstacle
has been studied rigorously by Serfaty and Serra
\cite{SerfatySerra}; our formula for the normal velocity is the
corresponding Coulomb-gas version, written in terms of the harmonic extension
of the perturbation \(f\).

\section{Julia's formula for the inverse map}
\label{app:inverseJulia}

We record a useful equivalent form of Julia's formula for the inverse Riemann
map. Let
\be
\varphi_s=\psi_s^{-1},
\qquad
z=\varphi_s(w),
\qquad
w\in\mathbb D^c .
\ee
Thus \(\varphi_s\) maps the exterior of the unit disk onto \(D_s^c\), and maps
\(\partial\mathbb D\) onto \(\partial D_s\). In particular, if
$
u=e^{i\theta}\in\partial\mathbb D,
$
then
$
t=\varphi_s(u)\in\partial D_s
$
is a parametrization of the moving boundary.

Differentiating the identity
$\psi_s(\varphi_s(w))=w$
with respect to \(s\), at fixed \(w\in\mathbb D^c\), gives
\be
\partial_s\psi_s(\varphi_s(w))
+
\psi_s'(\varphi_s(w))\,\partial_s\varphi_s(w)
=
0.
\ee
Since \(\psi_s'(\varphi_s(w))=1/\varphi_s'(w)\), we obtain
\be
\partial_s\log\psi_s(z)\big|_{z=\varphi_s(w)}
=
-\frac{\partial_s\varphi_s(w)}{w\,\varphi_s'(w)} .
\label{inversejuliaderivative}
\ee

We now rewrite Julia's formula \eqref{julia} in the \(w\)-plane. In the
boundary integral we change variables
\[
t=\varphi_s(u),
\qquad
u=e^{i\theta}\in\partial\mathbb D.
\]
Then
\be
\psi_s(t)=u,
\qquad
|\psi_s'(t)|=\frac{1}{|\varphi_s'(u)|},
\qquad
d\ell(t)=|\varphi_s'(u)|\,d\theta .
\ee
Using also the obstacle expression
\be
v_s
=
\frac{K_s}{\beta\Delta U_s},
\qquad
K_s=[\partial_n f^{D_s}],
\ee
where the jump is exterior minus interior, Julia's formula becomes
\be
-\frac{\partial_s\varphi_s(w)}{w\,\varphi_s'(w)}
=
\frac{1}{2\pi}
\int_0^{2\pi}d\theta\,
\frac{u+w}{u-w}\,
\frac{
K_s(\varphi_s(u))
}
{
\beta\Delta U_s(\varphi_s(u))\,|\varphi_s'(u)|
},
\qquad
u=e^{i\theta}.
\label{inversejuliaK}
\ee
This is the inverse-map form of Julia's formula.

Equivalently, we can express the jump \(K_s\) directly in the unit-disk
variables. For \(u\in\mathbb D^c\), write
\[
u=\rho e^{i\theta},
\qquad \rho>1,
\]
and define
\be
G_s(u,\bar u)
=
g_s^{\mathbb D}(u,\bar u)
-
f(\varphi_s(u),\overline{\varphi_s(u)}).
\label{defGsInverseJulia}
\ee
Here \(g_s^{\mathbb D}\) is the exterior harmonic extension in
\(\mathbb D^c\), while the second term is the original function \(f\), pulled
back by the inverse map. The two terms have the same boundary value on
\(\partial\mathbb D\), but their normal derivatives need not agree.

On \(\partial\mathbb D\), radial motion in the \(u\)-plane is mapped by
\(\varphi_s\) to normal motion in the \(z\)-plane. More precisely,
\[
\left|\partial_\rho\varphi_s(\rho e^{i\theta})\right|_{\rho=1}
=
|\varphi_s'(e^{i\theta})|,
\]
and therefore
\be
\partial_\rho G_s(u,\bar u)\big|_{\rho=1^+}
=
|\varphi_s'(u)|\,K_s(\varphi_s(u)),
\qquad
u\in\partial\mathbb D.
\label{radialGK}
\ee
Here \(\partial_\rho\) denotes the outward radial derivative in the exterior
disk \(\mathbb D^c\), evaluated at \(\rho=1^+\).

Using \eqref{radialGK}, the inverse Julia formula \eqref{inversejuliaK} can be
rewritten as
\be
-\frac{\partial_s\varphi_s(w)}{w\,\varphi_s'(w)}
=
\frac{1}{2\pi}
\int_0^{2\pi}d\theta\,
\frac{u+w}{u-w}\,
\frac{
\partial_\rho G_s(u,\bar u)\big|_{\rho=1^+}
}
{
\beta\Delta U_s(\varphi_s(u))\,|\varphi_s'(u)|^2
},
\qquad
u=e^{i\theta}.
\label{inversejuliaradial}
\ee

Finally, one may also write this in complex notation. On
\(\mathbb D^c\), with \(u=\rho e^{i\theta}\), one has
\be
\partial_\rho
=
u\partial_u+\bar u\partial_{\bar u}
\qquad \text{on } \rho=1,
\ee
where
\[
\partial_u=\frac12\left(\partial_{u_1}-i\partial_{u_2}\right),
\qquad
\partial_{\bar u}=\frac12\left(\partial_{u_1}+i\partial_{u_2}\right).
\]
Moreover, along \(\partial\mathbb D\), \(du=iud\theta\), and hence
\be
{\rm Im}\!\left(du\,\partial_uG_s\right)
=
\frac12
\left(
u\partial_u+\bar u\partial_{\bar u}
\right)G_s\,d\theta
=
\frac12\,\partial_\rho G_s\,d\theta .
\ee
Therefore \eqref{inversejuliaradial} is equivalently
\be
-\frac{\partial_s\varphi_s(w)}{w\,\varphi_s'(w)}
=
\frac{1}{\pi}
\oint_{\partial\mathbb D}
\frac{u+w}{u-w}\,
\frac{1}
{
\beta\Delta U_s(\varphi_s(u))\,|\varphi_s'(u)|^2
}
\,{\rm Im}\!\left(du\,\partial_uG_s(u,\bar u)\right).
\label{inversejuliacomplex}
\ee
In \eqref{inversejuliacomplex}, the boundary value of
\(\partial_uG_s\) is understood as the exterior limit from \(\mathbb D^c\).

\section{Green function and Hadamard's formula}
\label{app:greenhadamard}

We recall a Green-function formulation of the exterior harmonic extension and
of its shape derivative. This gives another way to view the boundary operators
appearing in the main text. The use of Green functions and Hadamard
variational formulae in domain perturbation theory is classical; see for
example \cite{Schiffer1946,BergmanSchiffer1953,SokolowskiZolesio1992,
HenrotPierre2018}.

Let \(D\subset\mathbb R^d\) be a bounded smooth domain, and let \({\bf n}\) be
the unit normal on \(\partial D\), pointing outward from \(D\) to \(D^c\). We
denote by \(G_D({\bf x},{\bf y})\) the Dirichlet Green function in the exterior
domain \(D^c\), normalized by
\be
\Delta_{\bf x}G_D({\bf x},{\bf y})
=
-\Omega_d\delta({\bf x}-{\bf y}),
\qquad
{\bf x},{\bf y}\in D^c,
\label{GreenDirichlet}
\ee
with
\be
G_D({\bf x},{\bf y})=0,
\qquad
{\bf x}\in\partial D.
\ee
{At infinity we impose decay for $d>2$ and boundedness for $d=2$.}
It is symmetric,
\be
G_D({\bf x},{\bf y})=G_D({\bf y},{\bf x}).
\ee

The associated exterior Poisson kernel is
\be
P_D({\bf x},\xi)
=
\frac{1}{\Omega_d}\partial_{n_\xi}G_D({\bf x},\xi),
\qquad
{\bf x}\in D^c,
\qquad
\xi\in\partial D,
\label{PoissonKernelGreen}
\ee
where \(\partial_{n_\xi}\) is taken with respect to the normal pointing from
\(D\) to \(D^c\). If \(h\) is a boundary function, then
\[
\int_{\partial D}dS(\xi)\,h(\xi)P_D({\bf x},\xi)
\]
is the exterior harmonic function with boundary value \(h\) and {with decay at infinity for $d>2$, or with boundedness for $d=2$.}

In \(d>2\), this decaying extension is not quite the convention used in the
main text, because it gives a monopole term for constant boundary data. In the
paper we use the no-monopole convention: constants remain constants at
infinity. Thus the harmonic extension used in the main text may be written as
\be
h^D({\bf x})
=
h_\infty
+
\int_{\partial D}dS(\xi)\,
\big(h(\xi)-h_\infty\big)P_D({\bf x},\xi),
\qquad
{\bf x}\in D^c,
\label{GreenHarmonicExtension}
\ee
where \(h_\infty\) is the unique constant for which the last term has no
monopole contribution at infinity. In \(d=2\), the same formula is understood
with the corresponding no-logarithm, or boundedness, condition at infinity. For
the unit ball, \(h_\infty\) is simply the uniform surface average of \(h\).

The exterior Dirichlet-to-Neumann operator used in the paper is
\be
\Lambda^+h
=
\partial_n^+h^D\big|_{\partial D}.
\label{GreenDtN}
\ee
Since the outward normal of the exterior domain \(D^c\) is \(-{\bf n}\) on the
inner boundary \(\partial D\), Green's identity gives
\be
\int_{D^c}d^d{\bf x}\,|\vec\nabla h^D|^2
=
-\int_{\partial D}dS\,h\,\Lambda^+h .
\label{GreenEnergyDtN}
\ee
Thus \(\Lambda^+\) is non-positive.

Hadamard's variational formula gives the shape derivative of the Green
function \cite{Schiffer1946,BergmanSchiffer1953,SokolowskiZolesio1992}. Let
\(D_\epsilon\) be obtained from \(D\) by moving the boundary point
\(\xi\in\partial D\) to
\[
\xi+\epsilon p(\xi){\bf n}(\xi).
\]
Thus \(p>0\) means that the droplet \(D\) expands and the exterior domain
shrinks. Then, for fixed \({\bf x},{\bf y}\in D^c\),
\be
\left.
\partial_\epsilon G_{D_\epsilon}({\bf x},{\bf y})
\right|_{\epsilon=0}
=
-\frac{1}{\Omega_d}
\int_{\partial D}dS(\xi)\,
p(\xi)\,
\partial_{n_\xi}G_D({\bf x},\xi)\,
\partial_{n_\xi}G_D({\bf y},\xi).
\label{HadamardGreen}
\ee
The minus sign is a consequence of our normal convention: for \(p>0\), the
exterior domain is reduced.

This formula is another way to understand the basic boundary condition used in
the main derivation. 
Let \(D_s\) move with normal velocity \(v_s\), and let
\(h_{\rm ext}^{D_s}\) denote the exterior harmonic extension of a fixed smooth
bulk function \(h\). We also write \(h^{D_s}\) for the function equal to \(h\)
inside \(D_s\) and equal to \(h_{\rm ext}^{D_s}\) in \(D_s^c\).
The boundary condition is
\be
h_{\rm ext}^{D_s}=h
\qquad\text{on }\partial D_s.
\ee
Differentiating this condition along the moving boundary gives
\be
\left(\partial_s h_{\rm ext}^{D_s}\right)^+
+
v_s\partial_n^+h^{D_s}
=
v_s\partial_n^-h^{D_s}.
\ee
Therefore
\be
\left(\partial_s h_{\rm ext}^{D_s}\right)^+
=
-v_s
\left(
\partial_n^+h^{D_s}
-
\partial_n^-h^{D_s}
\right)
=
-v_s[\partial_n h^{D_s}]
\qquad\text{on }\partial D_s.
\label{GreenBoundaryShapeDerivative}
\ee
This is precisely the boundary shape-derivative identity used in the main text.

Hadamard's formula could in principle be used to compute the shape derivatives
of the Dirichlet energy, and hence the higher cumulants, directly from the
Green-function representation. However, this route involves boundary limits of
normal derivatives of \(G_D\), leading to hypersingular kernels on
\(\partial D\). Making the resulting cancellations and finite-part
regularizations systematic would be interesting, but we leave this approach for
future work.

For the unit ball, the Green-function representation gives an explicit form of
the exterior Dirichlet-to-Neumann operator. Writing points on the unit sphere
as \(\chi,\xi\in S^{d-1}\), one obtains the singular-integral representation
\be
\Lambda^+h(\chi)
=
(2-d)\big(h(\chi)-\bar h\big)
+
\frac{2}{\Omega_d}\,
{\rm PV}\int_{S^{d-1}}dS(\xi)\,
\frac{h(\xi)-h(\chi)}{|\xi-\chi|^d},
\label{SphereDtNSingular}
\ee
where
\be
\bar h
=
\frac{1}{\Omega_d}
\int_{S^{d-1}}dS\,h .
\ee
The integral is understood in the principal-value sense. The subtraction
\(h(\xi)-h(\chi)\) removes the leading singularity: the linear term is odd and
cancels in principal value, while the quadratic remainder gives a locally
integrable singularity.

Equivalently, if \(Y_{\ell m}\) denotes a spherical harmonic of degree
\(\ell\), then
\be
\Lambda^+Y_{\ell m}
=
-(\ell+d-2)Y_{\ell m},
\qquad
\ell\geq1,
\qquad
\Lambda^+Y_{00}=0.
\label{SphereDtNSpectral}
\ee
For \(d=2\), this reduces to the familiar unit-circle formula
\be
\Lambda^+
=
-H\partial_\phi .
\ee
Thus the Hilbert transform on the circle { (combined with angular derivative)} is the two-dimensional version of the exterior Dirichlet-to-
Neumann operator.

\section{Variation of the surface element} \label{app:jacobian} 

For completeness, we recall the derivation of the variation formula for the
surface element. Let \(X_s({\bf y})\), with
\({\bf y}=(y^1,\ldots,y^{d-1})\), be a local parametrization of
\(\partial D_s\). We assume that the boundary moves with normal velocity
\(v_s\), namely
\be
\partial_s X_s({\bf y})
=
v_s(X_s({\bf y}))\,{\bf n}_s(X_s({\bf y})) .
\label{normalvelocityappendix}
\ee
Let
\be
e_i=\partial_i X_s,
\qquad
g_{ij}=e_i\cdot e_j,
\qquad
J_s=\sqrt{\det g},
\label{surfacegeomappendix}
\ee
so that
\be
d\sigma_s(X_s({\bf y}))=J_s({\bf y})\,d^{d-1}{\bf y}.
\ee
Differentiating \(g_{ij}\) with respect to \(s\), we obtain
\be
\partial_s g_{ij}
=
\partial_s e_i\cdot e_j
+
e_i\cdot\partial_s e_j
=
\partial_i(v_s{\bf n}_s)\cdot e_j
+
e_i\cdot\partial_j(v_s{\bf n}_s).
\ee
Since \({\bf n}_s\cdot e_i=0\), the terms containing tangential derivatives
of \(v_s\) vanish. Therefore
\be
\partial_s g_{ij}
=
v_s
\left(
\partial_i{\bf n}_s\cdot e_j
+
\partial_j{\bf n}_s\cdot e_i
\right).
\ee
With our sign convention for the second fundamental form,
\be
h_{ij}
=
\partial_i{\bf n}_s\cdot e_j,
\label{secondfundamentalappendix}
\ee
this becomes
\be
\partial_s g_{ij}=2v_s h_{ij}.
\label{metricvariationappendix}
\ee
Using the identity
\be
\partial_s \sqrt{\det g}
=
\frac12\sqrt{\det g}\,g^{ij}\partial_s g_{ij},
\ee
we find
\be
\partial_s J_s
=
v_s J_s\,g^{ij}h_{ij}.
\ee
Finally, with the outward-normal convention, the mean curvature is
\be
M_s
=
g^{ij}h_{ij}
=
\vec\nabla_{\partial D_s}\cdot{\bf n}_s .
\label{meancurvatureappendix}
\ee
Hence
\be
\partial_s J_s({\bf y})
=
M_s(X_s({\bf y}))\,v_s(X_s({\bf y}))\,J_s({\bf y}) .
\label{surfacejacobianappendix}
\ee
Equivalently,
\be
\partial_s d\sigma_s
=
M_s v_s\,d\sigma_s .
\label{surfaceelementappendix}
\ee
This is the formula used in the derivation of the fourth cumulant. With this
convention \(M_s=(d-1)/R\) for a sphere of radius \(R\) with outward normal,
so that an outward displacement \(v_s>0\) increases the surface element, as it
should.

\section{A Fourier check in the half-plane}
\label{app:halfspacefouriercheck}

We give an independent check of the third and fourth cumulants in the
half-plane geometry. This calculation is simpler than the general-domain
derivation because the boundary is flat at \(s=0\). One can therefore
parametrize the perturbed boundary as a graph and use the fixed Fourier basis
of the half-plane. In particular, the exterior Dirichlet-to-Neumann operator is
just the Fourier multiplier \(-H\partial_y\).

We take
\[
D_0=\{x<0\},
\qquad
\partial D_0=\{x=0\},
\qquad
{\bf n}={\bf e}_x,
\]
and assume that \(U(x,y)=\tilde U(x)\). We write
\[
\Delta U=\tilde U''(0),
\qquad
\partial_n\Delta U=\tilde U'''(0).
\]
The boundary is written as
\be
x=x_s(y),
\qquad
x_s(y)=s\,v(y)+\frac{s^2}{2}\,\alpha(y)+O(s^3),
\label{halfspacegraph}
\ee
where \(v=x'_0\) is the normal velocity and \(\alpha=x''_0\).

Let
\be
{\sf f}(y)=f(0,y),
\qquad
{\sf f}(y)=\int_{\mathbb R}\frac{dq}{2\pi}\,\hat{\sf f}_q e^{iqy}.
\ee
The exterior harmonic extension at \(s=0\) is
\be
F_0(x,y)
=
f^{D_0}(x,y)
=
\int_{\mathbb R}\frac{dq}{2\pi}\,
\hat{\sf f}_q e^{iqy-|q|x},
\qquad x>0.
\label{halfspaceF0}
\ee
Thus
\be
\partial_x^+F_0(0,y)
=
-H\partial_y{\sf f}(y),
\qquad
\widehat{Hg}_q=-i\,{\rm sign}(q)\,\hat g_q.
\label{halfspaceDtN}
\ee
We introduce the boundary quantity
\be
g(y)
=
\partial_x f(0,y)+H\partial_y{\sf f}(y).
\label{halfspaceg}
\ee
With our jump convention, exterior minus interior, one has
\be
K=[\partial_n f^{D_0}]
=
\partial_x^+F_0-\partial_x^-f
=
-g.
\label{halfspaceKg}
\ee
Hence the obstacle formula gives
\be
v(y)
=
\frac{K}{\beta\Delta U}
=
-\frac{g(y)}{\beta\Delta U}.
\label{halfspacevelocity}
\ee

We now define the dot notation used below. Let \(F_s=f^{D_s}\) denote the
exterior harmonic extension in \(D_s^c\). We write
\be
\dot F(x,y)
=
\left.\partial_sF_s(x,y)\right|_{s=0},
\qquad
\ddot F(x,y)
=
\left.\partial_s^2F_s(x,y)\right|_{s=0},
\qquad x>0,
\label{halfspacedotdef}
\ee
where the derivatives are taken at fixed point \((x,y)\) in the fixed
half-plane \(x>0\). The functions \(\dot F\) and \(\ddot F\) are harmonic in
\(x>0\). Expanding the boundary condition
\[
F_s(x_s(y),y)=f(x_s(y),y)
\]
to first order gives
\be
\dot F(0,y)
=
v(y)\left(
\partial_x f(0,y)-\partial_x^+F_0(0,y)
\right)
=
v(y)g(y).
\label{halfspacedotboundary}
\ee
Equivalently, using \eqref{halfspacevelocity},
\[
\dot F(0,y)=-\frac{g(y)^2}{\beta\Delta U}.
\]

Let
\be
E(s)
=
\int_{D_s}d^2{\bf x}\,|\vec\nabla f|^2
+
\int_{D_s^c}d^2{\bf x}\,|\vec\nabla f^{D_s}|^2.
\label{halfspaceE}
\ee
Since
\[
\frac{1}{N^2}\partial_s^2\chi(s,N)
=
\frac{1}{2\pi\beta}E(s),
\]
we have
\be
\langle{\mathcal L}_N^3\rangle_c
=
-\frac{1}{2\pi N\beta}E'(0),
\qquad
\langle{\mathcal L}_N^4\rangle_c
=
\frac{1}{2\pi\beta N^2}E''(0).
\label{halfspacecumsfromE}
\ee

Let us first compute \(E'(0)\). Differentiating the two domains gives
\be
E'(0)
=
\int_{\mathbb R}dy\,
v\left(
|\vec\nabla f|^2-|\vec\nabla F_0|^2
\right)_{x=0}
+
2\int_0^\infty dx\int_{\mathbb R}dy\,
\vec\nabla F_0\cdot\vec\nabla\dot F .
\label{halfspaceEprimefirst}
\ee
The tangential derivatives agree on the boundary, so the first term reduces to
\[
\int_{\mathbb R}dy\,
v\left[
(\partial_x f)^2
-
(\partial_x^+F_0)^2
\right]_{x=0}.
\]
Using \eqref{halfspaceDtN}, \eqref{halfspacedotboundary}, and the
self-adjointness of \(H\partial_y\), the second term gives
\[
2\int_{\mathbb R}dy\,
(H\partial_y{\sf f})\,v\,g .
\]
Combining the two contributions, one obtains the simple identity
\be
E'(0)
=
\int_{\mathbb R}dy\,v(y)g(y)^2
=
-\frac{1}{\beta\Delta U}
\int_{\mathbb R}dy\,g(y)^3.
\label{halfspaceEprime}
\ee
Therefore
\be
\langle{\mathcal L}_N^3\rangle_c
\simeq
\frac{1}{2\pi N\beta^2\Delta U}
\int_{\mathbb R}dy\,g(y)^3,
\label{halfspacecum3check}
\ee
which agrees with the half-plane formula stated in the main text.

We now turn to the fourth cumulant. Expanding the boundary condition to second
order gives the boundary value of \(\ddot F\):
\be
\ddot F(0,y)
=
\alpha(y)g(y)
+
v(y)^2\Delta f(0,y)
+
2v(y)H\partial_y\!\left(vg\right)(y).
\label{halfspacedoubleboundary}
\ee
Here we used that \(F_0\) and \(\dot F\) are harmonic in \(x>0\), so that
\[
\partial_x^2F_0(0,y)=-\partial_y^2{\sf f}(y),
\qquad
\partial_x\dot F(0,y)=-H\partial_y(\dot F(0,\cdot))(y).
\]
Differentiating the velocity relation \eqref{halfspacevelocity} once more,
one finds
\be
\alpha(y)
=
H\partial_y(v^2)(y)
-
\frac{2v(y)}{\beta\Delta U}\Delta f(0,y)
-
\frac{\tilde U'''(0)}{\Delta U}v(y)^2.
\label{halfspacealpha}
\ee
The first term is the variation of the exterior Dirichlet-to-Neumann part, 
the second combines the contributions of $\Delta f$ from the variation
of the normal-derivative jump and from the explicit $s$-dependence of
$\Delta U_s$, the last comes from the normal derivative of $\Delta U_s$
along the moving boundary.

A direct differentiation of \eqref{halfspaceE}, using
\eqref{halfspacedoubleboundary} and \eqref{halfspacealpha}, gives after
cancellations
\be
E''(0)
=
4\int_{\mathbb R}dy\,a\,v^3\Delta f
+
3a^2\int_{\mathbb R}dy\,v^2H\partial_y(v^2)
+
\beta\tilde U'''(0)
\int_{\mathbb R}dy\,a\,v^4,
\label{halfspaceEseconda}
\ee
where
\[
a=-\beta\Delta U,
\qquad
g=av.
\]
Equivalently, in terms of the boundary field \(g\), this becomes
\be
E''(0)
=
\frac{1}{\beta^2(\Delta U)^2}
\int_{\mathbb R}dy\,
\left[
3g^2H\partial_y(g^2)
+
4g^3\Delta f(0,y)
-
\frac{\tilde U'''(0)}{\Delta U}g^4
\right].
\label{halfspaceEsecondg}
\ee
Using \eqref{halfspacecumsfromE}, we obtain
\be
\langle{\mathcal L}_N^4\rangle_c
\simeq
\frac{1}{2\pi\beta^3N^2(\Delta U)^2}
\int_{\mathbb R}dy\,
\left[
3g^2H\partial_y(g^2)
+
4g^3\Delta f(0,y)
-
\frac{\tilde U'''(0)}{\Delta U}g^4
\right],
\label{halfspacecum4check}
\ee
again in agreement with the half-plane formula in the main text.

As a final check,  if \(f=f(x)\) is independent of \(y\), then
\(H\partial_y{\sf f}=0\), \(g=f'(0)\), and the nonlocal term vanishes. The
formula reduces to
\be
\langle{\mathcal L}_N^4\rangle_c
\simeq
\frac{1}{2\pi\beta^3N^2}
\int_{\mathbb R}dy\,
\left[
\frac{4f'(0)^3f''(0)}{\tilde U''(0)^2}
-
\frac{\tilde U'''(0)f'(0)^4}{\tilde U''(0)^3}
\right],
\ee
which also follows directly from the one-dimensional motion of the flat boundary.

\section{Perturbative harmonic extension around the disk}
\label{app:perturbativeharmonicdisk}

We record an elementary perturbative construction of the exterior harmonic
extension near a disk. Although it is not needed in the main derivation, it is
a useful way to compute shape perturbations explicitly. The method is specific
to perturbations of a circular droplet and should be distinguished from the
general-domain shape-derivative calculation.

Let the boundary of \(D_s\) be written as a radial graph
\be
z_s(\phi)=R_s(\phi)e^{i\phi},
\qquad
R_s(\phi)=R_0+s r_1(\phi)+O(s^2).
\label{pertboundarydisk}
\ee
At \(s=0\), the radial displacement \(r_1(\phi)\) is also the normal
displacement, since the unperturbed boundary is a circle.

Let \(u_s\) denote the exterior harmonic extension of \(f\) in \(D_s^c\). We
impose the boundedness, or no-logarithm, condition at infinity. In the exterior
of a perturbation of the disk, \(u_s\) may be represented perturbatively by the
multipole expansion
\be
u_s(r,\phi)
=
\sum_{n\in\mathbb Z} b_n(s)e^{in\phi}r^{-|n|}.
\label{pertmultipole}
\ee
For real \(f\), one has
\[
b_{-n}(s)=\overline{b_n(s)}.
\]
The boundary condition is
\be
\sum_{n\in\mathbb Z}
b_n(s)e^{in\phi}R_s(\phi)^{-|n|}
=
f(R_s(\phi),\phi).
\label{pertboundarycondition}
\ee

At \(s=0\), write
\be
{\sf f}(\phi)=f(R_0,\phi)
=
\sum_{n\in\mathbb Z}\hat f_n e^{in\phi},
\qquad
\hat f_n
=
\int_0^{2\pi}\frac{d\phi}{2\pi}\,
e^{-in\phi}f(R_0,\phi).
\label{pertboundaryfourier}
\ee
Then \eqref{pertboundarycondition} gives
\be
b_n(0)=R_0^{|n|}\hat f_n.
\label{pertbnzero}
\ee

We now expand
\be
b_n(s)
=
R_0^{|n|}\hat f_n
+
s b_n^{(1)}
+
O(s^2).
\label{pertbnexpansion}
\ee
Expanding \eqref{pertboundarycondition} to first order in \(s\), we obtain
\be
\sum_{n\in\mathbb Z}
e^{in\phi}
\left[
b_n^{(1)}R_0^{-|n|}
-
\frac{|n|}{R_0}\hat f_n r_1(\phi)
\right]
=
r_1(\phi)\partial_r f(R_0,\phi).
\label{pertlinearboundary}
\ee
Equivalently,
\be
\sum_{n\in\mathbb Z}
b_n^{(1)}R_0^{-|n|}e^{in\phi}
=
r_1(\phi)
\left[
\partial_r f(R_0,\phi)
+
\frac{1}{R_0}H\partial_\phi{\sf f}(\phi)
\right].
\label{pertlinearboundaryJ}
\ee
Here we used
\[
H\partial_\phi{\sf f}
=
\sum_{n\in\mathbb Z}|n|\hat f_n e^{in\phi}.
\]
Thus, if
\be
J(\phi)
=
\partial_r f(R_0,\phi)
+
\frac{1}{R_0}H\partial_\phi{\sf f}(\phi),
\label{pertJdisk}
\ee
then the first correction \(u_1=\partial_su_s|_{s=0}\) has boundary value
\be
u_1(R_0,\phi)=r_1(\phi)J(\phi).
\label{pertu1boundary}
\ee
This is the disk version of the general shape-derivative identity
\[
\left(\partial_s f_{\rm ext}^{D_s}\right)^+
=
-v[\partial_n f^{D_s}],
\]
since, for the disk,
\[
[\partial_n f^{D_0}]=-J,
\qquad
v=r_1.
\]

Let
\be
r_1(\phi)=\sum_{n\in\mathbb Z}\hat r_n e^{in\phi},
\qquad
\partial_r f(R_0,\phi)
=
\sum_{n\in\mathbb Z}\hat f^{(1)}_n e^{in\phi}.
\ee
Taking the Fourier coefficient of \eqref{pertlinearboundaryJ} gives
\be
b_n^{(1)}
=
R_0^{|n|}
\sum_{m\in\mathbb Z}
\hat r_{n-m}
\left[
\hat f^{(1)}_m
+
\frac{|m|}{R_0}\hat f_m
\right].
\label{pertbnfirstcompact}
\ee
Equivalently,
\be
b_n^{(1)}
=
R_0^{|n|-1}
\sum_{m\in\mathbb Z}
|m|\hat f_m\hat r_{n-m}
+
R_0^{|n|}
\sum_{m\in\mathbb Z}
\hat f^{(1)}_m\hat r_{n-m}.
\label{pertbnfirst}
\ee
Therefore the first-order correction to the harmonic extension is
\be
u_1(r,\phi)
=
\sum_{n\in\mathbb Z}
b_n^{(1)}e^{in\phi}r^{-|n|}.
\label{pertuone}
\ee

One may also formulate the perturbation theory to all orders as an infinite
linear system. Let
\be
F_m(s)
=
\int_0^{2\pi}\frac{d\phi}{2\pi}\,
e^{-im\phi}f(R_s(\phi),\phi).
\ee
The boundary condition \eqref{pertboundarycondition} becomes
\be
F_m(s)
=
\sum_{n\in\mathbb Z}
A_{mn}(s)b_n(s),
\qquad
A_{mn}(s)
=
\int_0^{2\pi}\frac{d\phi}{2\pi}\,
e^{-i(m-n)\phi}R_s(\phi)^{-|n|}.
\label{pertmatrixsystem}
\ee
Thus formally
\be
b_n(s)
=
\sum_{m\in\mathbb Z}
\left(A(s)^{-1}\right)_{nm}F_m(s).
\label{pertmatrixinverse}
\ee
Since
\[
A_{mn}(0)=R_0^{-|n|}\delta_{mn},
\]
it is often cleaner to introduce the rescaled coefficients
\[
c_n(s)=R_0^{-|n|}b_n(s).
\]
Then the boundary condition becomes
\be
F_m(s)
=
\sum_{n\in\mathbb Z}
\widetilde A_{mn}(s)c_n(s),
\qquad
\widetilde A_{mn}(s)
=
\int_0^{2\pi}\frac{d\phi}{2\pi}\,
e^{-i(m-n)\phi}
\left(\frac{R_0}{R_s(\phi)}\right)^{|n|},
\label{pertrescaledmatrix}
\ee
and now
\[
\widetilde A_{mn}(0)=\delta_{mn}.
\]
The first-order formula \eqref{pertbnfirst} is obtained by expanding
\(\widetilde A(s)^{-1}\) around the identity.

\section{Proof of a Hilbert-transform identity}
\label{app:Hilbertidentity}

We prove the identity
\be
H(aHb+bHa)
=
(Ha)(Hb)-ab
\label{Hilbertidentityappendix}
\ee
for functions \(a,b\) with zero average. More generally, one has
\be
H(aHb+bHa)
=
(Ha)(Hb)-ab+\langle a\rangle\langle b\rangle .
\label{Hilbertidentitygeneralappendix}
\ee

We use the Fourier convention
\[
a(\phi)=\sum_{k\in\mathbb Z}a_k e^{ik\phi},
\qquad
b(\phi)=\sum_{k\in\mathbb Z}b_k e^{ik\phi},
\]
and
\[
\widehat{Hg}_k=-i\,{\rm sgn}(k)\,\hat g_k.
\]
For brevity, set \(s_k={\rm sgn}(k)\), with \(s_0=0\). The \(k\)-th Fourier
coefficient of the left-hand side of \eqref{Hilbertidentitygeneralappendix}
is
\be
\left[H(aHb+bHa)\right]_k
=
-\sum_{k_1,k_2}
s_k(s_{k_1}+s_{k_2})
a_{k_1}b_{k_2}
\delta_{k,k_1+k_2}.
\ee
The \(k\)-th Fourier coefficient of \((Ha)(Hb)-ab\) is
\be
\left[(Ha)(Hb)-ab\right]_k
=
-\sum_{k_1,k_2}
\left(1+s_{k_1}s_{k_2}\right)
a_{k_1}b_{k_2}
\delta_{k,k_1+k_2}.
\ee
Thus it remains to compare
\[
s_k(s_{k_1}+s_{k_2})
\qquad\text{and}\qquad
1+s_{k_1}s_{k_2},
\qquad k=k_1+k_2 .
\]
If \(k_1\) and \(k_2\) have opposite signs, then
\[
s_{k_1}+s_{k_2}=0,
\qquad
1+s_{k_1}s_{k_2}=0,
\]
so the two expressions agree. If \(k_1\) and \(k_2\) have the same non-zero
sign, then \(s_k=s_{k_1}=s_{k_2}\), and both expressions are equal to \(2\).
If one of \(k_1,k_2\) is zero and the other is non-zero, both expressions are
equal to \(1\). The only exceptional case is
\[
k_1=k_2=k=0.
\]
In that case
\[
s_k(s_{k_1}+s_{k_2})=0,
\qquad
1+s_{k_1}s_{k_2}=1.
\]
Therefore the two sides differ only by the constant Fourier mode
\(a_0b_0=\langle a\rangle\langle b\rangle\), which proves
\eqref{Hilbertidentitygeneralappendix}. If \(a\) and \(b\) have zero average,
this constant term vanishes and \eqref{Hilbertidentityappendix} follows.

\end{document}